# Revisiting the Sunspot Number

*A 400-year perspective on the solar cycle*

Frédéric Clette
World Data Center SILSO, Observatoire Royal de Belgique, Brussels, Belgium

Leif Svalgaard
W.W. Hansen Experimental Physics Laboratory, Stanford University, Stanford, CA, USA

José M. Vaquero
Departamento de Física, Universidad de Extremadura, Mérida, Spain

Edward W. Cliver
National Solar Observatory, Sunspot, NM, USA

**Abstract:**

Our knowledge of the long-term evolution of solar activity and of its primary modulation, the 11-year cycle, largely depends on a single direct observational record: the visual sunspot counts that retrace the last 4 centuries, since the invention of the astronomical telescope. Currently, this activity index is available in two main forms: the International Sunspot Number initiated by R. Wolf in 1849 and the Group Number constructed more recently by Hoyt and Schatten (1998a,b). Unfortunately, those two series do not match by various aspects, inducing confusions and contradictions when used in crucial contemporary studies of the solar dynamo or of the solar forcing on the Earth climate. Recently, new efforts have been undertaken to diagnose and correct flaws and biases affecting both sunspot series, in the framework of a series of dedicated Sunspot Number Workshops. Here, we present a global overview of our current understanding of the sunspot number calibration.

After retracing the construction of those two composite series, we present the new concepts and methods used to self-consistently re-calibrate the original sunspot series. While the early part of the sunspot record before 1800 is still characterized by large uncertainties due to poorly observed periods, the more recent sunspot numbers are mainly affected by three main inhomogeneities: in 1880-1915 for the Group Number and in 1947 and 1980-2014 for the Sunspot Number.

After establishing those new corrections, we then consider the implications on our knowledge of solar activity over the last 400 years. The newly corrected series clearly indicates a progressive decline of solar activity before the onset of the Maunder Minimum, while the slowly rising trend of the activity after the Maunder Minimum is strongly reduced, suggesting that by the mid 18$^{th}$ century, solar activity had already returned to levels equivalent to those observed in recent solar cycles in the 20$^{th}$ century. We finally conclude with future prospects opened by this epochal revision of the Sunspot Number, the first one since Wolf himself, and its reconciliation with the Group Number, a long-awaited modernization that will feed solar cycle research into the 21$^{st}$ century.

## 1. Introduction: the sunspot number needs to be recalibrated

The sunspot number (SN) time series is the only direct record at our disposal to retrace the long-term evolution of the solar cycle and of the probable long-term influence of the Sun on the Earth environment. Therefore, it was and is still used as a key information in many fields of research, quite obviously in solar physics, but also in climate studies or even stock-market modeling. The relative sunspot number, as defined by R. Wolf (1851, 1856), is based on the total number of sunspots Ns and the number of sunspot groups Ng according to the well-known formula:

$$R = k\,(10 \times Ng + Ns) \qquad (1)$$

The k scaling coefficient, usually called the personal coefficient of the observer, allows compensating for the differences in the number of recorded sunspots by different observers. It depends mainly on the ability of the observer to detect the smallest sunspots (telescope aperture, local seeing, personal experience) and on how groups are split by the observer. As Wolf was the primary observer for the newly-created sunspot number, his k personal coefficient was set to 1, which defines the scale of the whole series. The sunspot number is thus a synthetic index defined on an absolute but arbitrary scale (no physical unit), which is why Wolf called it “relative”.

In most analyses and publications, the sunspot number series is assumed to be carved in stone, i.e. it is considered largely as a homogeneous, well-understood and thus immutable data set. This feeling was probably reinforced by the stately process through which it was produced by a single expert center at the Zürich Observatory during 131 years, from 1849 to 1980 (Waldmeier 1961).

Still, since the mid-20$^{th}$ century, e.g., with the introduction of the American sunspot number by Alan Shapley, the accuracy and validity of such a visual index has been regularly questioned (Shapley 1949). During the 1970's, the SN series even went through a crisis of confidence, with the advent of new modern measurements of solar activity like the $F_{10.7cm}$ background radio flux. The objectivity of such measurements was contrasted with the subjectivity of purely visual and manual sunspot counts, which at that time almost led to the termination of the SN production, considered as unreliable and old-fashioned. Fortunately, action took place in different Commissions of the URSI and IAU, ensuring the continuation of the series after the transfer of the World Data Center for sunspots from Zürich to Brussels in 1981 (For an historical account of this transition, see Berghmans et al. 2006, Clette et al. 2007).

Since then, this wave of skepticism has receded thanks to a few key findings. First, studies of the AAVSO data series and of its statistical method (Hossfield 2002, Schaefer 1997a, 1997b) showed that the differences between the American SN and the Zürich-International SN could be traced to flaws in the American SN and that both series could be brought to a close match after correction. Moreover, with the advent of new data series derived from advanced techniques like space-based solar spectral irradiances (Lyα, MgII core-to-wing ratio, etc.) or automated image-based feature recognition (e.g. sunspot areas, flare detection), statistical tests accumulated showing the high degree of correlation between the traditional SN and those modern impersonal indicators (e.g., Bachmann et al. 2004, Rybansky et al. 2005, Wilson and Hathaway 2006, Tapping et al. 2007, Bertello et al. 2010, Hempelmann and Weber 2012, Stenflo 2012). This high correlation indicates that sunspot and group counts give an accurate measurement of the emergence rate of the toroidal magnetic flux at the solar surface by the action of the subsurface dynamo process (Stenflo 2012, Petrovay 2010).

However, when considering the accuracy of the SN, the fact that this index is distributed as a single time series is misleading, as most users assume that the statistical properties of the series are constant with time. As we will show later in this paper, the SN series was actually built from successive blocks using different base data and processing techniques, often with rather abrupt transitions between them. Over the past decades, only a few authors delved into the series and its base input data. Most of those studies focused on specific time intervals or on single base observers. Those valuable efforts led to proposed revisions of specific segments of the SN record (Kopecky et al. 1980, Letfus 1993, Vaquero et al. 2011, Leussu et al. 2013, Lockwood et al. 2014) and to the still ongoing debate about a missing short cycle between cycles 4 and 5 (Usoskin et al. 2001, 2009, Arlt 2009, Zolotova and Ponyavin 2007, 2011). Overall, all those proposed corrections were mostly local in time and were dispersed over multiple publications. Moreover, they also often lacked an independent validation and are still a topic of scientific controversy. Therefore, until now, none of them were included in the master SN series.

The main effort undertaken in recent times was actually the production of an entirely new series: the group number (hereafter GN; Hoyt and Schatten 1998a, 1998b) that will be discussed in section 3 of this chapter (Figure 1). This work involved a revision of the original data used for the SN and the recovery of many additional observations, in particular in the early period, between the first telescopic observation in 1610 and the start of the systematic sunspot census initiated by Wolf in 1849. However, as we will show in more detail in section 3, the new series showed immediately a strong discrepancy with the SN before 1880, a disagreement that remained unexplained since the GN publication (Figure 1).

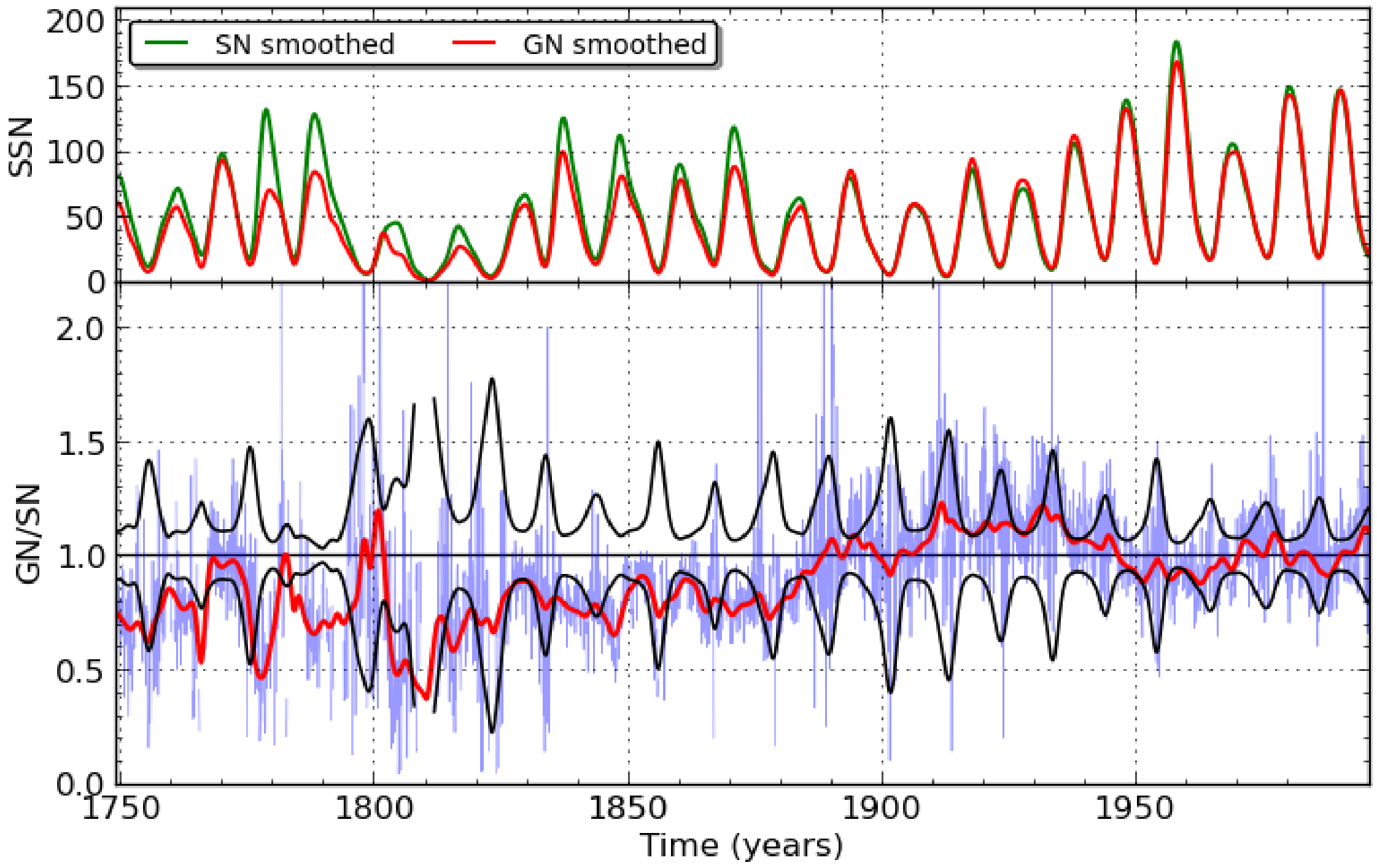

Figure 1: Top panel: time series of the SN (green) and GN (red), processed with a 12-month Gaussian smoothing. Middle panel: the ratio GN/SN (blue curve) and the same with a 12-month Gaussian smoothing (red). Confidence intervals (black) are based on uncertainties given by Hoyt and Schatten (1998a). The ratio deviates significantly and systematically from unity before ~1880.

This left the users of the sunspot number series in certain confusion, having to choose between two apparently equivalent and interchangeable but disagreeing series. In many cases, authors have settled for the GN series when studying phenomena before the mid-19$^{th}$ century, mainly assuming that the GN is then more reliable, as it benefits from a wider observation base and was derived from a single well-documented compilation. As this early period includes the Maunder Minimum (1645-1715; Spörer 1887, Eddy 1976), the GN number played a key role in many recent reconstructions of solar outputs (irradiance, solar wind, total open magnetic flux): e.g. Hathaway et al. 2002, Solanki and Krivova 2004, Wang et al. 2005, Krivova et al. 2007, Vieira et al. 2011, Shapiro et al. 2011 and Owens and Lockwood 2012. Hence, it was a key element in the resulting conclusions about the past solar forcing on the Earth climate since the Maunder Minimum. Given the importance of the reconstructed time series, the coexistence of two conflicting series is a highly unsatisfactory situation that should now be actively addressed.

Starting from this enduring problem, various studies have been recently undertaken on the initiative of the Sunspot Number Workshops (Cliver et al. 2013) to revisit the calibration of both series and as far as possible, to reconcile them by diagnosing the biases affecting them and identifying their underlying causes. By exploiting all the new sunspot data and the knowledge that were accumulated since Wolf and his successors established the official "heritage" SN series, those analyses have unraveled several major anomalies in both series.

For a large part, this chapter provides a review of the results harvested over the past three SN Workshops (Sacramento Peak, September 2011; Brussels, May 2012; Tucson, January 2013, with a 4th Workshop to be held in Locarno in May 2014). In section 2, we first synthesize the history of the construction of the SN and GN series, emphasizing the key eras and dates that can leave an imprint in the resulting index values. Then, we will analyze in chronological order different biases and trends identified in those series, providing the corresponding diagnostics of their causes. In section 3, we focus on the historical part of the series that was backward reconstructed, before the start of systematic observations in 1849. Section 4 is devoted to the Zürich era and finally section 5 focuses on the most recent part of the series, derived by a new method by the SIDC-Brussels. In this section, we will also consider some peculiarities of the last two solar cycles and discuss how they can help us better interpret past trends in the historical series. We then come to an overall discussion (section 6), bringing together the key corrections that have been established and assessing how the agreement of the SN with other data sets is improved after applying the corrections. As this is still work in progress, we finally conclude on the upcoming release of a fully revised SN series and on the possible implications of the corrected and reconciled SN and GN series on current solar issues.

# 2. The Sunspot Number in time

## *2.1. Wolf's historical sunspot number reconstruction*

Soon after Rudolph Wolf started the systematic census of sunspots in 1849, he turned to past observations in order to quickly extend his still very short series over several past solar cycles. First, he naturally turned to the very observations that triggered his interest in sunspots, namely the long record by Samuel Heinrich Schwabe (1789-1875), the discoverer of the (approximately) decadal cyclicity of the solar activity (Schwabe 1844). Schwabe was the most dedicated sunspot observer of his time (Johnson 1857, Hufbauer 1991, Cliver 2005), being active continuously from 1825 to 1868 (last preserved data in 1867). His original notes and sunspot drawings are preserved in the archives of the Royal Astronomical Society, London. Arlt (2011) provides a comprehensive inventory of the sunspot information from Schwabe's logbooks, which contain 8486 full-disk drawings with sunspots and 3699 additional verbal reports. It is interesting to note that, more than 140 years ago, De la Rue et al. (1869) used these drawings to estimate the time evolution of sunspot areas between 1832 and 1853 (Vaquero et al. 2002). As Schwabe was still observing in parallel with Wolf until 1867, his observations would later remain one of the main auxiliary set of observations used by Wolf to establish the daily Zürich sunspot number.

In 1857, Wolf extended further the historical reconstruction by using the longest continuous series of observations of the 18th century, produced by Johann Caspar Staudach (Wolf 1857, Svalgaard 2013a). Staudach, an amateur astronomer, made sunspot drawings from 15 February 1749 to 31 January 1796 (these drawings are currently stored in the library of the Leibniz Institute for Astrophysics Potsdam, Germany). There is a total of 1016 days giving sunspot positions, including dates with no spot observed. The average number of observations per year equals 21 but the distribution of observed dates is highly variable from year to year (Arlt 2008). Figure 2 shows two typical sunspot drawings by Staudach.

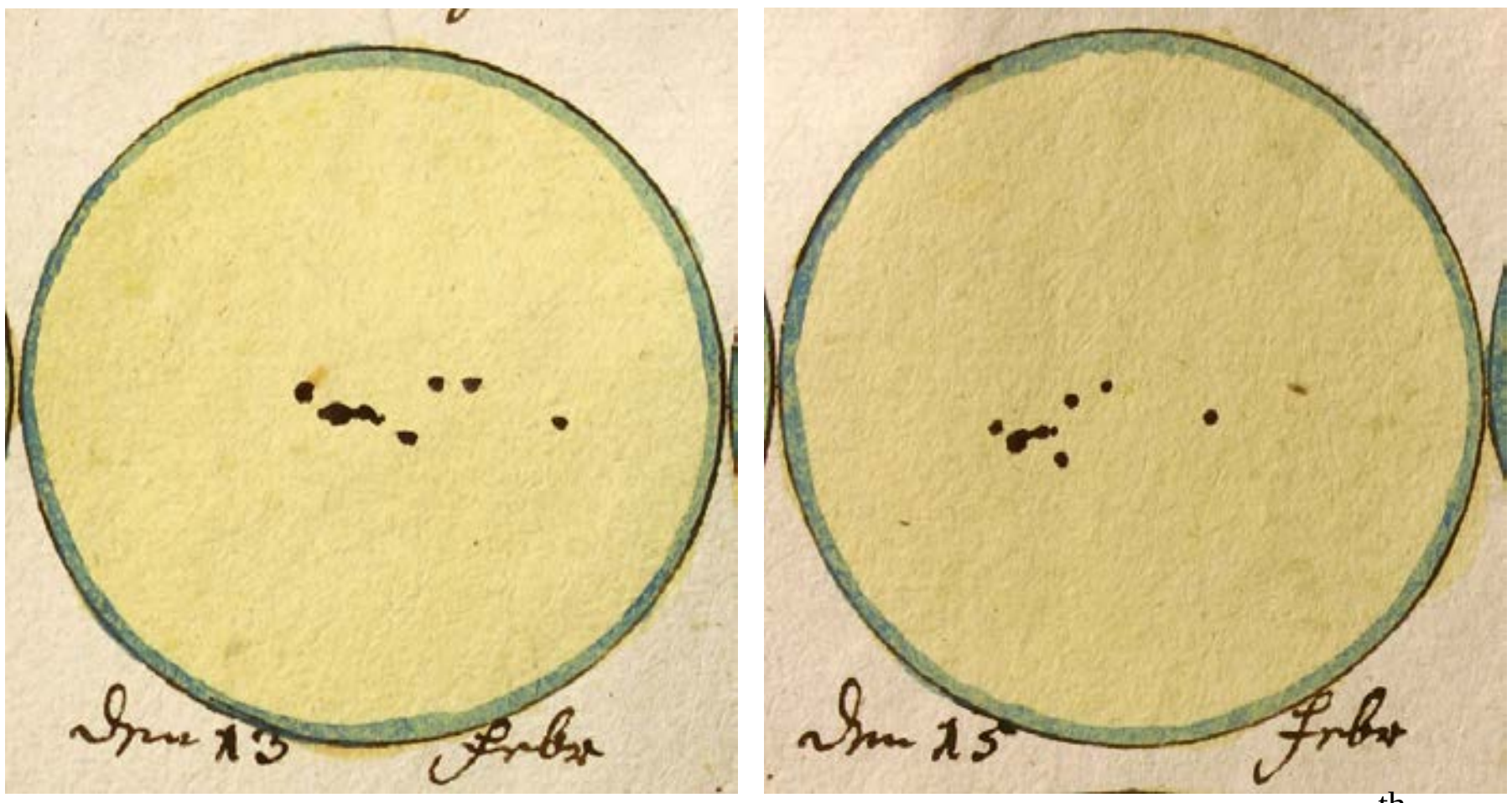

Figure 2: Sunspot drawings by J.C. Staudach corresponding to the (left) 13th and (right) 15th of February 1760 (Image source: R. Arlt).

However, Wolf soon realized that Staudach’s values were systematically lower than those for the recent cycles observed by Schwabe and he suspected that this was most probably due to the cruder instrument used by Staudach and the limited amount of details recorded in his small sketches (Figure 2; Arlt 2008). Therefore, Wolf concluded that Staudach's observations had to be multiplied by a k personal coefficient of 2 in order to match his own 19th century observations (Wolf 1861a).

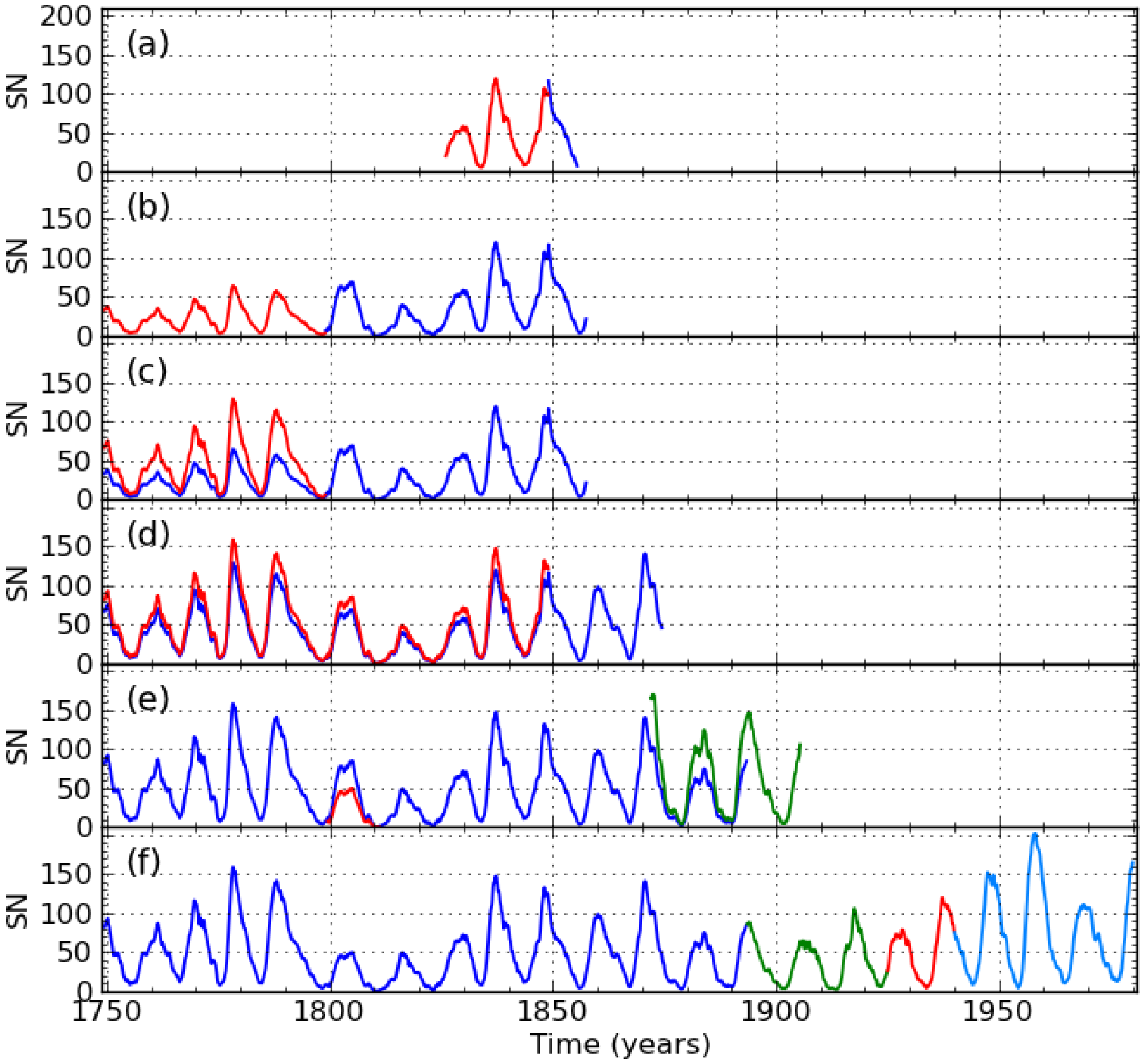

Figure 3: The progressive construction of the Zürich sunspot number series by Rudolf Wolf. Panel a: pre-pending Schwabe's observation (1849); Panel b: pre-pending Staudach's observations (1857); Panel c: rescaling Staudach's data by a factor 2 (1861); Panel d: correction due to comparisons of Schwabe-Carrington-Hornstein-Wolf (1882); Panel e: revision of cycle 5 by Wolfer and new Wolfer counts from 1877 (1902); Panel f: extension of the original Wolf series (blue), successively by Wolfer (green), Brunner (red) and Waldmeier (blue).

In 1874, Wolf added new observations mostly by Flaugergues, to bridge the gap between the two main series by Staudach and Schwabe (Wolf 1874). Indeed, the interval 1790 to 1826 is only sparsely covered by a few short sets of observations (Letfus 1999, 2000, Usoskin et al. 2003, Vaquero et al. 2012). Comparing observations by himself, Schwabe, Hornstein, and Carrington, Wolf had found already in 1861 that Schwabe's observations had to be increased by 25%. This increase was effectively included only in the 1880 series (Wolf 1880). As the scaling of the early historical sunspot numbers was based on those pre-1849 Schwabe numbers, Wolf thus applied a factor 1.25 to the entire reconstructed series before 1849. This was the last adjustment made by Wolf on the reconstructed part of the series. It is only much later in 1902 that his successor, Alfred Wolfer, applied a correction to cycle 5, based on new observations from Kremsmünster (1802-1830), lowering cycle 5 by a factor 0.58 (Wolfer 1902). Cycle 5 thus became, together with cycle 6, the weakest cycle of the Zürich series, forming what would later be named the Dalton minimum (see Figure 3).

Wolf considered another way to estimate the strength of past solar cycles: magnetic needle readings. Far ultraviolet (FUV) radiation from the Sun, enhanced by solar activity, creates and maintains the E-layer of the ionosphere, where dynamo action from moving air causes an electric current to flow above the dayside of the Earth at about 100 km altitude (Cf. Svalgaard 2014b, this volume). The magnetic effect of this current is readily measured by magnetometers on the ground and is best seen in the East Component of the geomagnetic field (Nevanlinna and Kataja 1993, Nevanlinna 1995, Svalgaard and Cliver 2007, Cliver and Svalgaard 2007). The current stays fixed with respect to the direction to the Sun and its magnetic effect, deflecting the “magnetic needle” at a right angle to the current, increases to a maximum at about $8^h$ local time, then disappears when the current is overhead, and finally increases again, but in the opposite direction, to a maximum at about $2^h$. The range, rY, from the morning deflection to the afternoon deflection, depends essentially on the solar zenith angle and the FUV flux. In the yearly average, the zenith angle dependence averages out and the resulting index variations then essentially reflect the long-term variations of the solar FUV irradiance, i.e. the level of solar activity.

This magnetic effect was discovered as early as 1722 by George Graham and as it varied in step with the recently discovered 11-year solar cycle, Wolf considered that it provided a valid independent check on the amplitude of past solar cycles. However, published documents (Wolf 1861b, 1861c, 1862, 1875, 1882) indicate that the key modifications to the Wolf sunspot series were based only on sunspot counts from various observers progressively collected by Wolf over many years and that magnetic needle data were only used as a validation (NB: scanned versions of the original “Mitteilungen” can be found at: http://adsabs.harvard.edu/historical.html).

Those various modifications left clear traces in the standard sunspot number series. By exploiting the sharp lower boundary in the Wolf number for the first spot (i.e. R = 10+1 = 11, cf. Eq. 1), histograms of the lower range of SN values provide a direct confirmation of the time and magnitude of the corrections reported in the Zürich publications, i.e. of the k personal coefficients effectively adopted for the early observers (Figure 4).

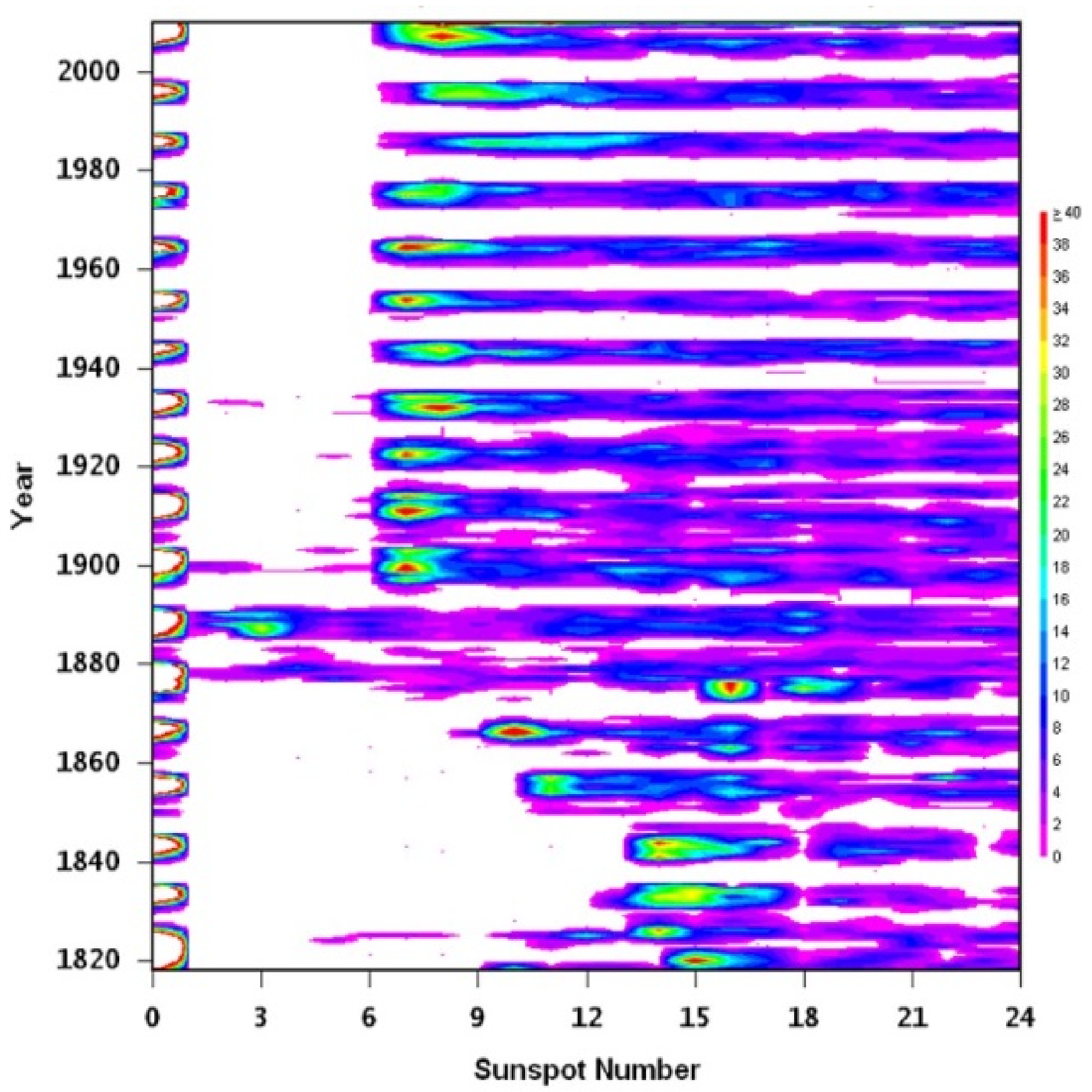

Figure 4: Contour plots of the lower part of the SN histogram (horizontally) as a function of time (vertically by cycle) showing the variations of the lower values corresponding to a single spot (Wolf number = 11). This lower cut-off is constant at 7 (11 x 0.6) after 1893, while earlier values show step-wise changes that match the reported corrections brought by R. Wolf to the historical SN that he recovered.

It should be noted that because of the scarcity of the data recovered by Wolf before the mid-18$^{th}$ century, he never extended his daily and monthly mean sunspot numbers before 1749. Only yearly means were derived back to 1700. The 17$^{th}$ century, including most of the Maunder minimum, was left out.

## 2.2. The Zürich era

Until his death in 1893, Wolf established the base principles to calculate the Zürich sunspot number and guarantee its long-term stability. The base Zürich number was simply the raw Wolf number based on the sunspot and group counts made on the aerial image of the standard 80mm "4-foot" Fraunhofer refractor installed on the grounds of the Zürich Observatory. As Wolf was often traveling across Switzerland for official duties, he also used smaller portable instruments, while an assistant was making counts with the standard 80mm refractor (Figure 5). A mean ratio between those simultaneous observations allowed deriving a k personal coefficient for the portable telescopes (Wolfer 1895, Svalgaard 2013a). Those k ratios were then applied to all observations made by Wolf with his travel telescopes to bring them to the standard scale. As those telescopes still exist and are still in regular use (Friedli and Keller 1993, Friedli 1997), a recent unpublished statistical analysis made by the WDC-SILSO over one year in 2012, shows that the scaling ratio for those instruments relative to the current sunspot number matches within 5% the values derived by Wolf.

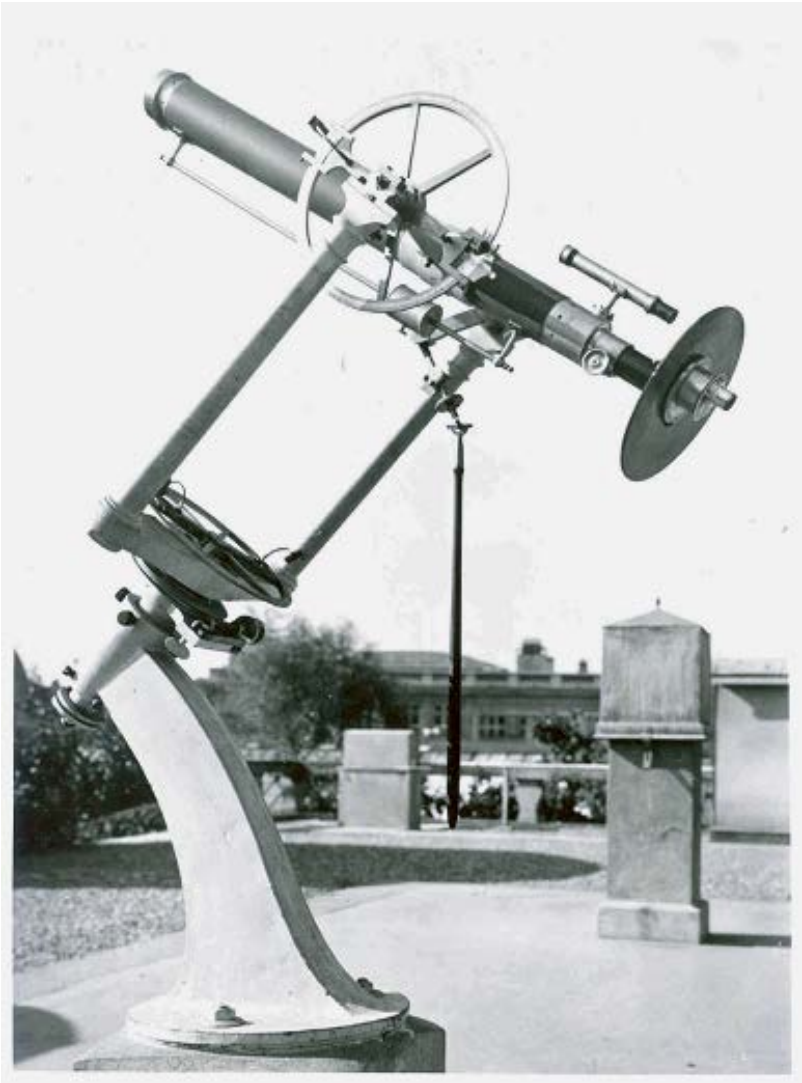 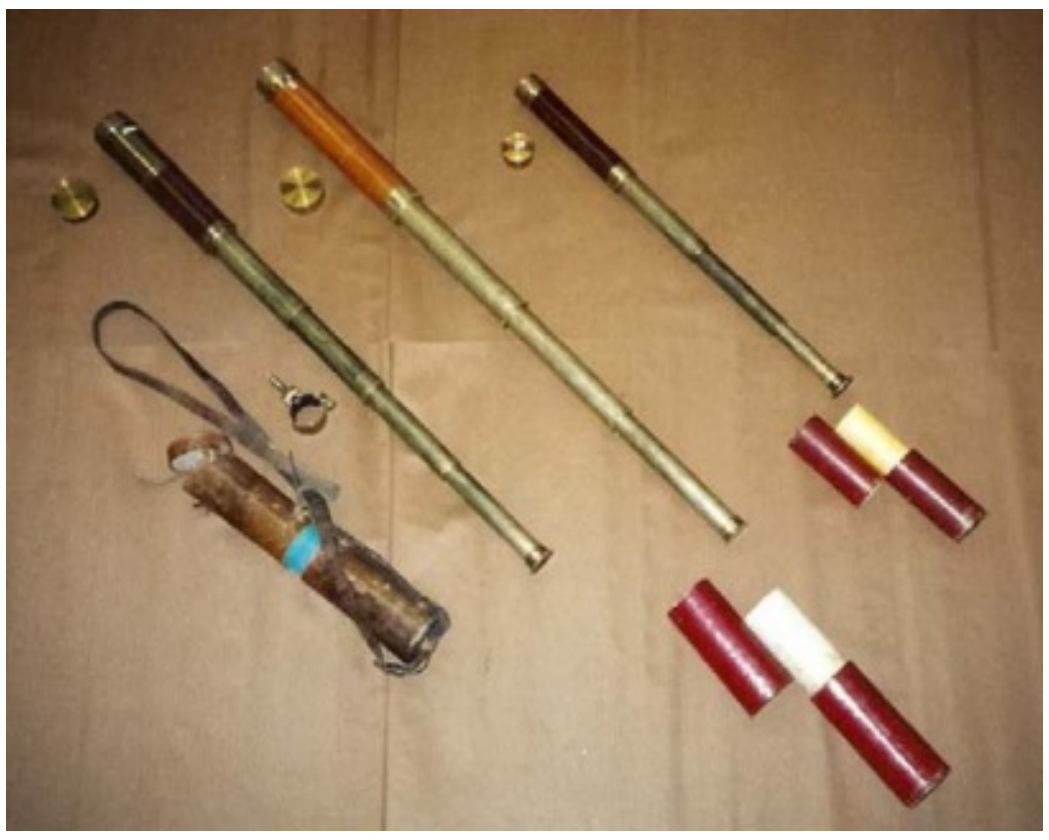

Figure 5: The original instruments used by R. Wolf. On the left, the standard “4-foot” 80mm Fraunhofer refractor, pictured here as it was set up at the Zürich Observatory (image source: Library of the ETH Zürich, Image Archive, http://www.e-pics.ethz.ch, Record N° Ans-05063-001). On the right, the three smaller portable refractors (apertures of 30 and 40 mm) used by Wolf while he was travelling.

In addition, in order to fill in the daily gap due to bad weather in Switzerland, Wolf used a set of auxiliary observers in order to derive a sunspot number for the missing days. For this purpose, average k scaling coefficients were derived by yearly means of the ratios between the raw numbers from each station and the corresponding Zürich number. The missing sunspot number was then computed by an average of all auxiliary values, each multiplied by its k factor. Wolf's successors continued to follow those principles without much change except for a steady increase of the external collaborating stations, until the production of the Zürich sunspot number came to an end in 1980.

However, other very important aspects of the method for determining the sunspot number underwent changes following Wolf's death. Until then, as Wolf wanted to match his sunspot number with the early historical values and as he realized that his predecessors had smaller and cruder instruments, he deliberately applied restrictive rules to his counts, trying to mimic early observers, namely:

- small short-lived sunspots without penumbra (“pores”) were not counted
- multiple umbrae within a common penumbra were counted as single sunspot

This lowered the resulting counts compared to what the standard 80mm instrument could actually show. Although those reduced counts matched rather well what Wolf would also naturally count with his small portable telescopes, it imposed a sunspot selection process when using the 80mm refractor and thus an additional interpretation prone to personal subjectivity. This “censored” count also dropped useful sunspot information that was plainly shown by the more advanced telescopes of the 19$^{th}$ century.

This is why a new counting series was started by Alfred Wolfer in 1877 in parallel with Wolf, where the above criteria were dropped and all spots (including pores) were included. Based on 16 years of parallel observations (1877-1893) between Wolfer at the 80mm telescope and Wolf using his smaller portable telescopes and taking into account the average k ratio between the portable and standard telescopes (estimated by Wolf at 1.5), the average ratio between the two counts was established at 0.6 (Wolfer 1895). Once Wolfer took over the position of Director in 1893, he continued with the new counting method but applying the fixed factor 0.6 to the raw counts in order to match the original Wolf series. Ever since, all raw Zürich sunspot numbers were multiplied by this 0.6 factor. At that time, doing so was probably considered more convenient and less time-consuming than manually rescaling 150-years worth of past sunspot numbers. Although the 0.6 Zürich factor was initially equivalent to the Wolfer k personal coefficient relative to Wolf's numbers, it was later applied as a fixed factor allowing to join all recent SNs to the original Wolf series (thus assuming that all Zürich SNs produced since Wolfer have by construction a fixed k coefficient equal to 1 relative to Wolfer). Therefore, this fixed factor should not be confused with usual k coefficients in Equation 1, which are used to rescale raw Wolf numbers from auxiliary stations to the Wolf numbers from the primary station, initially Zürich and later, Locarno. Those k coefficients can vary in time and are based on a continuous statistical recalculation.

The new Wolfer counting method was apparently applied uniformly during the following decades, at least by the next Director, William Brunner. Only when the last Director Max Waldmeier took over the SN compilation in 1945, a last major modification was introduced in the counting method, where individual sunspots are weighted according to their sizes (Waldmeier 1968, 1948). The consequences of this change are discussed in more detail in section 4.2. While the timing is uncertain, the method itself can be properly reconstructed as it is still in use nowadays at the Specola Solare Ticinese station in Locarno. Indeed, this station was set up by Waldmeier in 1957 in order to complement the primary Zürich station. Sergio Cortesi, the main observer who was then recruited, was fully trained to count according to the Zürich method and has been continuously observing ever since (Figure 6). He is thus a living witness of the Zürich "school". Part of the diagnostics presented in this paper is based on direct information from the Locarno station (private communications, archives, drawings).

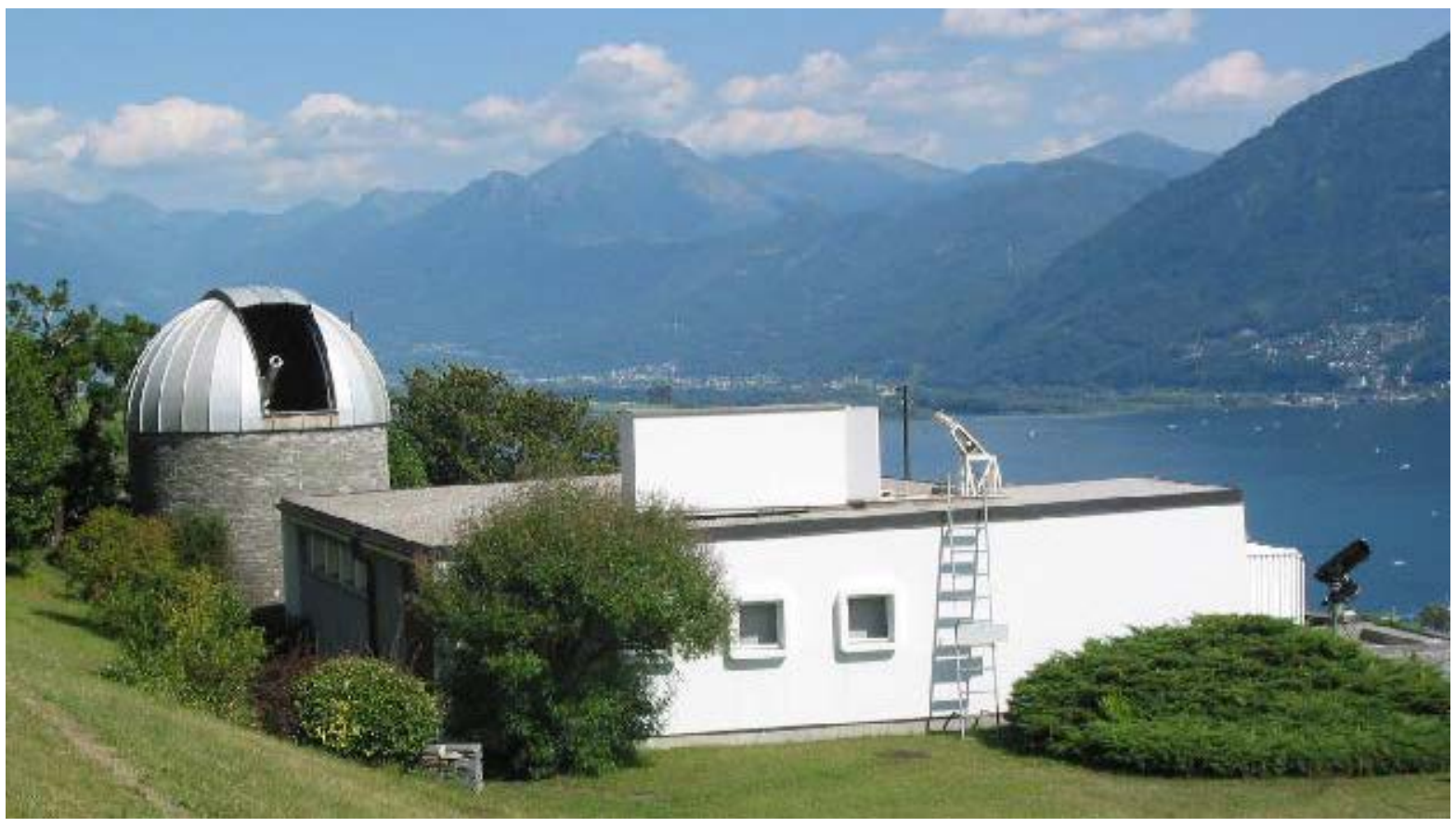
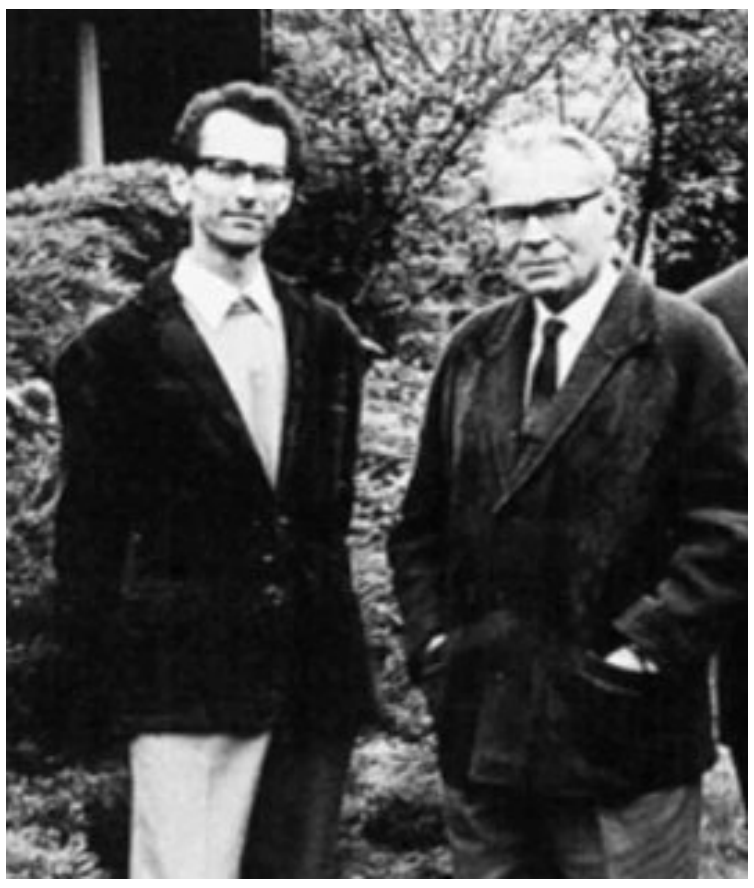

Figure 6: Left image: view of the Specola Solare Ticinese Observatory overlooking the Lake Major in Locarno. Right image: M. Waldmeier (right) and Sergio Cortesi (left), the primary observer at Locarno, in 1970 (Images courtesy Specola Solare Ticinese, Locarno).

## 2.3. The Brussels era

In 1980, when Max Waldmeier retired, the Zürich Observatory ceased to produce the sunspot number and this activity was taken over by the Royal Observatory of Belgium, leading to the foundation of the SIDC, "Sunspot Index Data Center". The context and circumstances of this transition are described by Berghmans et al. (2006) based on the original documents, including the correspondence between André Koecklenbergh, the founder of the SIDC, and A. Zelenka, one of Waldmeier's assistants. At that time, Zelenka was preparing a computerization of the Zürich processing and this codification was useful to ensure the continuity in the processing of the sunspot number, as it provided the base for the new SIDC method. The latter was developed in Brussels with two main objectives. First, the new computation should take advantage of a larger base of contributing stations. Second, it should include a new mechanism ensuring the stability of the index after the loss of the primary Zürich station.

The resulting method, which is summarized by Clette at al. (2007), features 3 key steps:

- Determination of the monthly average k coefficients relative to a pilot station, namely the Specola Locarno station, with statistical elimination of individual values when the daily k deviates abnormally relative to the monthly average. All values for each station are normalized by applying the corresponding monthly average k factor.
- Rejection of outlying values from the pilot station: daily values from the reference station are compared to the network average and values are rejected when they deviate significantly based on the standard deviation of all values for that day.
- Iterative calculation of the average daily sunspot index: the final sunspot number is computed by taking the average of k-normalized Wolf numbers from all available stations. It is done iteratively with a final elimination of outliers.

Part of the station statistics is a rather direct translation of what was done manually at Zürich, but the new method differs from the previous one in two main aspects:

- The network averaging: while most of Zürich sunspot numbers were simply the raw Wolf number of the Zürich station, the SIDC international sunspot number includes the information from all contributing stations.
- A validity control of the daily values from the pilot station: each daily reference value is compared against the values of all stations for the same day and can be eliminated in favor of the network average.

This new procedure leads to a reduction of the RMS dispersion of daily sunspot numbers, compared to daily Wolf numbers from a single station, which amounts to about 8% rms (1.5% rms for monthly averages). The number of observations available each day typically varies between 10 and 30. In order to mark this change of method, the resulting index was renamed to “International Sunspot Number”, noted $R_i$, in order to make the distinction with the former Zürich Sunspot Number, noted $R_Z$.[1]

In spite of those differences and the higher complexity of the SIDC calculation, the adopted mechanism still ensures a close similarity with the earlier sunspot number. Indeed, as all values for each station are scaled relative to the pilot station though a monthly average of the k ratio, the final average of all normalized numbers leads to a value that is close to the Wolf number of a single station, namely the pilot station. The latter thus defines entirely the absolute scale of the index over timescales longer than one month, just like Zürich did previously. On the other hand, the statistical elimination process improves the precision of daily values, in particular by rejecting the outlying Locarno values and thus avoids biasing the network statistics due to a bad reference.

The outcome is illustrated in Figure 7. Over the last 32 years, the ratio between the raw Locarno Wolf numbers and the international SN $R_i$ remained perfectly flat around unity (or 0.6 taking into account the historical Wolfer scaling factor). The $R_i$ scale is thus fully defined by the pilot station. On the other hand, small random deviations in the monthly mean k (~1.5% rms) correspond to the statistical cleaning brought by the systematic use of the entire network compared to a single station. It gives also a measure of the gain in the RMS error of daily sunspot numbers

[1] In this paper, we will follow this conventional notation for the SN, with $R_Z$ and $R_i$ corresponding to the periods before and after 1981, respectively.

associated with the global network statistics.

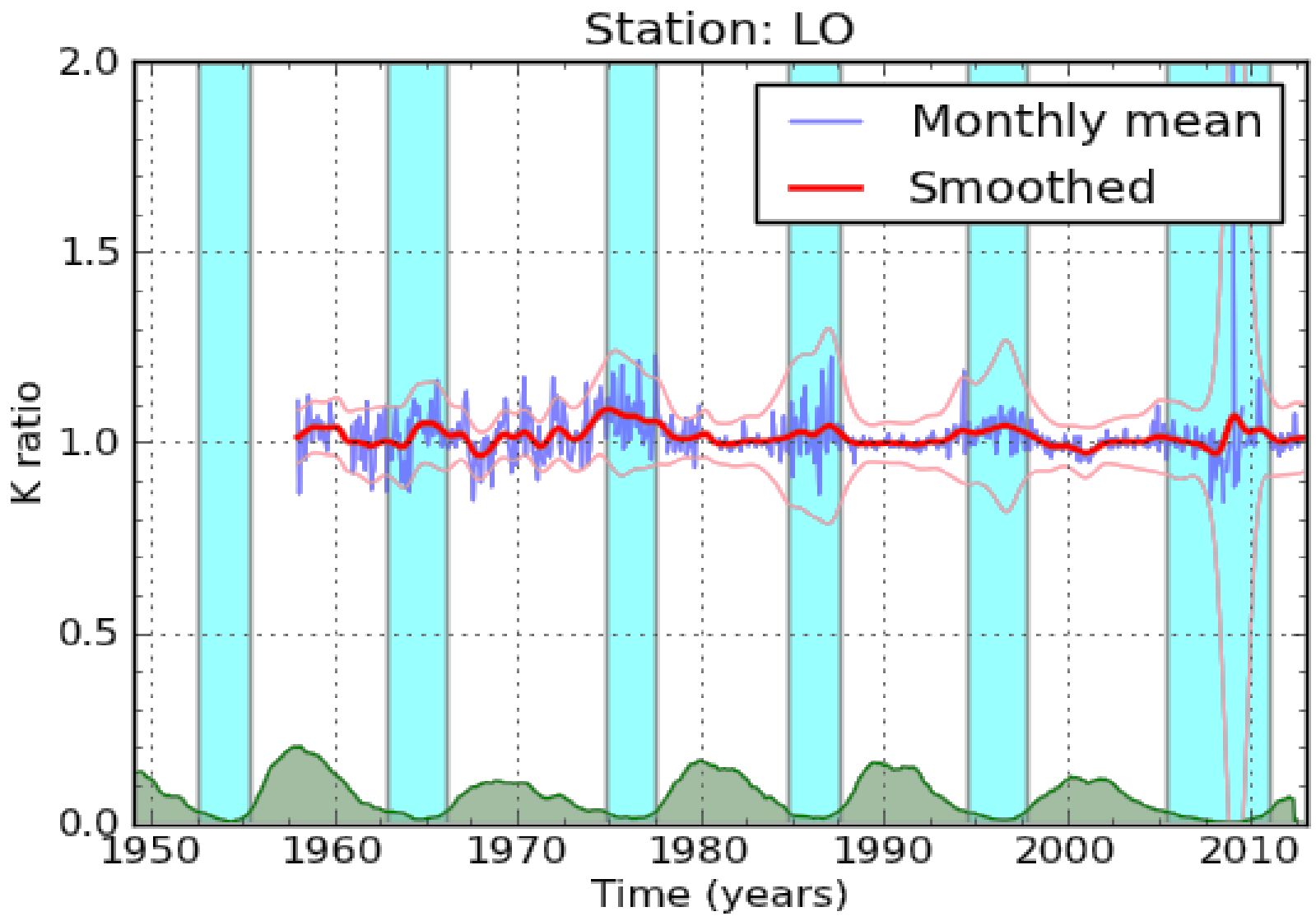

Figure 7: Ratio between the raw Wolf number from the Locarno station (multiplied by the standard factor 0.6) and the SN. The thin lines indicate the 3 sigma confidence limits for the monthly average values (blue). The red line is the 13-month smoothed ratio. The sunspot number (shaded green) and the periods of minima (shaded blue), when the ratios are less accurate, are overlaid as time references. The ratio remains close to 1, showing that the scale of the SN followed closely the Locarno reference, in particular after 1981 when the Locarno station took the role of pilot station for the international sunspot index $R_i$ (note the lower dispersion after 1981).

Finally, the choice of Locarno for the pilot station was quite natural. Its main observer, Sergio Cortesi, was trained into the Zürich observing method and in 1981, he had already carried out parallel observations with Zürich for more than 20 years. Locarno's equivalence to Zürich had thus been extensively checked before Zürich ceased observing. Moreover, as the Specola Observatory could continue observing for the SIDC-Brussels, it provided an uninterrupted reference straddling the critical transition period between the Zürich and International SN around 1980. We will see in section 6.2 that this choice indeed ensured a clean transition around 1980, although it brought other problems later on.

Figure 8 gives global statistics of the evolution of the world-wide sunspot network managed by WDC-SILSO in Brussels over the last 32 years. Since 1981, there have been a total of 270 contributing stations distributed over 30 countries. About 2/3 of stations are individual amateur astronomers and 1/3 professional observatories. All are subjected to the same quality control for stability and continuity. While many stations contributed only for a few years, there are 80 long-duration stations that provided data for more than 15 years (Fig. 8 right). After an initial rise of the number of stations, starting from the ~40 auxiliary stations used formerly by the Zürich Observatory, the yearly number of contributing stations has remained between 80 and 100, providing from 15,000 to 20,000 observations each year (Fig 8 left). Each station collects an average of 175 daily observations per year. The distribution of yearly averages is rather broad (Fig 8 right), with a peak around 240 days/year (a typical value for dedicated stations limited only by weather conditions) and up to 360 days/year for a few stations, i.e. close to a 100% time coverage. In total, since its creation in Brussels, the WDC has accumulated more than 450,000 sunspot counts that are now all preserved in a single database.

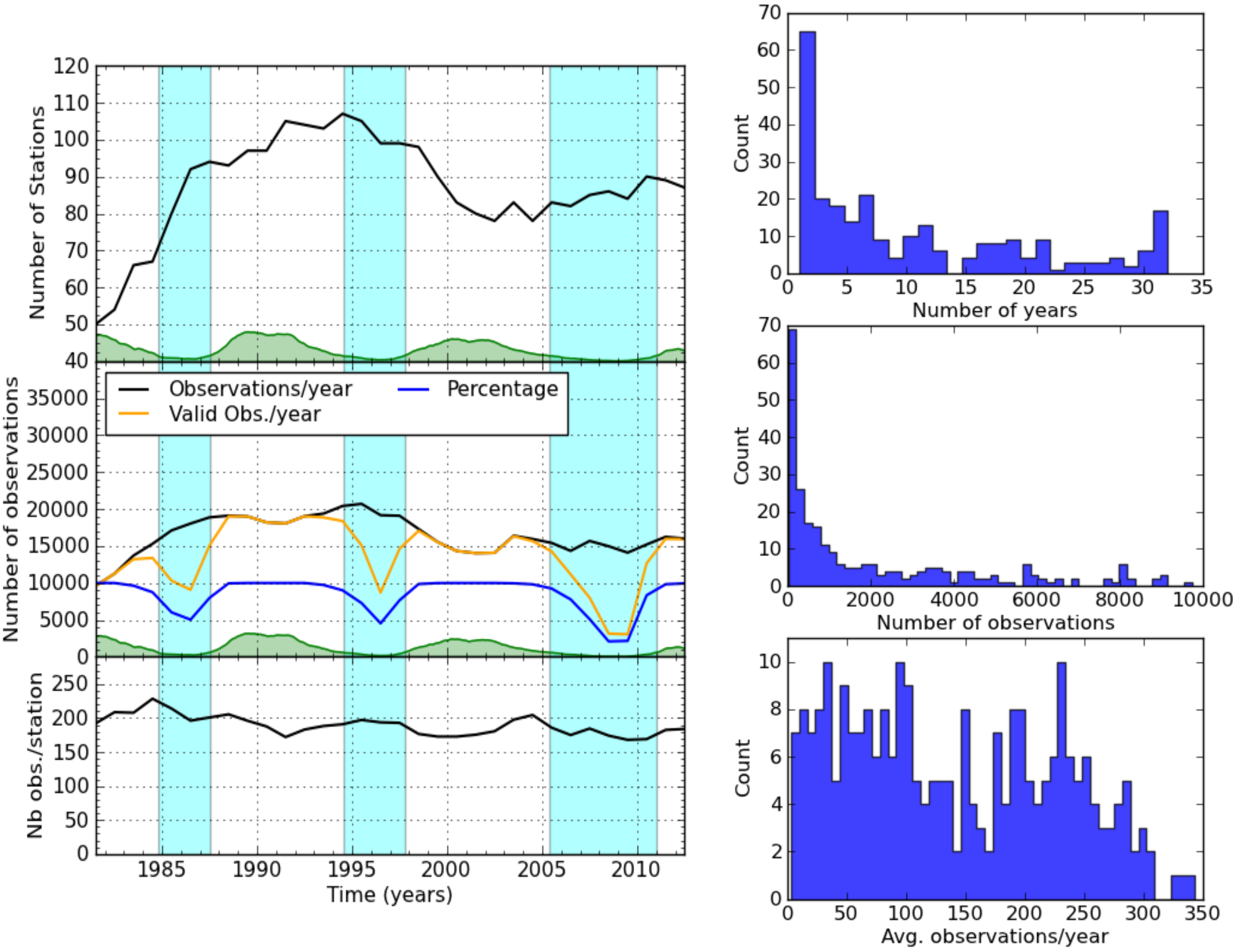

Figure 8: (Left) Plots retracing the evolution of the worldwide sunspot network: number of contributing stations (top panel), annual number of collected observations (middle panel) and average number of observations per station per year (lower panel). The solar cycles (shaded green) and minima intervals (shaded blue) are overlaid as time reference. In the middle panels, valid observations are days when both a given station and the reference pilot station give a non-null Wolf number (i.e. when a personal k coefficient can be established).

(Right) Overall statistical distributions of the contributions from all stations: duration over which stations have contributed (top panel), total number of contributed observations (middle panel) and average number of observations per year (lower panel).

# 3. The group sunspot number: a better number?

## *3.1. Group counts*

As can be seen from the previous section, the standard SN series is a composite time series assembled from very inhomogeneous material. Its accuracy decreases as we go back in time, in particular as the observer base shrinks dramatically before the 19$^{th}$ century. Moreover, Wolf's backwards reconstruction of the early sunspot numbers does not include at all the important Maunder minimum episode because of the dearth of sunspots prior to 1700.

Therefore, given the interest of extending the series and the knowledge of additional archived observations that were unknown to Wolf, a new long-term series was constructed by Hoyt and Schatten (1998a,b). As early observations were rather sparse and crude, obtaining detailed counts of small spots proved to be difficult or even impossible. Moreover, a study by Schaefer (1993) suggested that the SN is largely proportional to the group count alone. This is why the new index included only the total count of sunspot groups. Hoyt and Schatten defined the group number as:

$$R_G = \frac{12.08}{N} \sum k_i Ng_i \qquad (2)$$

where $Ng_i$ is the number of sunspot groups recorded by the $i^{th}$ observer, $k_i$ is the $i^{th}$ observer's personal scaling factor, $N$ is the number of observers used to compute the daily value and the 12.08 constant is a normalization factor chosen to bring the GN to the same scale as the Zürich sunspot number.

As a base for this normalization, they chose the Greenwich sunspot catalog based on the Greenwich photographic plate collection that spans the 1874-1976 interval (Willis et al. 2013) and the visual USAF/SOON group counts for 1976-1995. Starting from this reference, the scaling of earlier historical group counts is derived by working backward in time, using the average ratios between counts of parallel observers where the series are overlapping. Following this methodology, Hoyt and Schatten derived daily, monthly and yearly means from 1610 to 1995.

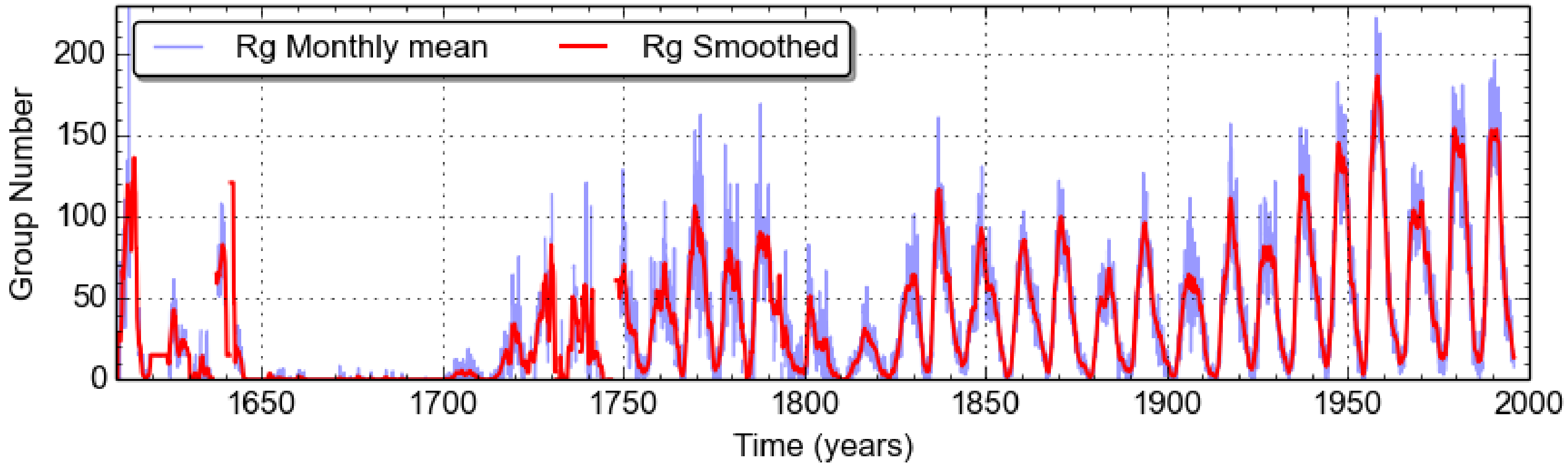

Figure 9: The original group sunspot number series from 1610 to 1995 as published by Hoyt and Schatten (1998): monthly averages in blue, 12-month Gaussian smoothed values in red). It shows the Maunder Minimum (1645-1715) followed by a progressive rise up to the mid-20$^{th}$ century.

### 3.2. A new extended and improved data set

By approximately doubling the number of recovered observations and cleverly interpolating between sparse observations (Hoyt et al. 1994), it was possible to reduce the gaps in Wolf's original SN series and to extend it back to the very first telescopic observations (Figure 9).

The first observations of sunspots were made by the Englishman Thomas Harriot on December 18, 1610 (Shirley 1983). Three other scientists made telescopic observations of sunspots independently about this time: Johannes Fabricius, Christoph Scheiner and Galileo Galilei. Thereafter, telescopic observations of sunspots were generalized using different techniques such as solar filters or projection of the solar disk on white screens. Preserved documents about those early observations allow us to retrace the evolution of sunspots since 1610, though with some gaps and indeterminacy (see the reviews by Hoyt and Schatten (1997) and by Vaquero and Vázquez (2009)).

Hoyt and Schatten (1998a) made a huge effort to recover historical observations preserved in archives and libraries around the world. In general, they achieved a good coverage of the series during the period 1610-1750: the mean number of sunspot observers per year is 5 and the mean number of days with records per year equals 237. However, these values have a large dispersion. Figure 10 shows the coverage of the original GN series in the time period 1610-1750, including the annual GN values (red line). The number of days with records per year is shown as blue bars and the number of sunspot observers per year as a green line.

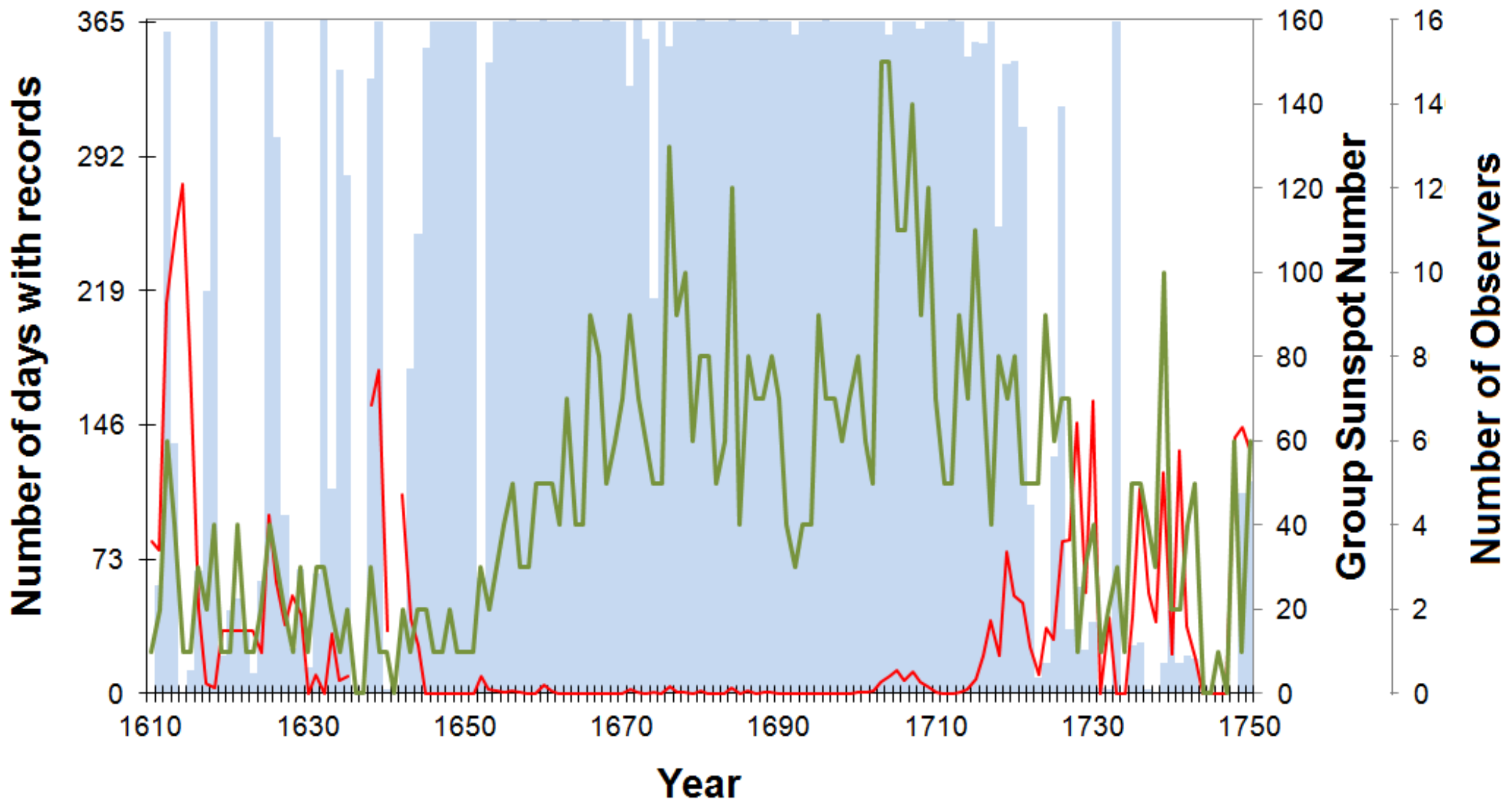

Figure 10: Coverage of the original GN series in the time period 1610-1750: number of days with records per year (blue bars), number of sunspot observers per year (green line) and GN (red line).

In order to trace correctly the solar activity in the first decades of the 17th century, it is necessary to obtain more information from the most important observers. Based on the number of years with records, the most regular observers were Hevelius (39 years), La Hire (37), Flamsteed (32), Eimmart (25), Picard (23), Siverus (17), Kirch (17), Scheiner (16) and Derham (13). Moreover, we should note that some observers are contemporaries. For example, La Hire, Einmart and Flamsteed observed almost during the same period. Therefore, the main observers are Scheiner for the period before the Maunder Minimum (1610-1642), Hevelius for the first half of the Maunder minimum period (1643-1679) and La Hire for the second half of the Maunder minimum period (1680-1718). Note that there are very few records for the second quarter of the 18th century. However, we could highlight Plantade (423 records from 1704 to 1726), Kirch (394 records from 1716 to 1736), Alischer (524 records from 1719 to 1727) and Adelburner (367 records from 1730 to 1733).

We have additional information on some of the leading observers of this early period thanks to an encyclopedic work on sunspots, the book entitled *Rosa Ursina* by Christoph Scheiner published in 1630 (for a biographical account, see Kant 2007). In his book, Scheiner describes for instance the first telescope with equatorial mounting, called “Heliotelescopium”, specially designed to observe sunspots. Unfortunately, this book was written in Latin and there is no translation to other modern language. However, Mitchell (1916) provides essential fragments translated into English. Recently, E. Reeves and A. van Elden have also translated into English some published documents with sunspot observations written by Scheiner and Galileo, with introduction and notes (Galilei and Scheiner 2010).

Sunspot observations by Johannes Hevelius (see MacPike (1937) for details about his life and astronomical background during this time period) are very important to obtain a picture of solar activity in the first decades of the Maunder minimum, because these are the only daily listings of solar observations during this time. Hevelius lists his daily solar observations, intended for the determination of the solar meridian altitude, in the last part of his book *Machina Coelestis*, published in 1679. Hoyt and Schatten (1995a) mention that Hevelius reports 19 sunspot groups during the period 1653-1679 and also provides the best record of the sunspot maximum of 1660 (when one sunspot group was active for seven solar rotations).

The most important observer after Hevelius was the French astronomer Philippe La Hire (1640-1718). The latter conducted solar observations for at least 35 years (from 1683 to 1718) at the Paris Observatory. These observations were studied by Ribes and Nesme-Ribes (1993) from the records and sunspot drawings preserved in the historical archive of this institution.

Another important observer supplementing the observations of La Hire is John Flamsteed. Hoyt and Schatten (1995b) studied the sunspot records that appear in his book *Historia Coelestis Britannica* which was published in three volumes in 1725, in the *Philosophical Transactions of the Royal Society* and in his letters to William Derham preserved in the Cambridge University Library. According to Hoyt and Schatten (1995b), in the period 1676-1700, Flamsteed only observed sunspots in the years 1766 (24 active days and 19 inactive days) and 1684 (17 active days and 32 inactive days).

In the second half of the 17th century, we can note that there is a large number of observations and observers documenting the Maunder Minimum (1645-1715; Eddy 1976). By contrast, the periods before and after the Maunder minimum are generally poorly covered (Fig. 9). In fact, there are even six years without any sunspot record (1636, 1637, 1641, 1744, 1745 and 1747) and there are many years with very few observations. The years 1610, 1614, 1640, 1723, 1731, 1732, 1734, 1737, 1738, 1746 and 1748 contain less than ten observations.

The high number of sunspot observations made during the Maunder Minimum needs a clarification. Hoyt and Schatten (1996) examined how well the Sun was observed during this Grand Minimum of solar activity. They compiled the specific dates of observations by Hevelius, Picard, La Hire, Flamsteed and others and they derived an estimate of the minimum fraction of the time over which the Sun was observed: namely, 52.7% of all days have specific recorded observations. Moreover, they also compiled general comments mentioning the absence of sunspots during specific years or time intervals, obtaining an upper estimate of 98%. Figure 11 shows the number of days with records per year, either for dates with explicit sunspot observations (blue line) or for all observations, including the general comments (red line), according to Hoyt and Schatten (1996, 1998a). We can see that the number of days with explicit observations is very low in the beginning of the Maunder Minimum and only becomes large in its late part. In any case, observers listed with ~365 days of observations per year should be removed from any new reconstruction because these values (usually zero values) are based on general indirect comments and not on well-documented observations.

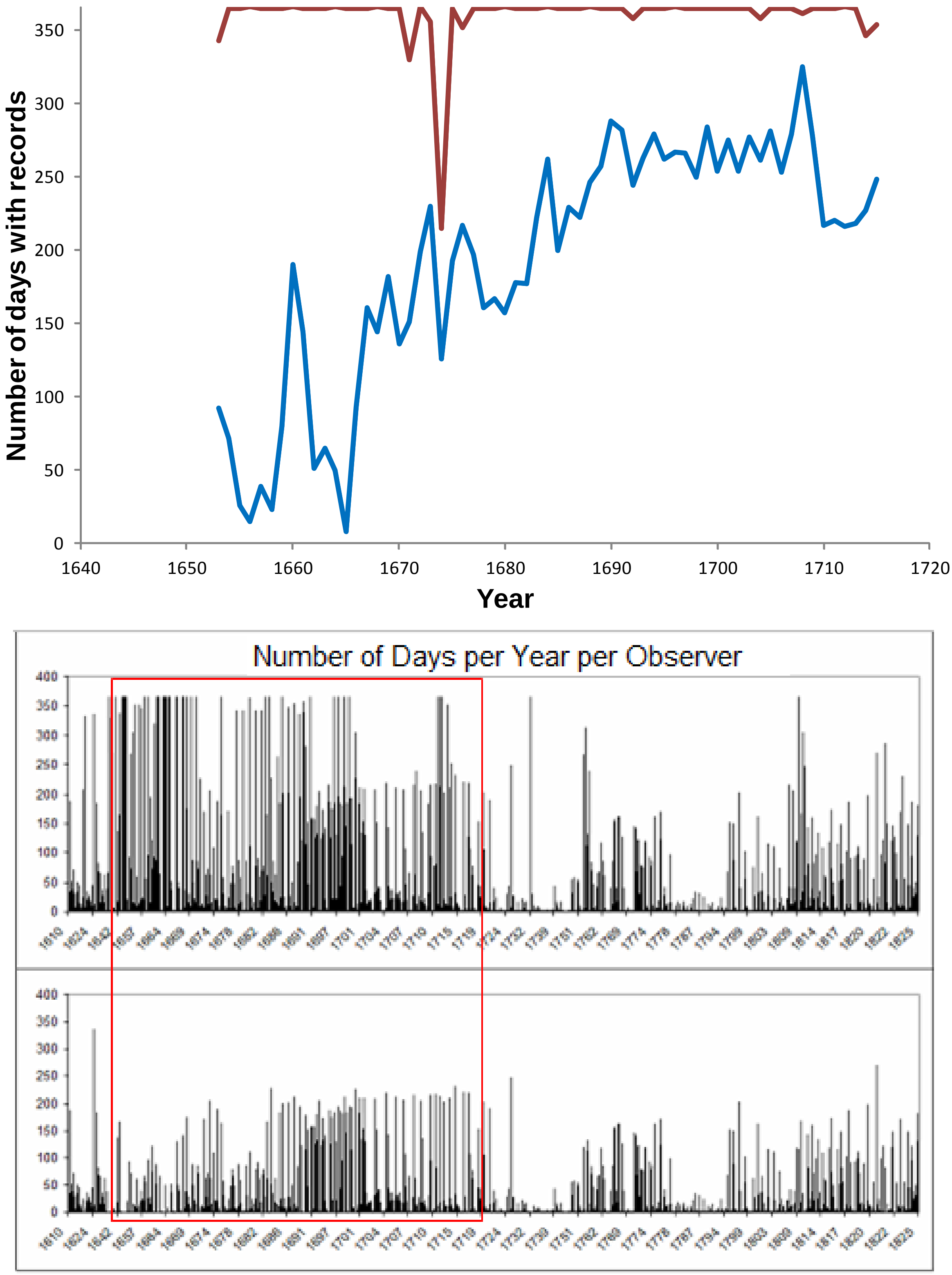

Figure 11: (Top) Coverage of the GN series during the Maunder Minimum: number of days with records per year for specific dates of observation (blue line) and for all the observations including the general comments (red line).

(Bottom) The number of days per observer for each year as assumed by Hoyt and Schatten (upper panel) and after removal of the observers who did not report specific observations. The red box indicates the Maunder Minimum corresponding to the top part of the figure.

After the Maunder minimum, in the first half of the 18th century, the data coverage is particularly bad in the fifteen years from 1734 to 1748, during which the average annual number of days with records is only 12. This lack of interest in systematic observations of sunspots is common during the 18th century. Significantly, Lalande (1771) gave a very low priority to the regular surveying of sunspots in a list of astronomical observations (14th out of 18 duties). After the years 1744-1747 (without sunspot records), astronomers of the second half of the 18th and early 19th century also showed a limited interest for sunspot observations. Consequently, the yearly number of observations fluctuates around intermediate values, except for a new poorly observed interval from 1779 to1794, before finally rising to almost 100% coverage in the early 1810s. Thus, the annual average of days with sunspot records in the Hoyt and Schatten (1998a) database amounts to only 30.8 and 176.9 in the periods 1779-1794 and 1750-1810, respectively. Two observers stand out in the period 1750-1850: J.C. Staudach and S.H. Schwabe. Both observers were used by Wolf in his first reconstruction. Arlt (2008, 2011) recently localized, digitized and analyzed the original sunspot drawings made by those two observers. Other important observers of the second part of 18th and early 19th century were Herschel, Pastorff and Christian Horrebow.

Christian Horrebow and his colleagues of Copenhagen, Denmark, were active sunspot observers from 1761 to 1777. These observations were examined both by Thiele (1859) and by d'Arrest, who gave very different interpretations. Hoyt and Schatten (1995c) reexamined Horrebow's original notebooks and provided a more coherent interpretation of Horrebow's observations.

Another important set of sunspot observations is contained in the notebooks of the famous astronomer William Herschel (Crowe and Lafortune 2007). This documentation is preserved in the Churchill College in Cambridge, England. Hoyt and Schatten (1992a) edited and reproduced these solar observations made from 1779 to 1818 (although most of the observations were made in the period 1799-1806). Hoyt and Schatten (1992b) showed that these observations provide a better reconstruction of solar activity around solar cycle 5.

One of the main anomalies in the Wolf reconstruction was the exceptional length of solar cycle 4: 17 years, from the peak of solar cycle 4 to peak of solar cycle 5. Hoyt and Schatten (1992b) have shown that the date of the cycle 5 maximum falls in 1801-1803 instead of 1805 as originally proposed by Wolf. This reduced the cycle length from 17 to 14 years and partially solved the anomaly of solar cycle 4, although the new length was still the longest in the entire SN record.

The German astronomer J. W. Pastorff made about 1477 observations of sunspots between 1819 and 1833. Hoyt and Schatten (1995d) examined Pastorff's original observations providing more nearly correct values for the number of sunspot groups than the values published by Wolf (who got them indirectly in 1874 from A. C. Ranyard, who misinterpreted the original records).

### 3.3. Group Number assets and limitations

Compared to the sunspot number series, the new group number brought several key assets:

- The series rests on a larger and more refined observational base, as shown in the previous section.
- The group count is less dependent on the visibility of small spots in the early observations, reducing and hopefully eliminating the expected downward biases.
- The series was processed as a single batch by the same scientists, thus avoiding possible changes in practices over several generations of sunspot observers and compilers.
- The calibration is exclusively based on the backward propagation of cross-scaling factors between group counts from successive observers, without involving external indirect indicators like Wolf's magnetic needle corrections.

This is why the group number is often considered as the most reliable reference to retrace the past solar activity, in particular before the 19th century.

However, the group number has also some specific weaknesses:

- Methodological weaknesses:
  - The interpretation of some observing records is questionable. This will be shown in section 4.
  - The base calibration of the series rests on a unique non-visual reference, the RGO photographic plates. Any flaw in the RGO group counts will thus affect the whole series before 1874.
- Truncated sunspot information:
  - The constant 12.08 scale factor assumes a constant average number of spots per group, namely: Ns = 10.13 Ng, as $R_S$= 0.6 (10 Ng + Ns). Yet, this ratio may actually vary over time, in the course of a solar cycle and over successive cycles, or according to the level of activity, as shown later in section 6.6. The information on the actual group size, which is contained in the SN, is thus lost here.

Therefore, other global biases may affect the Group number in a different way and at other times than the SN, which suggests that both series can and should be used for a mutual cross-validation rather than assuming that one of them is the only true accurate reference. That is not to say that the two time series are equally well-constructed. As we shall show, the flaws in the Hoyt and Schatten GN series are more significant than those in the international SN.

Even more importantly, once the $R_G$ series was published, it appeared that while the SN and GN series nicely agreed over the 20th century, there was a large discrepancy before ~1880, with $R_G$ falling about 40% below $R_Z$ (Figure 1). In the next sections, we will review in a chronological order the different flaws or ambiguities in both series that we have established or that have been identified by other studies.

# 4. Biases and uncertainties in the early group and sunspot numbers

## 4.1. The early sunspot number and Maunder Minimum (1610-1749)

After the publication of the GN series, a scenario for the occurrence of the Maunder Minimum emerged and was clearly based on the fact that solar activity dropped sharply at the onset of this Grand Minimum. However, the discovery and recovery of sunspot observations made by G. Marcgraf in 1637 and the revision of some earlier uncertain data for the period 1636–1642 now indicate a different picture of the onset of the Maunder Minimum (Vaquero et al. 2011). In particular, the additions and changes that Vaquero et al. (2011) made can be summarized as follows:

1. The Marcgraf sunspot records preserved in the Leiden Regional Archive were added. These observations were made in 1637, including three drawings of the solar disk. Note that there is no data for the year 1637 in the original GN series.
2. The estimated (not observed) values from Crabtree's comments (1638-1639) were eliminated. This problem has been discussed in Vaquero (2007).
3. The dates and numbers of sunspot groups from Horrox observations in Hoyt and Schatten (1998a,b) (from Julian Calendar to Gregorian Calendar) were corrected.
4. One spurious observation by Gassendi on 1 Dec 1638 was eliminated, because this record does not appear in his astronomical observations published in his *Opera Omnia* (1658, Tome IV).
5. The sunspot record by Rheita for 1642 were changed, after consulting the original source (pp. 242–243 of his book *Oculus Enoch et Eliae* published in 1645).
6. One sunspot record by Horrox in 4 December 1639 was added.

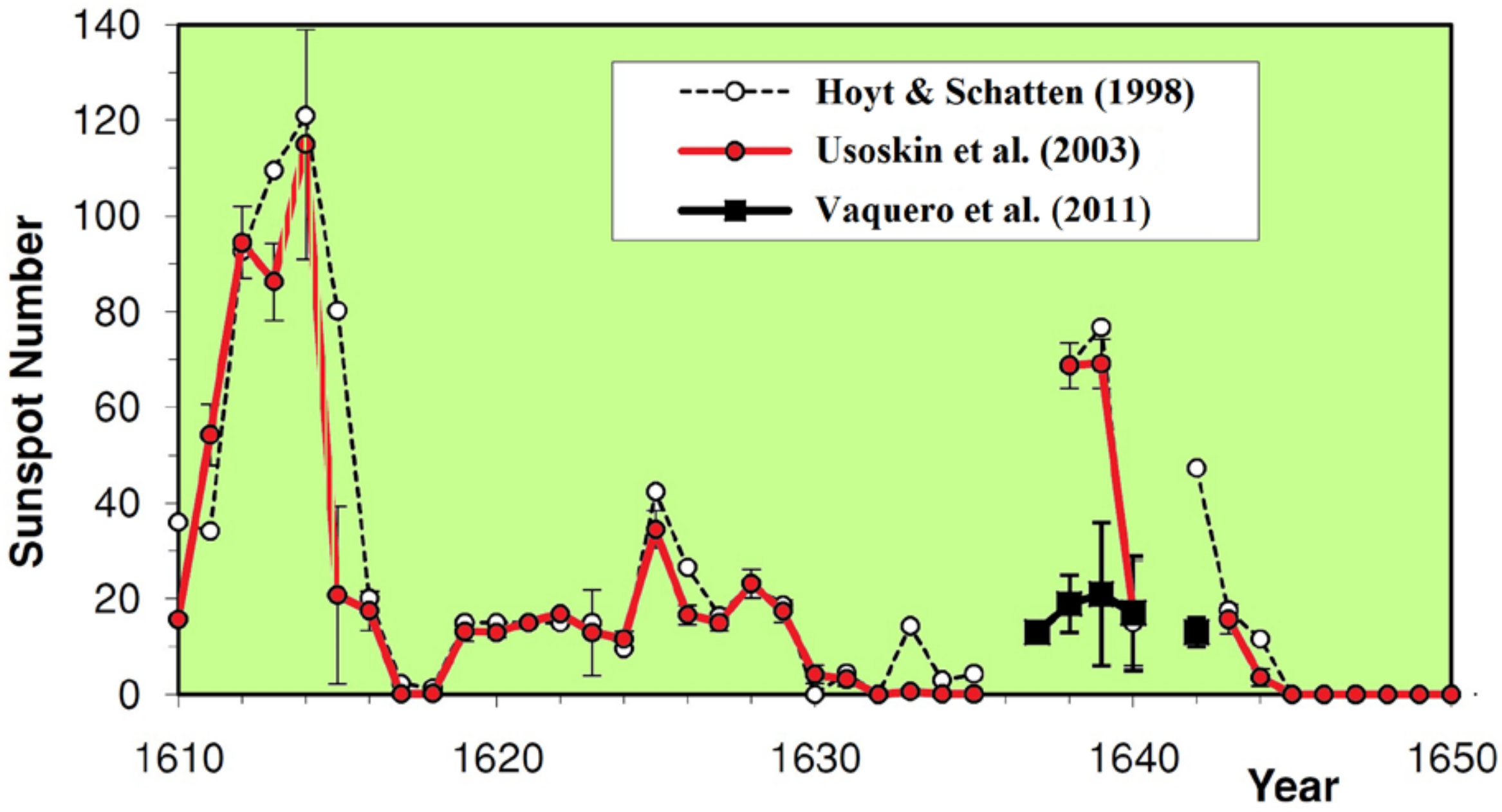

Figure 12: The sunspot number before the Maunder minimum according to Hoyt and Schatten (1998a,b) (dashed line, white circles), Usoskin et al. (2003) (red line and circles) and Vaquero et al. (2011) (black line and squares). Note that the peak value in 1614 was computed only with one sunspot observation made by F. Colonna in Naples (who reported 8 sunspot groups the day 3 October). Therefore, we have used a broken line around this date.

The final result of all these additions and corrections to the Hoyt and Schatten (1998a,b) data series is shown in Figure 12. The original GN series shows a high maximum for the solar cycle just preceding the Maunder Minimum in 1639. The new estimates by Vaquero et al. (2011) show a maximum in the same year but the amplitude of the cycle is much lower. Following those changes, the last two cycles preceding the Maunder minimum have only small amplitudes, in contradiction with the abrupt transition between states of normal solar activity and grand minimum that was assumed previously.

Despite the enormous effort by Hoyt and Schatten to locate all sunspot observations made during the Maunder minimum, it is still possible to find some additional records. As an example, Casas et al. (2006) analyzed hitherto ignored sunspot drawings by Nicholas Bion made in October and November 1672. Surprisingly, in October 1672, other astronomers also observed the Sun (Picard, Hevelius and Montanari) but did not detect any sunspots. A plausible explanation for this difference may be based on the fact that some of the observations compiled by Hoyt and Schatten (1998a,b) were meridian observations of the Sun and, strictly speaking, these were not meant for establishing the presence or absence of sunspots.

For example, Montanari observed the Sun in 21 October 1672. However, he was using the great meridian line located at the Basilica of San Petronio in Bologna (Italy) (see Heilbron (1999) for more information about this gigantic camera obscura). His astrometric results were published by E. Manfredi (1736) and Hoyt and Schatten interpreted this observation as zero sunspot groups. However, the tables that we can consult in Manfredi (1736) never include information on sunspots. The negative records of Picard and Hevelius during October 1672 probably are also related to solar meridian observations. In fact, the information about sunspots reported by Hevelius on these dates are included in a table of solar meridian observations published in the last part of his book

*Machinae Coelestis* (1679). Figure 13 shows a fragment of this table, where the date, the solar meridian altitude, the instrument, the state of the sky, the quality of the observation and additional comments are listed. Note how, in this table, comments about sunspots are not exactly associated with the reported solar meridian altitude.

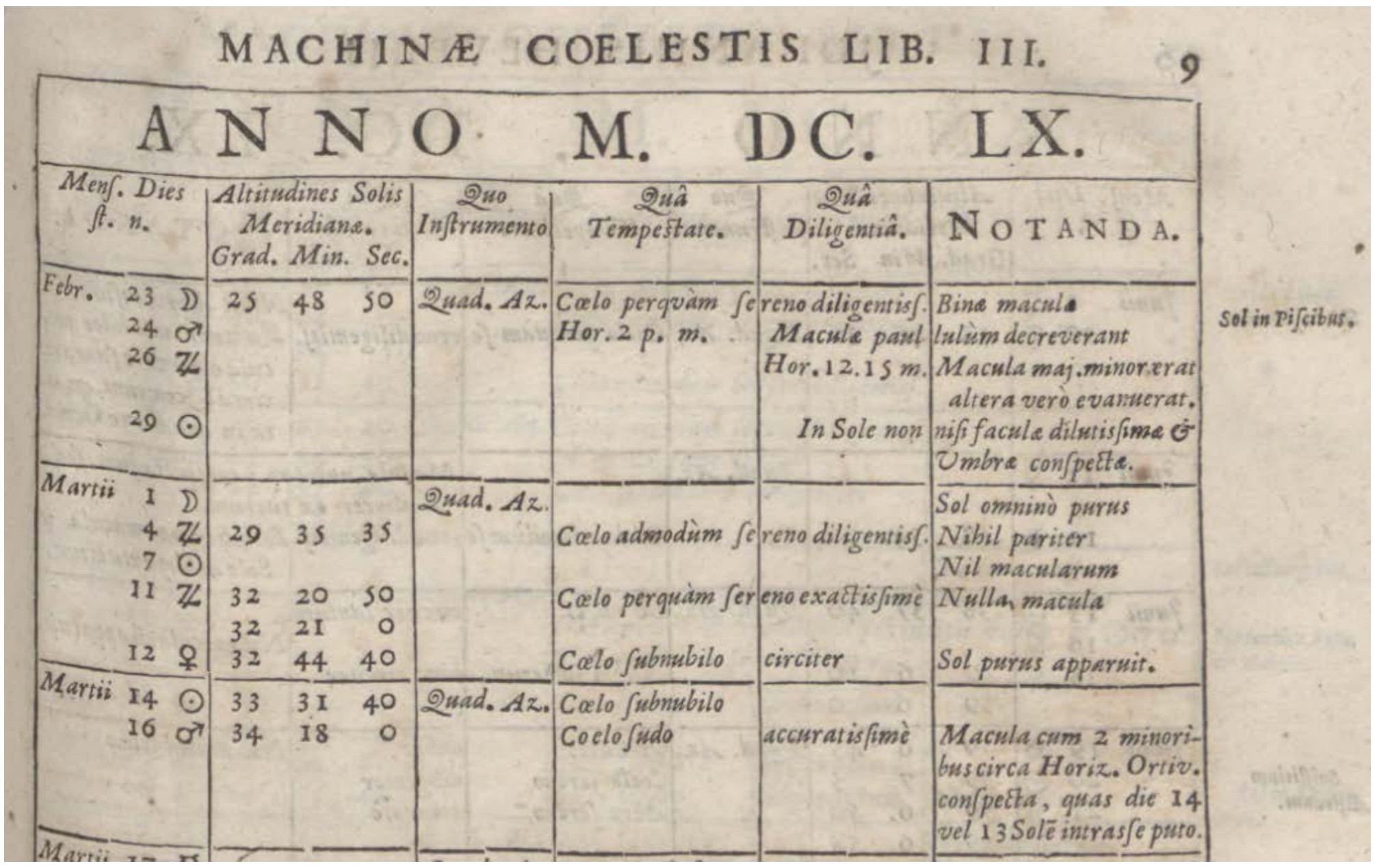

MACHINÆ COELESTIS LIB. III. 9

ANNO M. DC. LX.

| Mens. Dies st. n. | Altitudines Solis Meridianæ. Grad. | Min. | Sec. | Quo Instrumento | Quâ Tempestate. | Quâ Diligentiâ. | NOTANDA. | |
|---|---|---|---|---|---|---|---|---|
| Febr. 23 ☽ | 25 | 48 | 50 | Quad. Az. | Cœlo perqvàm se | reno diligentiss. | Binæ maculæ | Sol in Piscibus. |
| 24 ♂ | | | | | Hor. 2 p. m. | Maculæ paul | lulùm decreverant | |
| 26 ♃ | | | | | | Hor. 12.15 m. | Macula maj. minor erat altera verò evanuerat. | |
| 29 ☉ | | | | | | In Sole non | nisi faculæ dilutissimæ & Umbræ conspectæ. | |
| Martii 1 ☽ | | | | Quad. Az. | | | Sol omninò purus | |
| 4 ♃ | 29 | 35 | 35 | | Cœlo admodùm se | reno diligentiss. | Nihil pariter | |
| 7 ☉ | | | | | | | Nil macularum | |
| 11 ♃ | 32 | 20 | 50 | | Cœlo perquàm ser | eno exactissimè | Nulla macula | |
| | 32 | 21 | 0 | | | | | |
| 12 ♀ | 32 | 44 | 40 | | Cœlo subnubilo | circiter | Sol purus apparuit. | |
| Martii 14 ☉ | 33 | 31 | 40 | Quad. Az. | Cœlo subnubilo | | | |
| 16 ♂ | 34 | 18 | 0 | | Coelo sudo | accuratissimè | Macula cum 2 minoribus circa Horiz. Ortiv. conspecta, quas die 14 vel 13 Solē intrasse puto. | |

Figure 13: A fragment of the table of solar meridian altitudes published by Hevelius in his *Machinae Coelestis*, showing records of sunspots from 23 February to 16 March 1660.

Accordingly, Vaquero and Gallego (2014) have reached the conclusion that solar meridian observations should be used with extreme caution to evaluate past solar activity. If no sunspots were mentioned in a meridian observation, it does not necessarily mean that they were absent. Vaquero and Gallego (2014) presented, as an example, the marginal notes about sunspot observations that were included in the manuscripts of the meridian solar observations made at the Royal Observatory of the Spanish Navy during the "modern" period 1833–1840. These meridian observations were made by Thomas Jones (meridian telescope: D = 0.125 m, f = 3.05 m). A simple analysis shows that there is not a clear relationship between the timing of recorded spots and solar activity indices. Additionally, no notable periodicities in these observations were detected. It is noteworthy that these records of sunspots only appear in the manuscripts of the observations and do not appear in the printed version of the meridian observations. Thus, the sunspots recorded in the manuscripts of the meridian solar observations made in the Royal Observatory of the Spanish Navy illustrate the difficulty to reconstruct the past solar activity from this kind of historical record.

The other important modification of the early part of the GN series is related to the anomalous shape of solar cycle -1 (approximately from 1733 to 1744). The original GN series shows three peaks in 1736, 1739 and 1741 (see Figure 14, red line), i.e., a cycle shape completely different from all other observed solar cycles. The last of those peaks has the highest amplitude and should be considered as the maximum of the cycle. Although we still do not have a fully satisfactory solution, several recent publications find a more plausible shape for this solar cycle.

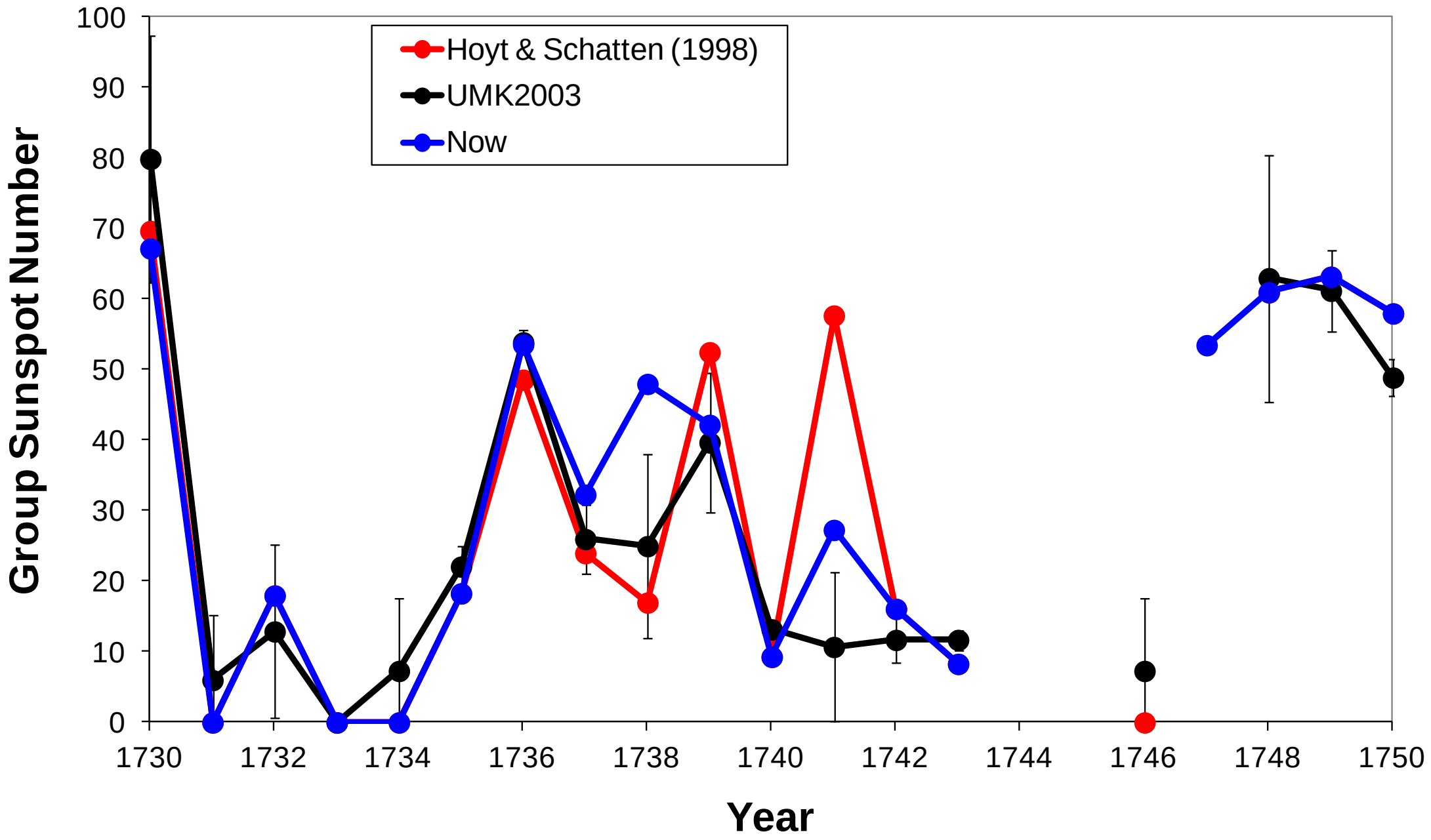

Figure 14: Solar activity around the solar cycle -1: original GN series (red), weighed values by Usoskin et al. (2003) (black line) and the latest values from Vaquero et al (2007a, 2007b), Vaquero and Trigo (2014) and Arlt (personal communication) in blue.

Vaquero et al. (2007a) improved the reliability of the GN series for the period 1736-1739 using the information about sunspot observations published in three of the most important scientific journals of that epoch: "Philosophical Transactions", ''Histoire de l'Académie Royale des Sciences", and "Nova Acta Eruditorum". They identified 42 papers containing solar observations, including 30 papers with relevant information on sunspots, and from these, they provided corrected GN values for the years 1736-1739.

Moreover, Vaquero et al. (2007b) reviewed the sunspot observations included in the compilation by Maximiliano Hell (1768) of astronomical observations made by Jesuits in China during the period 1717-1752. Here, the sunspot information is based exclusively on solar eclipse observations. From Hell's compilation, Hoyt and Schatten (1998a,b) deduced that there were no spots on the Sun on the day of an eclipse in 1731 and during another eclipse in 1746. Conversely, Vaquero et al. (2007b) pointed out that the correct interpretation is that no sunspot observations were made on these days. This is especially important in the case of 1746, because it could create a gap from 1744–1747 during which sunspot observations are not available. Fortunately, Rainer Arlt has found recently a "new" sunspot observer (Pehr Wargentin, from Uppsala) who left 17 drawings for 1747. The average group number is 3.53 (Arlt, personal communication). Therefore, we obtain a group sunspot number for this year equal to 53.5, assuming a calibration constant equal to 1.255 used by Hoyt and Schatten (1998) for similar cases. In any case, an important data gap remains in the period 1744-1746.

Usoskin et al. (2003) indicated the anomalous GN value for the year 1741 using a statistical analysis (see Figure 14, black line). Recently, Vaquero and Trigo (2014) have shown that the original manuscript (preserved in the Harvard University Archives) of the report by John Winthrop dated 10 January 1741 is crucial to solve this problem. Interpreting this manuscript, they provide a new value significantly lower than the original GN value. With this revision, solar cycle -1 adopts a much more normal shape, more comparable to all the other observed cycles (see Figure 14, blue line).

### 4.2. The Zürich reconstruction and anomalous cycle 4 (1750-1849)

Corrections to the original GN series have also been proposed for the period 1775-1795, when there are very few sunspot observations. For instance, Vaquero (2004) and Arlt (2009b) revisited sunspot observations in the years 1784 and 1795-1797 respectively. However, the main issue is related to an anomaly in the late part of cycle 4, preceding the Dalton minimum, diagnosed by several independent studies (Loomis 1870, Gnevyshev and Ohl 1948, Sonett 1983). The most striking hypothesis invokes the presence of a "lost" cycle between solar cycles 4 and 5 (according to the numbering by Wolf), as proposed by Usoskin et al. (2001). More recently, using recovered solar drawings by Staudach (Arlt 2008) and Hamilton (Arlt 2009b), Usoskin et al. (2009) managed to build the solar butterfly diagram for that period and found a sudden, systematic occurrence of sunspots at high solar latitudes in 1793–1796 (Fig. 15). This strengthened the case of a new cycle started in 1793, which would be missing in the original Wolf SN.

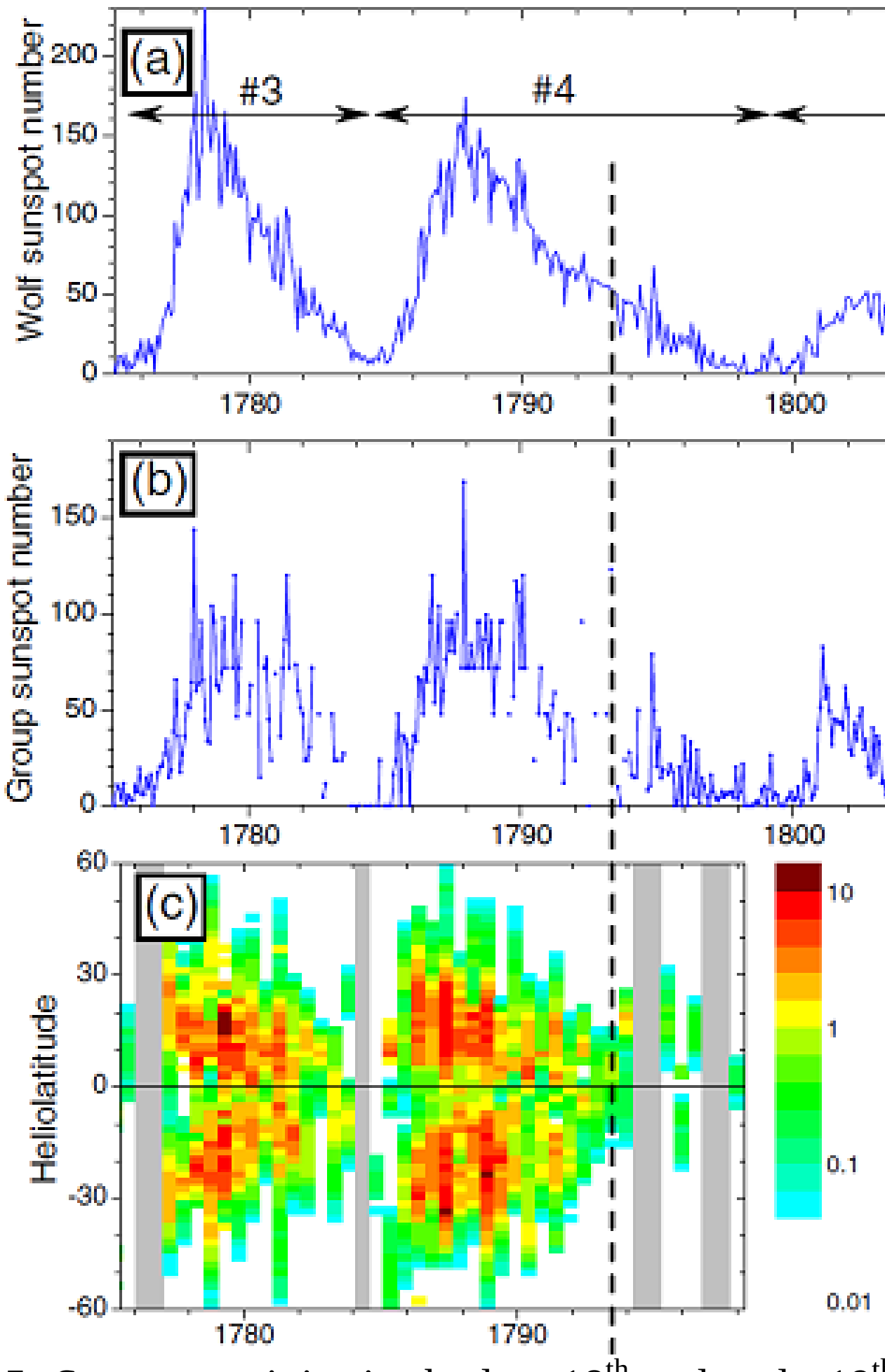

Figure 15: Sunspot activity in the late 18$^{th}$ and early 19$^{th}$ century: (a) Monthly mean Wolf SN. (b) Monthly mean GN. (c) Reconstructed sunspot butterfly diagram. The color scale on the right gives the density (in year$^{-1}$ deg$^{-1}$) of sunspots in latitude-time bins and gray bars indicate that no latitudinal information is available. Each bin covers 2° in latitude and six months in time. The vertical dashed line marks the start of the "lost" cycle, late in 1793 (from Usoskin et al. 2009).

However, using another approach, Zolotova and Ponyavin (2011) conclude that the unusual length of cycle 4 can be the result of a late pulse of activity in the northern hemisphere during the declining phase, similar to what occurred in more recent cycles, like cycle 20. They conclude that splitting cycle 4 in two brief cycles leads to anomalously short durations, even less compatible with statistical properties of all known cycles than a long 14-year cycle 4. Therefore, this issue still remains open and new findings in historical archives and libraries are necessary to improve the database of historical solar observation and, thus, our knowledge on solar activity over this peculiar period.

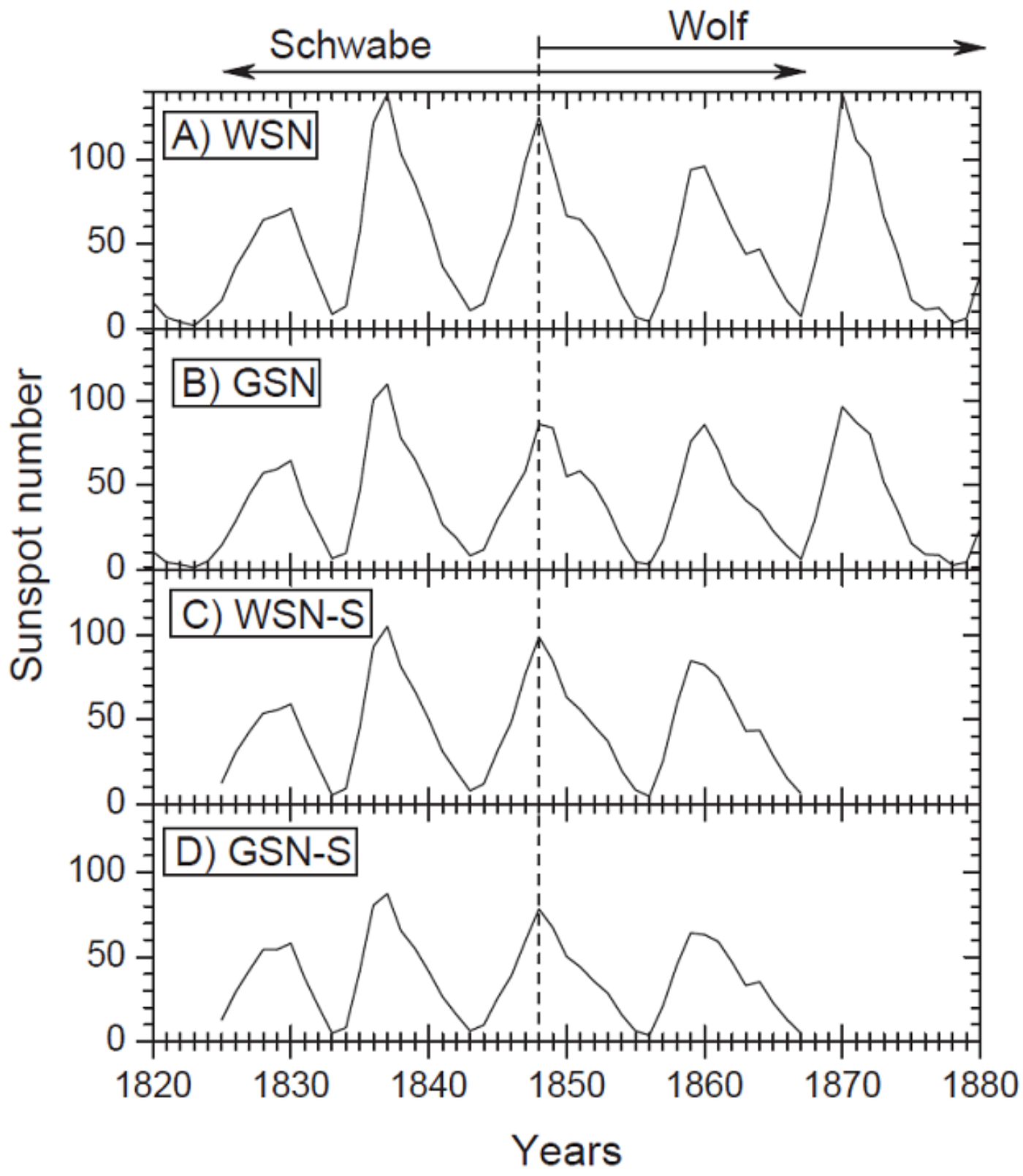

Figure 16: Yearly mean SNs: (a - WSN) the original Wolf SN, (b - GSN) the GN, (c – WSN-S) the Wolf SN calculated from Schwabe data and (d – GSN-S) the GN calculated from Schwabe data (Figure from Leussu et al. 2013).

A last issue was raised recently that questions the calibration of all early observations versus Wolf's own systematic sunspot numbers, which rests mostly on Schwabe's series. Using the Schwabe original sunspot data collected by Arlt (2011) and Arlt et al. (2013), Leussu et al (2013) recently compared the original Wolf SN and the GN series with the long homogeneous sunspot record by Schwabe (1835–1867). Figure 16 gives a comparison of those yearly SNs and indicates the intervals when the main observers (Schwabe and Wolf) were active. They find that while the GN series is homogeneous and consistent with the Schwabe data throughout the studied period, the scale of the Wolf SN shows a jump around 1848. Based on their results, they conclude that all values in the original Wolf SN need to be lowered by 20% before 1848. As we will see in the next section, the series resulting from other well established corrections does not support this conclusion. Further analysis will thus be required to clarify this apparent contradiction.

# 5. Biases in the Zürich era: causes and diagnostics

## *5.1. The 1880 SN-GN divergence (1850-1930)*

A powerful way of comparing the two time series is to form the ratio between them, like in Figure 1. In Figure 17, we show again the ratio between the (Hoyt and Schatten) Group Number and the “Wolf” Sunspot Number, as preserved by the WDC-SILSO, but giving the mean ratio for each year in the interval 1700-1995 and only keeping the most accurate values by restricting the yearly ratio to years when the yearly values are above a suitable threshold, thus avoiding the smallest values, e.g. zeros. From this plot, it is clear that over the last two centuries, there are essentially two discontinuities in this ratio, as well as minor, short-lived drifts.

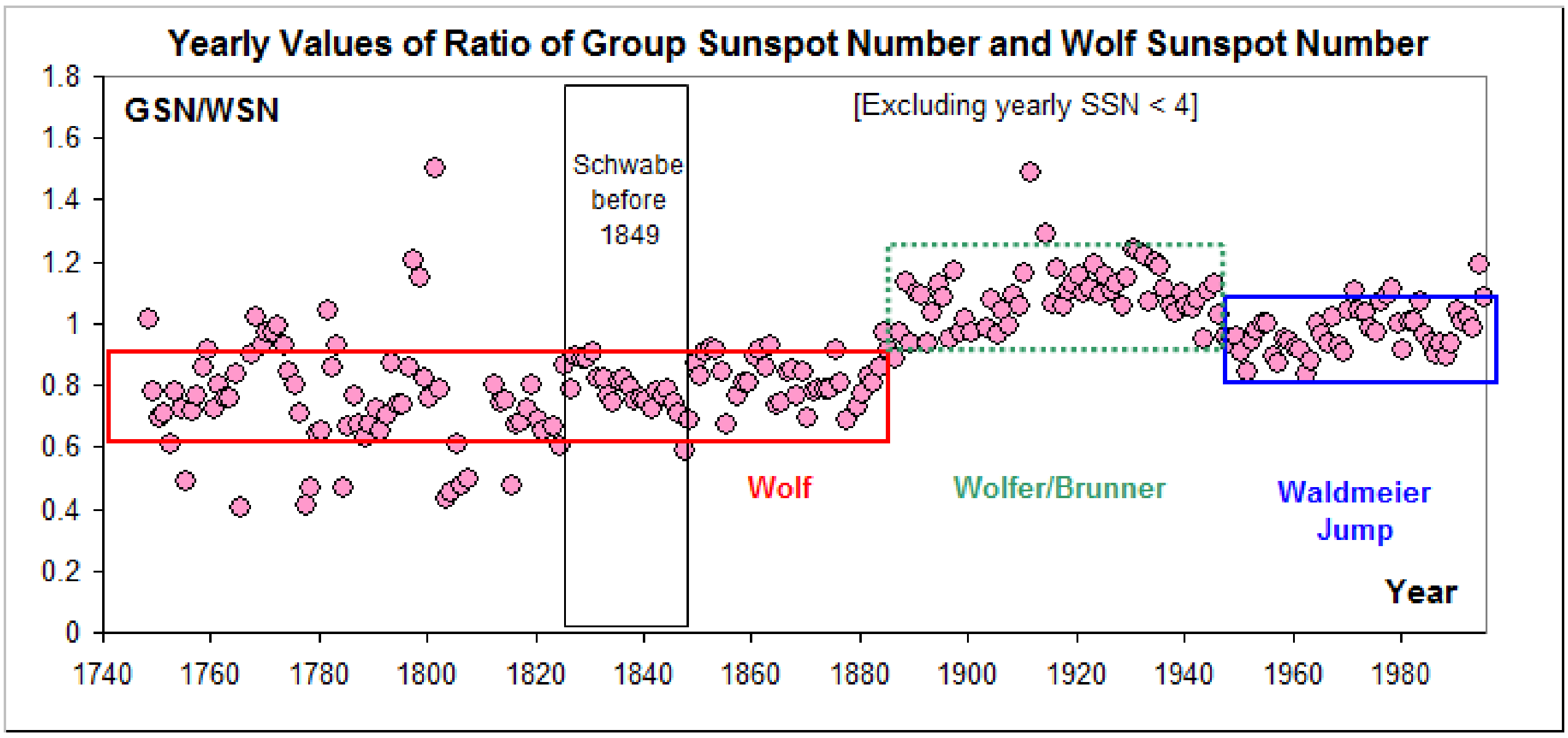

Figure 17: Ratio between yearly averages of Group and "Wolf" Sunspot Numbers (when both are not less than 4).

In this section we shall explore the first jump around 1885. We shall use yearly averages of the original Group numbers as reported by Hoyt and Schatten, calculated by averaging for each year all monthly values for which there is data.

The *backbone-method* developed by Svalgaard (2013b) for reconstructing the GN series starts by selecting a single "primary" observer for an interval of time. The selection should be based both on the length of the observational series (as long as possible) and on the perceived "quality" of the observations, taking into consideration such factors as regularity of observing, suitable telescope, and lack of obvious problems. Two backbones will first be discussed here, those of Schwabe (1794-1883) and Wolfer (1841-1944). The Schwabe backbone is centered on the observing interval for Schwabe and includes all "reliable" observers who overlap in time with Schwabe. The reliability is judged by how high the correlation is between simultaneous (on a yearly basis) observations by the observer and by Schwabe. Similarly, the Wolfer backbone includes all reliable observers who overlap with Wolfer. The two backbones overlap by 42 years so can be cross-calibrated with confidence. Figure 18 gives an overview of the time intervals observed by the observers listed.

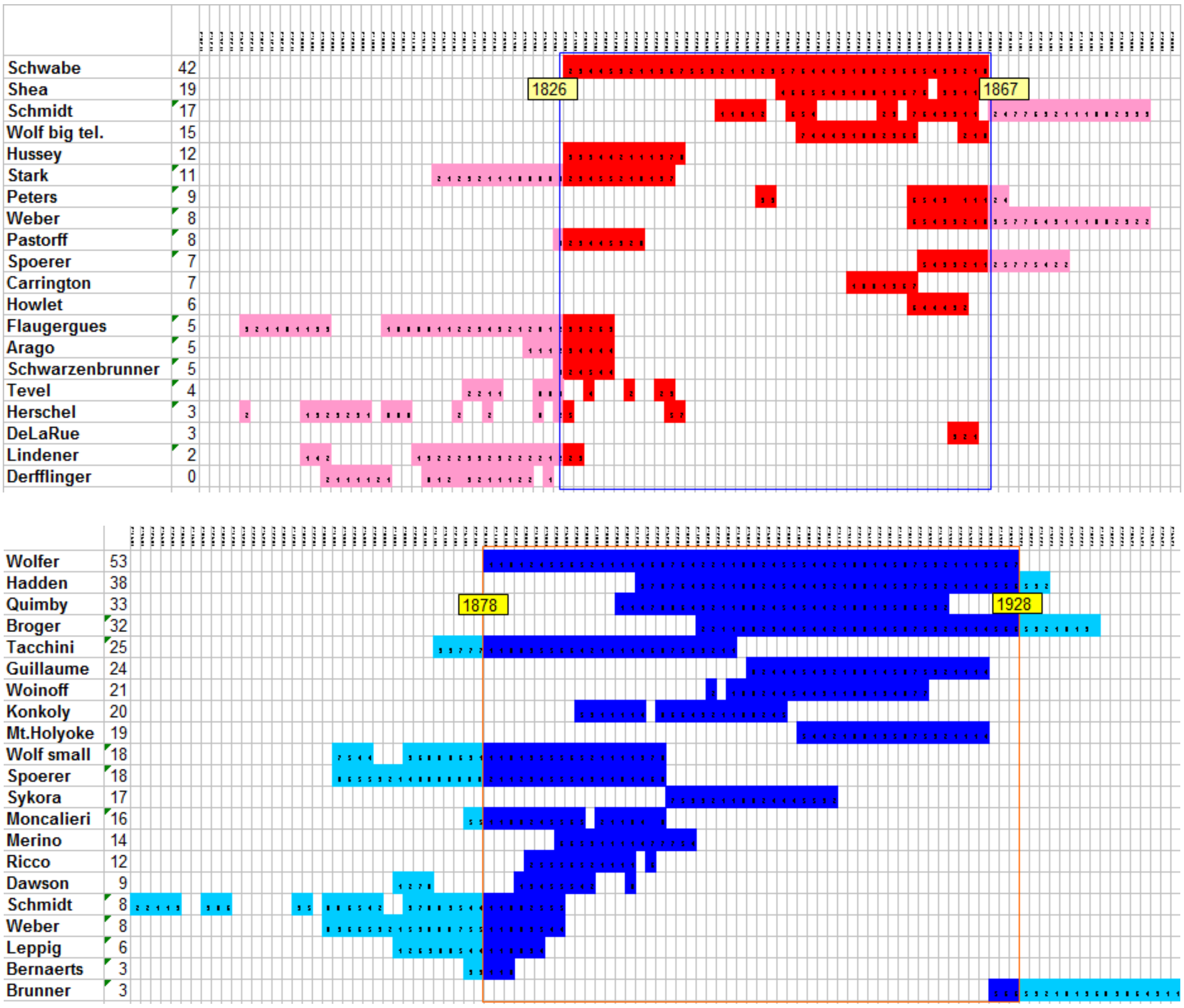

Figure 18: (Top) Coverage and observers for the Schwabe Backbone (1794-1883). (Bottom) Coverage and observers for the Wolfer Backbone (1841-1944).

For each Backbone, we regress the primary observer's group count against each observer's count for each year and we plot the result (for an example see Figure 19). Experience shows that the regression line almost always very nearly goes through the origin, so we force it to do so and calculate the slope and various statistics, such as 1-σ uncertainty and the F-value.

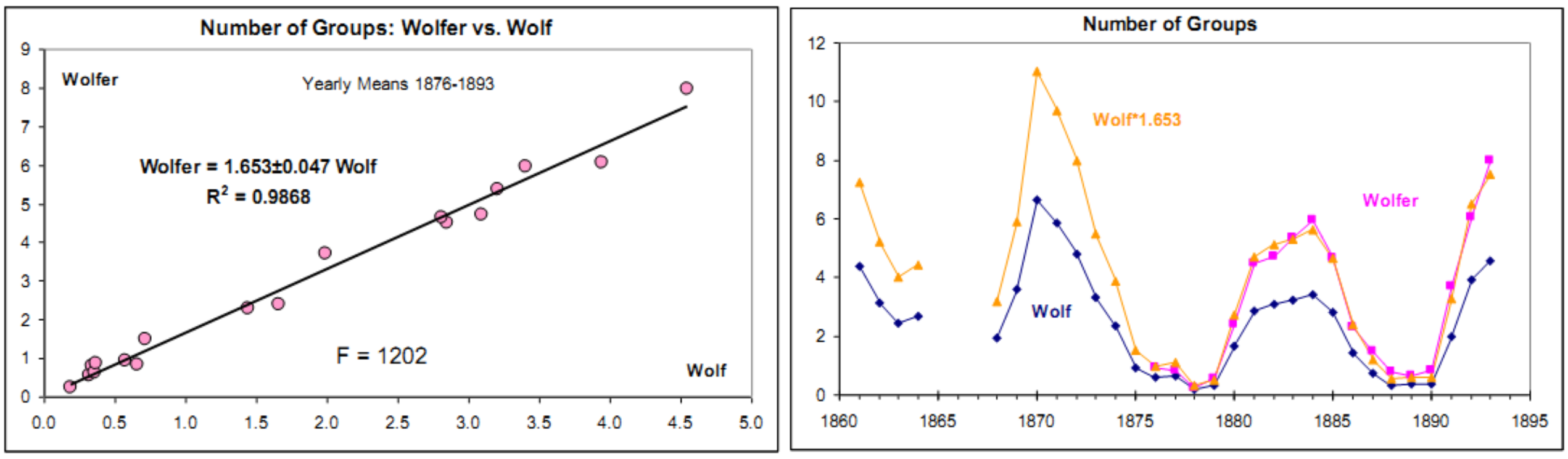

Figure 19: Regression of number of groups observed by Wolfer (with standard telescope) against the number of groups observed by Wolf (with small telescope).

The slope gives us what factor to multiply the observer's count by to match the primary's count. The right panel shows a result for the Wolfer Backbone: blue is Wolf's count (with his small telescope), pink is Wolfer's count (with the larger telescope), and the orange curve is the blue curve multiplied by the slope, bringing Wolf's observations on the same scale as Wolfer's. It is clear that the harmonization works well and that it shows that Wolfer with the larger telescope saw 65% more groups than Wolf did with the small, handheld telescope (Figure 5) as we would rightly expect. Applying this methodology yields the two backbones (Figure 20). We stress that the backbones are independent and are based purely on solar observations with no empirical or *ad hoc* adjustments apart from the (necessary) harmonization just described.

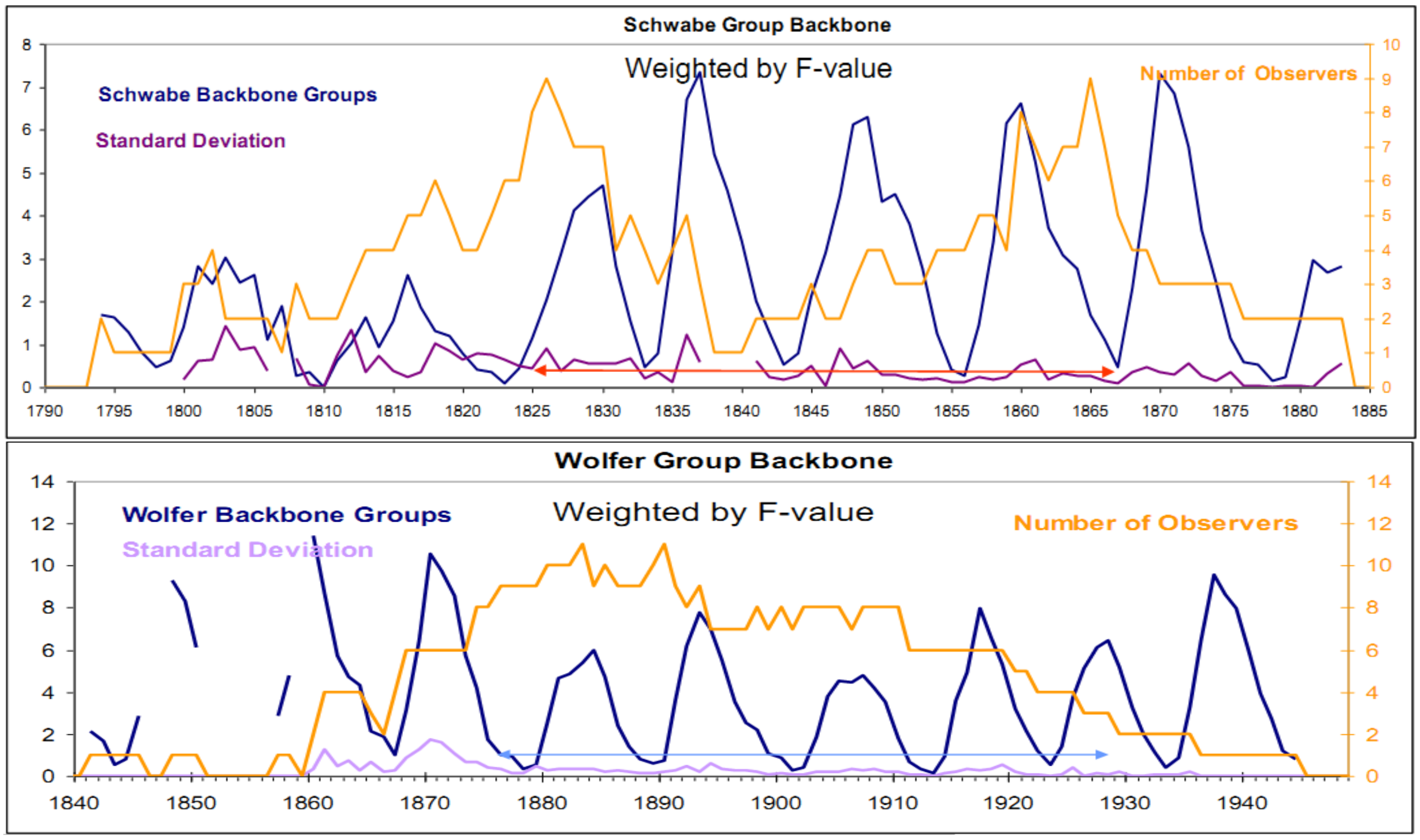

Figure 20: (Top) Schwabe (interval of observation: red arrow) and (bottom) Wolfer (interval of observation: blue arrow) backbones (blue curves) and the numbers of observers (orange curves) contributing to the mean weighted by their goodness of fit. The standard deviation is shown by the purple curves.

It is of considerable interest to compare our "Schwabe" backbone with the Group Counts compiled by Hoyt and Schatten (Figure 21). Apart from the very noisy period before 1815, the agreement is very good, as would be expected as the series are based on the same data. The minor disagreement around ~1838 for the maximum of solar cycle 8 needs to be resolved, and then, of course, we can see the beginning of the drift after 1882.

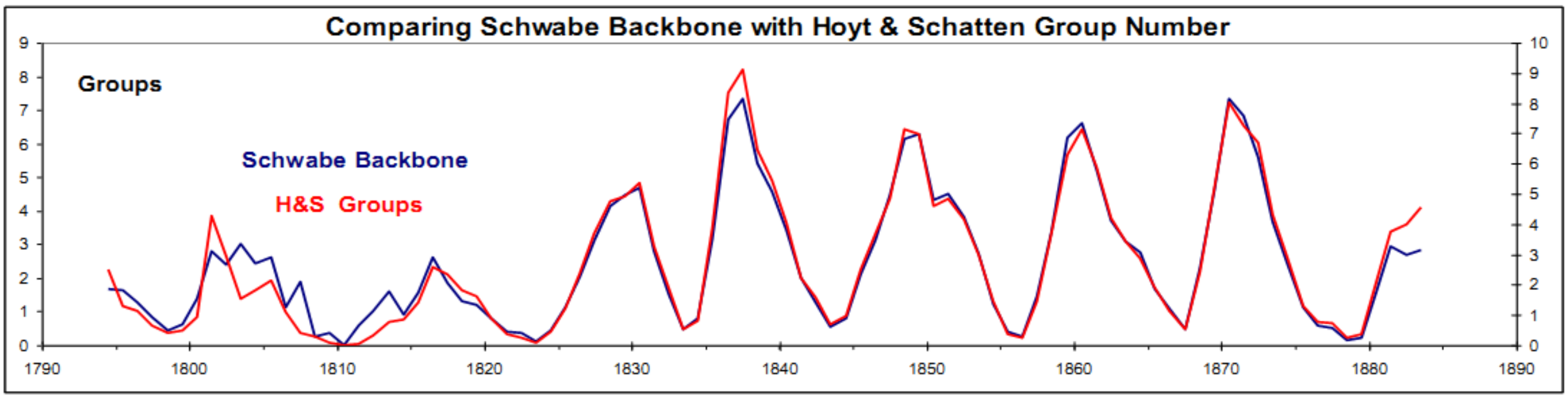

Figure 21: Number of Groups per year reported by Hoyt and Schatten (1998b) (red curve; right-hand scale) and resulting from the Schwabe backbone (blue curve; left-hand scale). The scale of the backbone is at this point “free floating” and, in fact, 9/10 that of Hoyt and Schatten’s.

The next order of business is to harmonize the two backbones, i.e. bring them onto the same scale. We shall use the Wolfer scale as the base scale, because of its larger number of (better?) observers and choose the common interval 1860-1883 as the basis for the normalization (Figure 22).

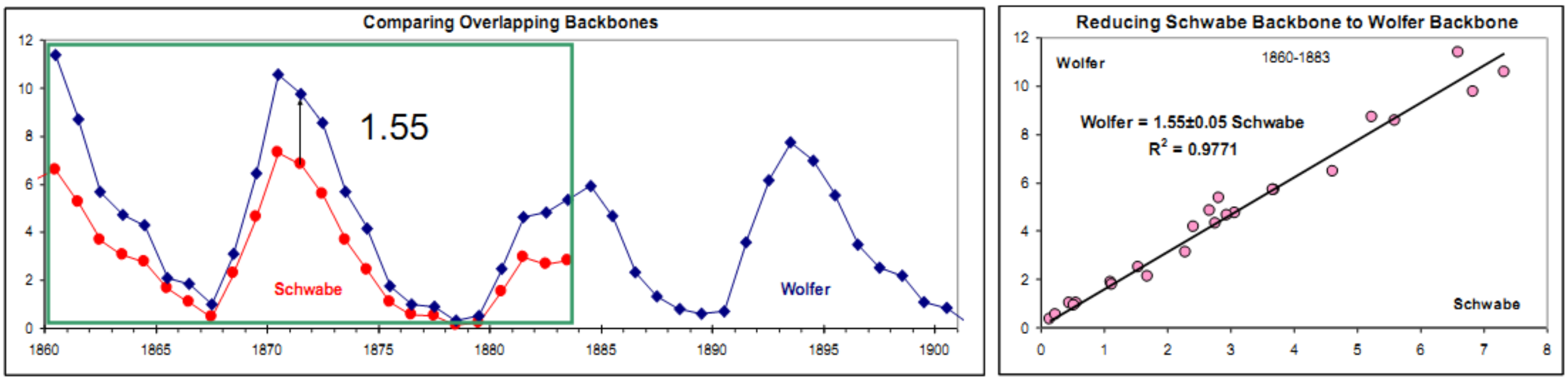

Figure 22: (Left) overlapping backbones: Wolf (blue) and Schwabe (red). (Right) Linear correlation showing the best (least-square) fit for the interval 1860-1883.

Assuming a normalization factor of 1.55 “explains” 98% of the difference between the two backbones, with no clear systematic variation with time, we can thus produce a composite series by multiplying the Schwabe backbone values by 1.55 and then simply average the resulting, normalized Schwabe backbone and the Wolfer backbone (Figure 23).

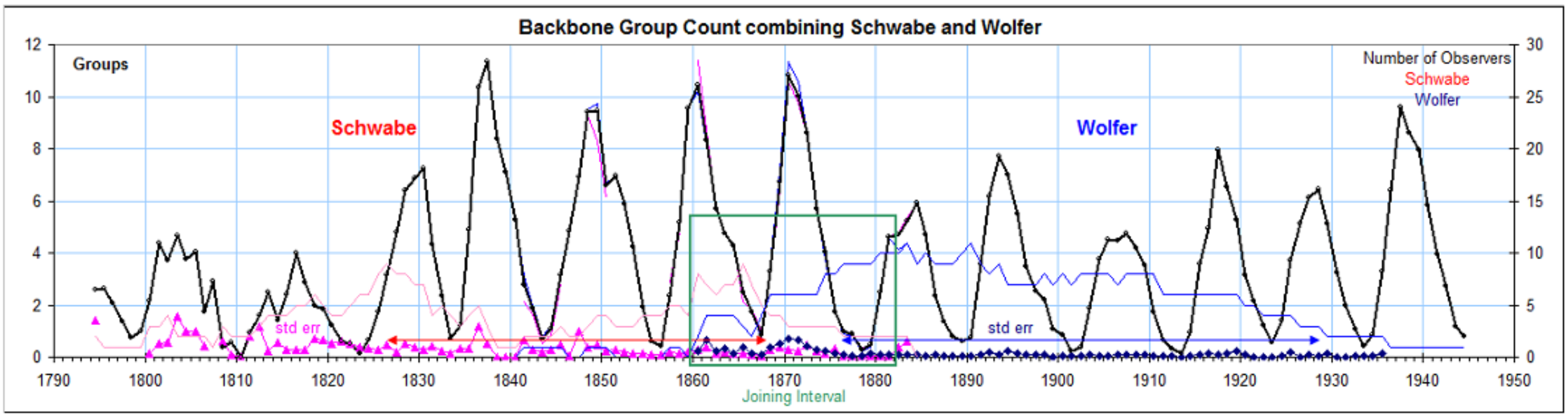

Figure 23: Composite backbone 1794-1944, with the standard error of the mean.

Hoyt and Schatten used the Group Count from RGO (Royal Greenwich Observatory) as their Normalization Standard. However, the ratio between the RGO group count and the Wolfer backbone count is not stable (Figure 24).

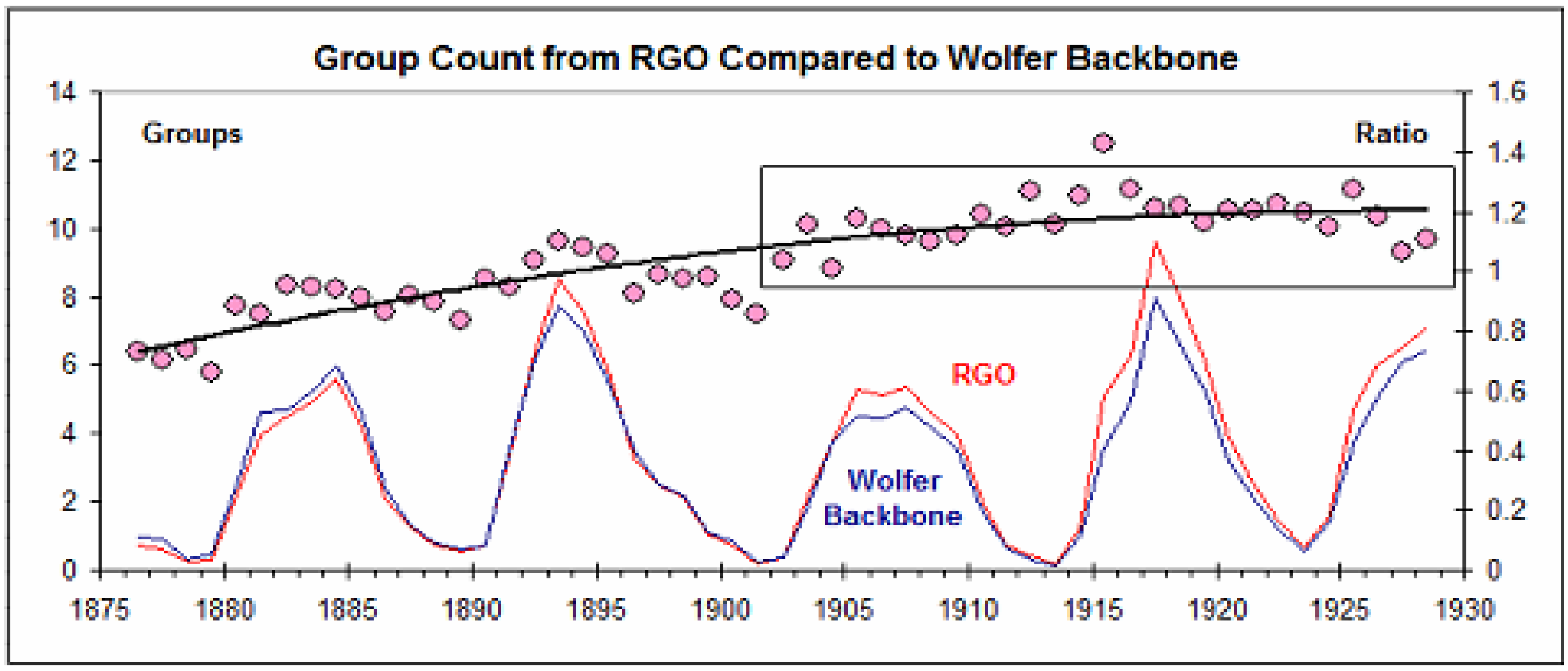

Figure 24: Ratio (right-hand scale) between Group Count from RGO (red curve) and Wolfer Backbone (blue curve). After ~1900, the ratio is approximately constant, but before there is a clear progressive change. This change translates into the progressive drift of the SN/GN ratio over 1875-1915 seen in Figure 17. Over this time interval, the number of observers per year is high (>5) and the standard deviation is comfortably low, as shown by Figure 20.

This and the discrepancy between the Wolfer/Wolf ratio (Figure 19) that we find (1.65) and that used by Hoyt and Schatten (1.02) seem to be the main reasons behind the large difference between the GN and the Zurich SN before and after ~1885. It is not clear why Hoyt and Schatten report an almost equal normalization factor for Wolf and for Wolfer with respect to the RGO group count, in spite of the fact that Wolf used the much weaker, handheld 37mm telescope compared to the standard 80mm telescope used by Wolfer.

At this point, the composite backbone is still "free floating". We wish to connect it to "modern" observations so construct yet another backbone (from 1921-2000), based of the group counts by the National Astronomical Observatory of Japan (We name the backbone in honor of the principal observer, Ms. Hisako Koyama, 小山 ヒサ子 (1916-1997)). See Figure 25.

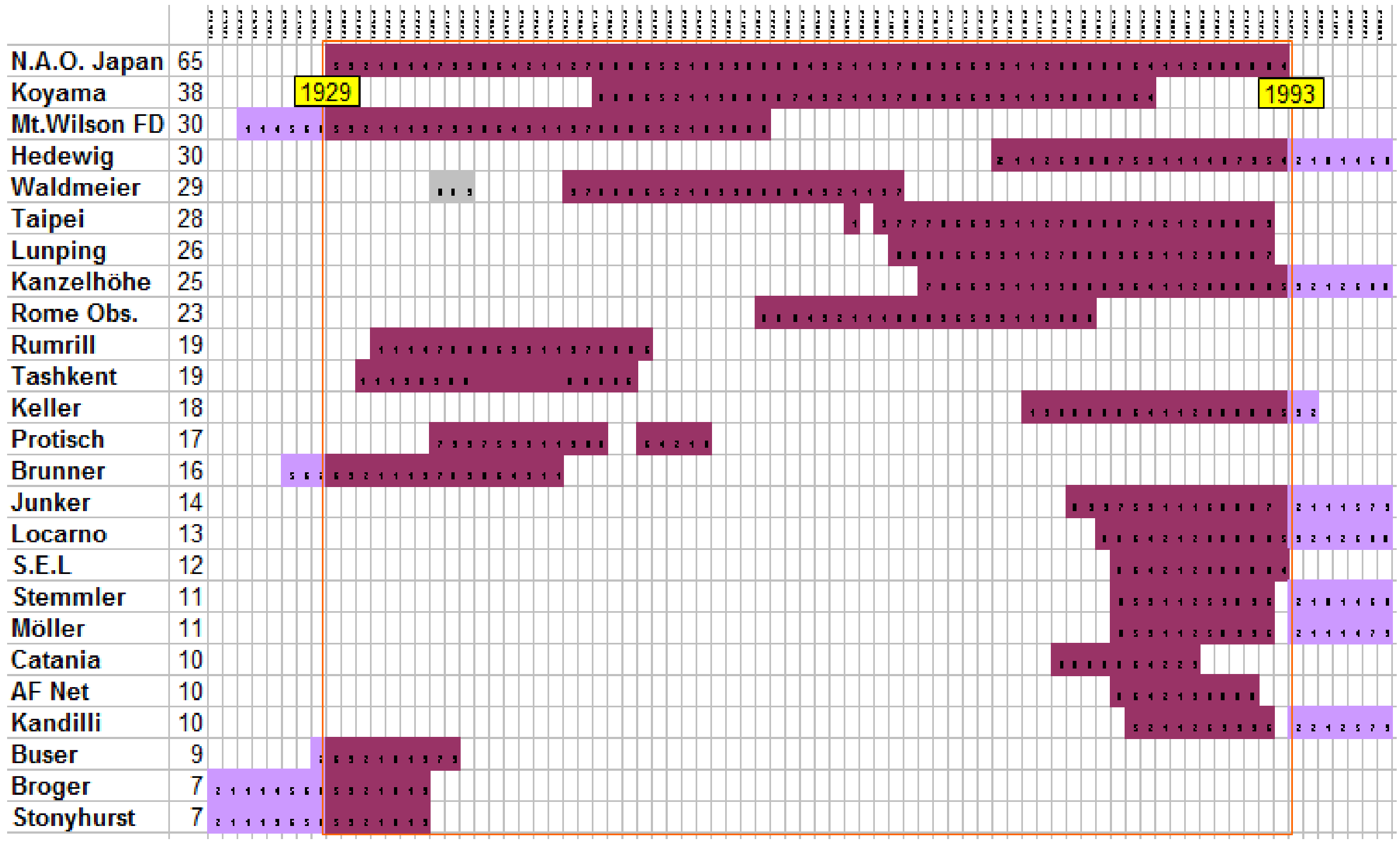

Figure 25: Coverage and observers for the Koyama Backbone (1921-2000).

The resulting backbone is shown in Figure 26 in the same format as used for Figure 20. The correlation with the Wolfer backbone for the 24 years of overlap (1921-1944) is very high: Wolfer = 1.0002 Koyama ($R^2$ = 0.9952), i.e. indicating that the two backbones already match exactly and no re-normalization is needed to join them.

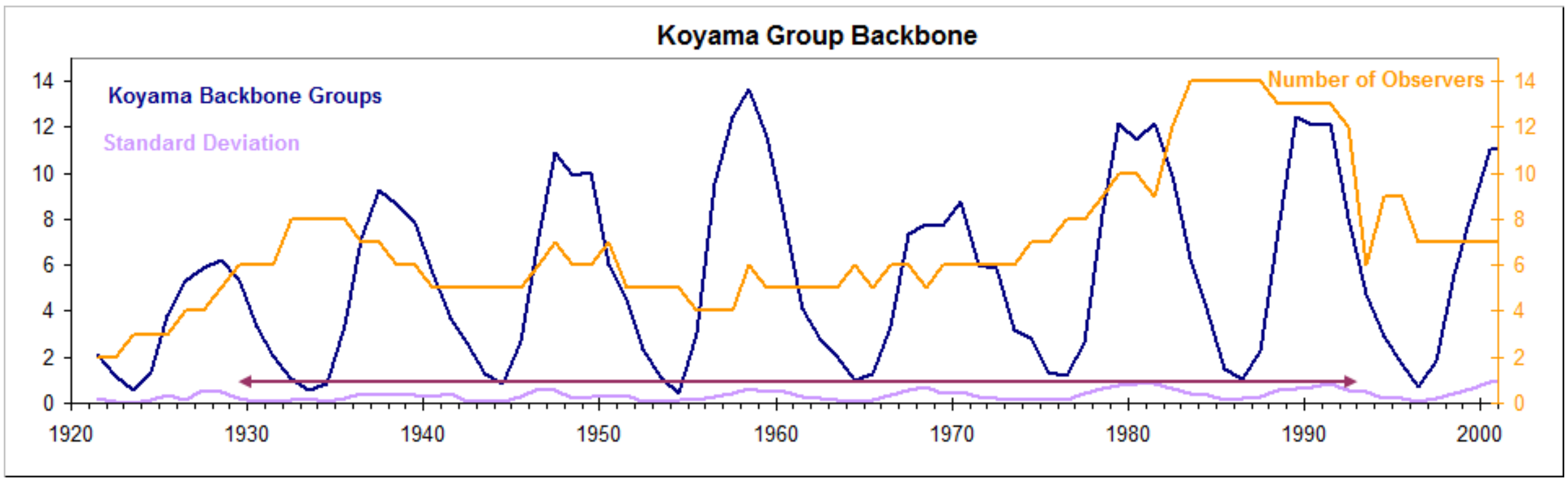

Figure 26: The Koyama backbone (interval of observation: purple arrow) and its number of observers (orange curve). The standard deviation is shown by the bottom purple curve.

Finally, we also wish to extend the backbone reconstruction backwards to the 18th century. The original set of drawings constituting the long series of observations (1749-1796) by J. C. Staudach was examined by Wolf (1857), who determined group counts and sunspot counts for each drawing. Wolf's counts form the basis for the Staudach Backbone. The analysis of this is still ongoing but we shall here report a preliminary result (Figure 27).

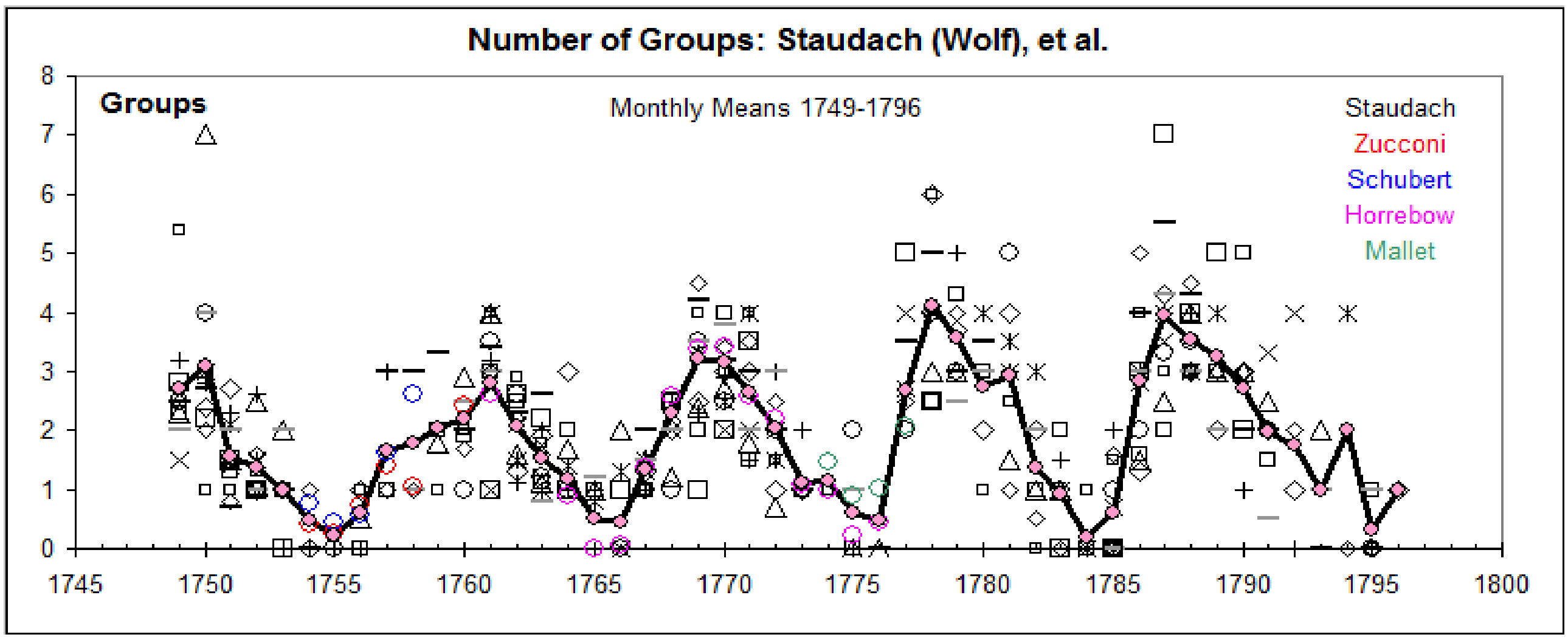

Figure 27: Open black symbols show the counts of groups by month (by Staudach as determined by Wolf (1857)). Yearly group counts by Zucconi (Venice), Schubert (Danzig), Horrebow (Copenhagen), and Mallet (Berlin) scaled to the yearly counts by Staudach are shown by colored open circles. The full black curve with small pink circles shows the yearly group counts, averaged over all observers, forming the Staudach Backbone, which is here "free-floating". A "group count" for an interval of time is the average of the number of groups observed by an observer on each day within the interval.

In his 1861 series, Wolf (1861) effectively doubled the counts that he had derived for Staudach, followed by a further factor of 1.25 increase in the 1882 series. Arlt (2008) suggests that Staudach missed all A- and B-groups (on the modern Waldmeier classification) on account of the relatively low quality of his telescope, perhaps justifying the doubling assumed by Wolf, as A&B-groups make up about 40% of all groups. Comparisons with the geomagnetic and cosmic ray records are consistent with an overall (but still highly uncertain) factor of ~3 to match the combined Schwabe-Wolfer backbone. A task for the Fourth Sunspot Number Workshop is dedicated to improving the determination of that scaling factor, taking into account the recent digitization of Staudach's drawings by Arlt (2008).

Adding the raw group counts from SIDC's database to the Koyama backbone, allows us to present a preliminary synthesis of the evolution of a composite of the average number of groups per year back to 1749 derived from all four backbones, each backbone shown with a different color (Figure 28).

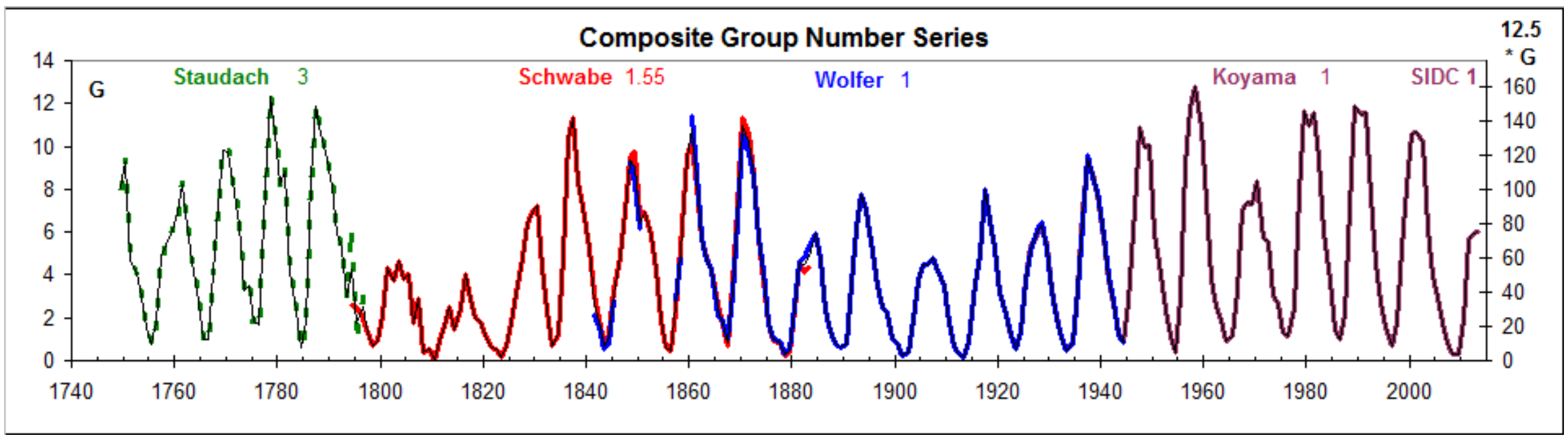

Figure 28: A composite record of the yearly mean number of groups, back to 1749, derived from the four backbones constructed as described in this section. The numbers to the right of the backbone designations indicate the scale factor applied to each "raw" backbone to harmonize them to the Wolfer scale. The green dashed line marking the Staudach backbone reflects that it is uncertain. The right-hand scale is for the thin black curve showing the quantity 12.5 times the average group count, as an “equivalent” GN.

In this section, we have striven to build a series based solely on solar observations. In another chapter of this volume (Svalgaard 2014b), we review what Geomagnetism can tell us about solar activity, but it is instructive already here to compare the number of groups with the range rY of the diurnal variation of the geomagnetic field, an indirect but fully independent tracer of solar activity introduced in section 2 (Wolf's magnetic needle readings). Figure 29 shows the excellent agreement between our reconstructed number of groups and this diurnal range throughout the entire interval 1845-2013.

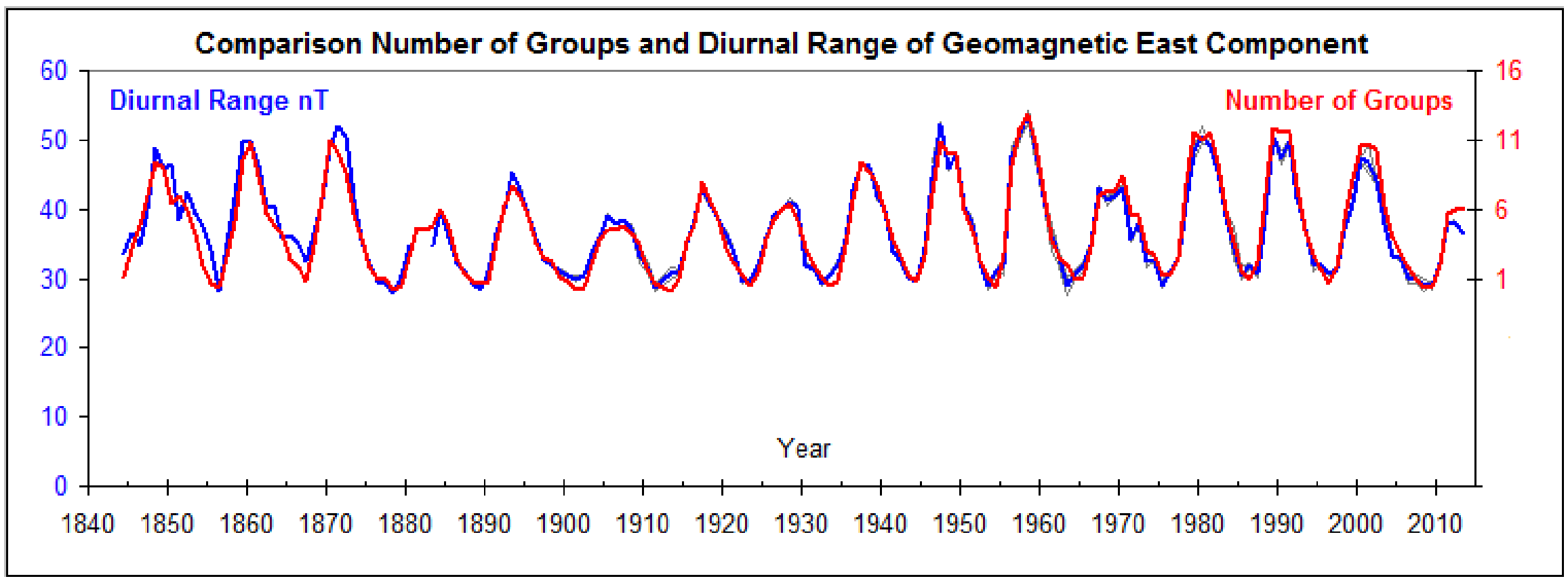

Figure 29: Number of groups (red curve) compared with the range of the diurnal variation of the East Component (blue curve) of the geomagnetic field.

We can now compare the composite Group Number (GSN*) series with the official Zürich Sunspot Number ($R_Z$) and several geomagnetic indicators (Figure 30).

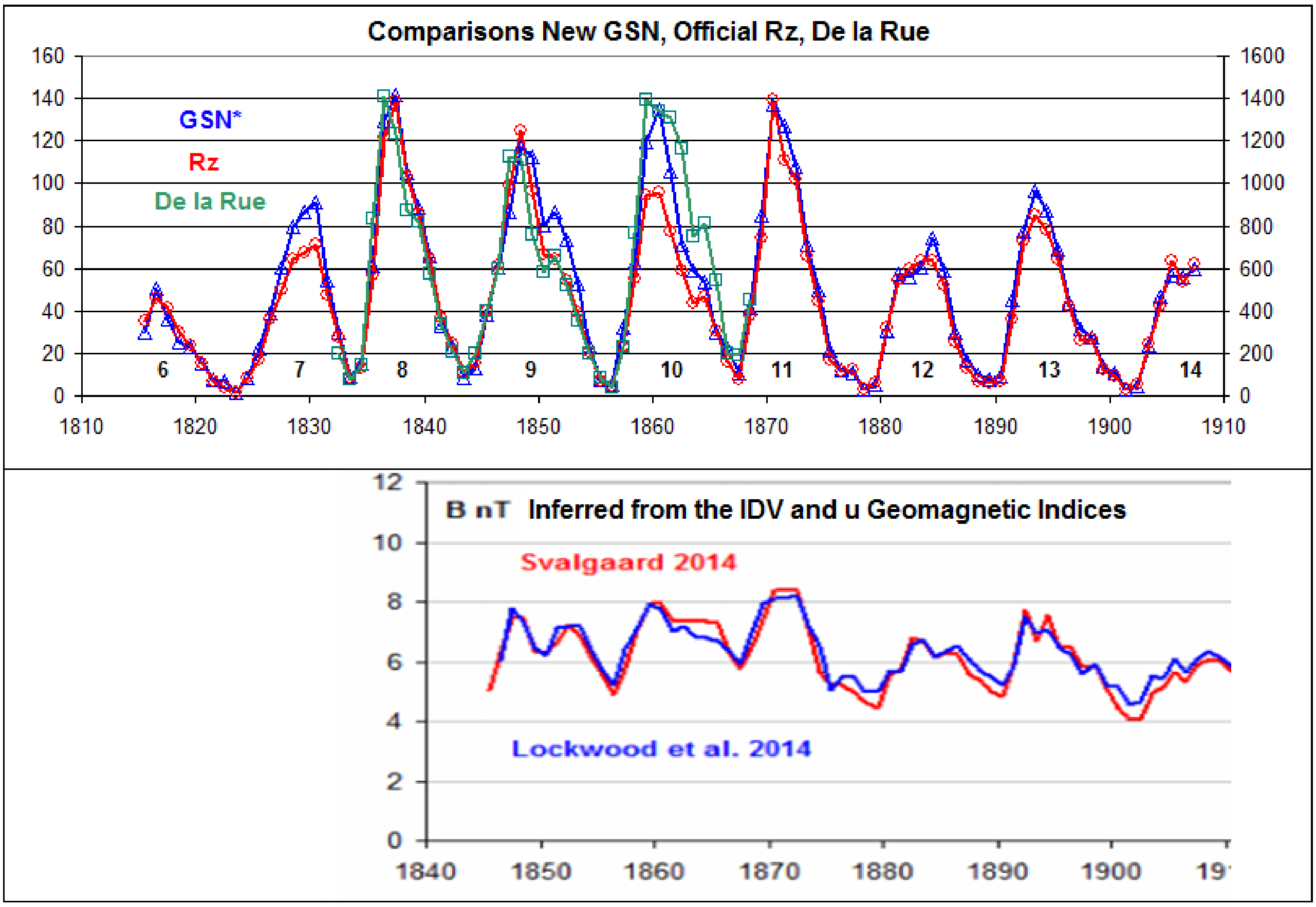

Figure 30: (Top) The composite GSN* matches $R_Z$ well, except for cycles 7 and 10. There is also a good match with the (scaled) sunspot areas determined by De la Rue (Vaquero et al. 2002). The geomagnetic record (bottom) of the inferred Heliospheric magnetic field (Svalgaard 2014a) also shows that cycle 10 was not lower than cycles 9 and 11.

Based on these comparisons, it does not seem reasonable to apply a wholesale decrease of the Wolf Numbers by 20% before 1849, as advocated by Leussu et al. (2013) in section 6.3. If anything, $R_Z$ is already too low (e.g. cycle 7). The size of cycle 10 (maximum in 1860.1) now becomes an important research problem, possibly for resolution at the Locarno 2014 Sunspot Number Workshop. This serves as a reminder that much work remains to be done.

The corrections derived thus far involved the Group number. After 1876, when the GN relies entirely on photographic data and is scaled to the SN, the latter becomes the actual reference. It turns out that other inhomogeneities were found in that more recent part of the SN. Those inhomogeneities, described in the following sections, will thus influence the scale of both the SN and GN over the entire period preceding the 20th century, as the calibration is established by working backwards in time.

## 5.2. The 1950 Waldmeier jump (1930-1980)

In this section, we address the discontinuity at ~1947 in Figure 17, which we have termed the “Waldmeier Jump” (Svalgaard 2010), because we attribute it to changes in the SN counting procedure implemented when Max Waldmeier became the Director of Zürich Observatory. In 1961, Max Waldmeier published the definitive Zürich sunspot numbers up until 1960 (Waldmeier 1961). He noted that “Wolf counted each spot – independent of its size – but single. Moreover, he did not consider very small spots, which are visible only if the seeing is good. In about 1882 Wolf’s successors changed the counting method, which since then has been in use up to the present. This new method counts also the smallest spots, and those with a penumbra are weighted according to their size and the structure of the umbra". In 1968 Waldmeier (1968, 1948) codified the weighting scheme as follows "Später wurden den Flecken entsprechend ihrer Größe Gewichte erteilt: Ein punktförmiger Fleck wird einfach gezählt, ein größerer, jedoch nicht mit Penumbra versehener Fleck erhält das statistiche Gewicht 2, ein kleiner Hoffleck 3, ein größerer 5". [2] However, Wolfer in 1907 (Wolfer, 1907) explicitly states: "Notiert ein Beobachter mit seinem Instrumente an irgend einem Tage *g* Fleckengruppen mit insgesamt *f* Einzelflecken, ohne Rücksicht auf deren Grösse, so ist die daraus abgeleitete Relativzahl jenes Tages $r = k(10g+f)$".[3] We can verify that Wolfer, contrary to Waldmeier's assertion that the Zürich observers began to use weighting "around 1882", did not weight the spots according to Waldmeier's scheme by comparing Wolfer's recorded count with sunspot drawings made elsewhere, e.g. Figure 31.

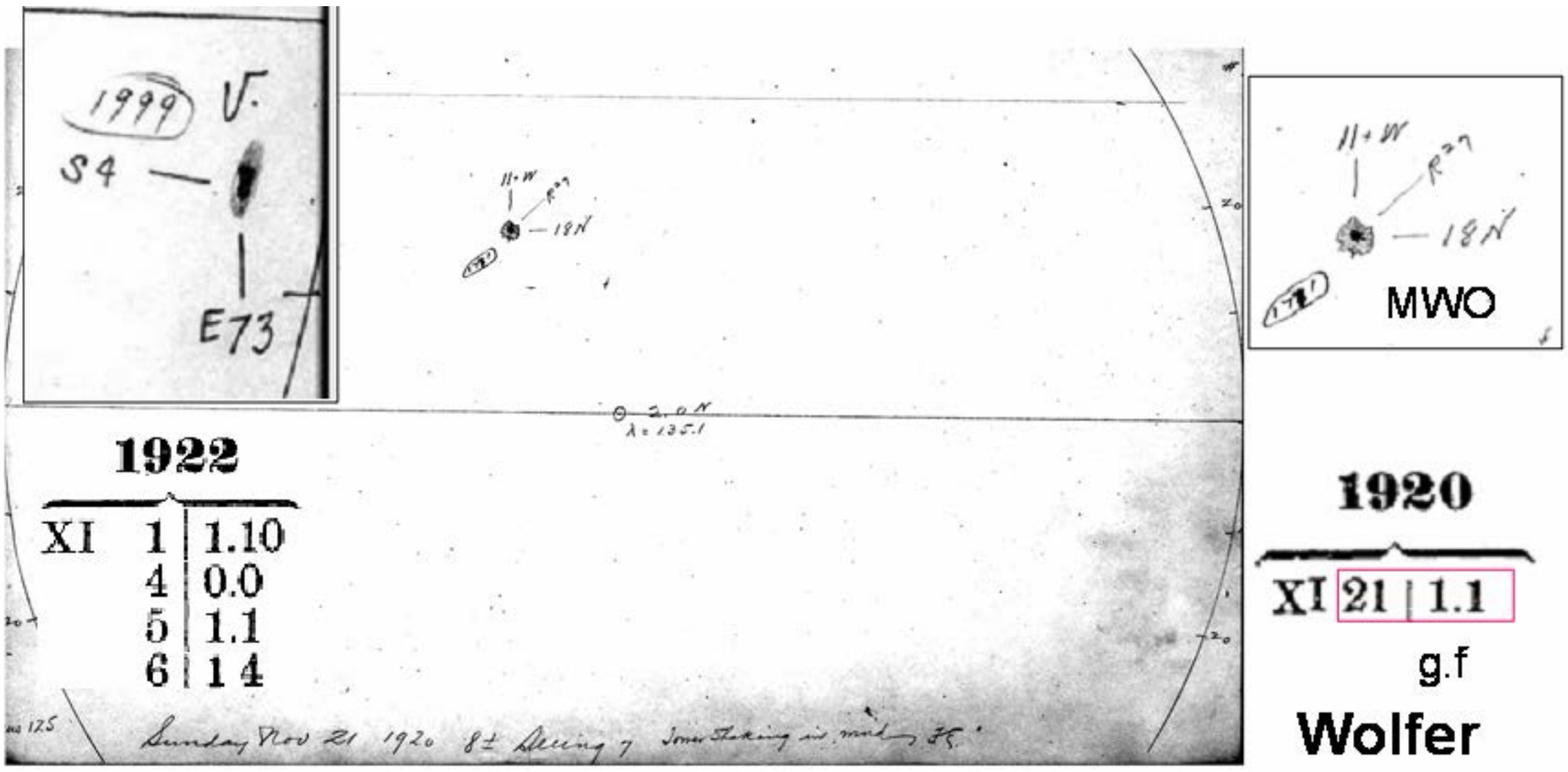

Figure 31: Drawing from Mount Wilson Observatory (MWO) of the single spot with penumbra on 21st Nov. 1920. The insert at the left shows a similar group observed at MWO on 5th Nov., 1922. For both groups, Wolfer should have recorded the observation as "1.3" if he had used the weighting scheme, but they were recorded as "1.1", thus counting the large spots only once (with no weighting).

L.Svalgaard uncovered many other such examples (e.g. 16th September, 1922 and 3rd March, 1924) in a systematic search of MWO drawings (from 1917-1925) for single sunspots with penumbra. He found no such spots during this period for which weighting was applied at Zürich. We thus consider it established that Wolfer did not apply the weighting scheme. This is consistent with the fact that nowhere in Wolf's and Wolfer's otherwise meticulous yearly reports in the *Mittheilungen über Sonnenflecken* series is there any mention of a weighting scheme.

We shall not here speculate about the motive or reason for Waldmeier ascribing the weighting scheme to Wolfer. Waldmeier himself was an assistant to Brunner since 1936 and performed routine daily observations with the rest of the team so would presumably have known what the rules were. Figure 32 shows that Brunner and Waldmeier were observing very close to the same scale in 1937:

[2] A spot like a fine point is counted as one spot; a larger spot, but still without penumbra, gets the statistical weight 2, a smallish spot within a penumbra gets 3, and a larger one gets 5.

[3] When an observer at his instrument on any given day records *g* groups of spots with a total of *f* single spots, without regard to their size, then the derived relative sunspot number for that day is $r = k(10g+f)$.

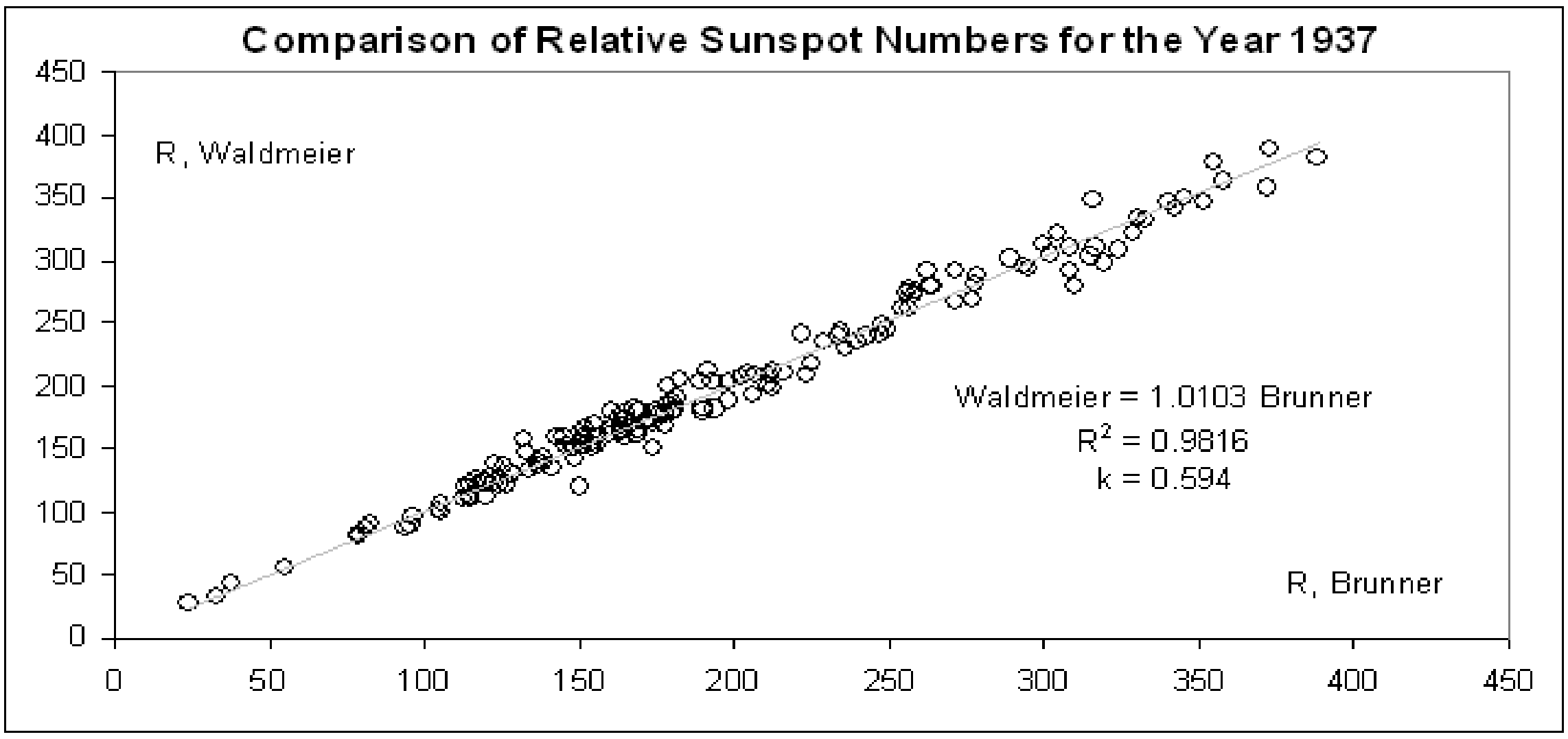

Figure 32: Comparison of daily "raw" (i.e. with no *k*-factor applied) relative sunspot numbers derived by Waldmeier and Brunner for the year 1937. The *k*-factor for Waldmeier comes to 0.594=0.6/1.0103 (Brunner reports 0.59).

All the Zürich SN observer records after 1937 appear to be lost, but in spite of the lack of original material, it is possible to perform a statistical analysis as follows. From the RGO series of sunspot group areas (Hathaway 2014), we select days where only one group was recorded on the disk. If that group had precisely one spot, the sunspot number for that day would be recorded as 11 by Wolf and as 7 (0.6*(10*1+1) = 6.6) by Wolfer and later observers, if there were no weighting by size and complexity. Figure 33 shows the distribution of solitary large spots over time. During the Wolf period, the largest single-spot groups had a sunspot number of 11 (there were scattered lower values in the 1880s due to averaging with Wolfer). Starting with Wolfer, there were many large groups with a single spot counted as just one spot (sunspot number 7), i.e. no weighting. With Brunner and later, the 7s effectively disappear. This seems to indicate that some weighting was done already by Brunner, explaining why Waldmeier matched Brunner's counts. On the other hand, there are many 8s, so any weighting must have been slight and there simply were very few solitary spots during the active 1940s, so it is difficult to draw a definite conclusion from this analysis about the amount of weighting done by Brunner.

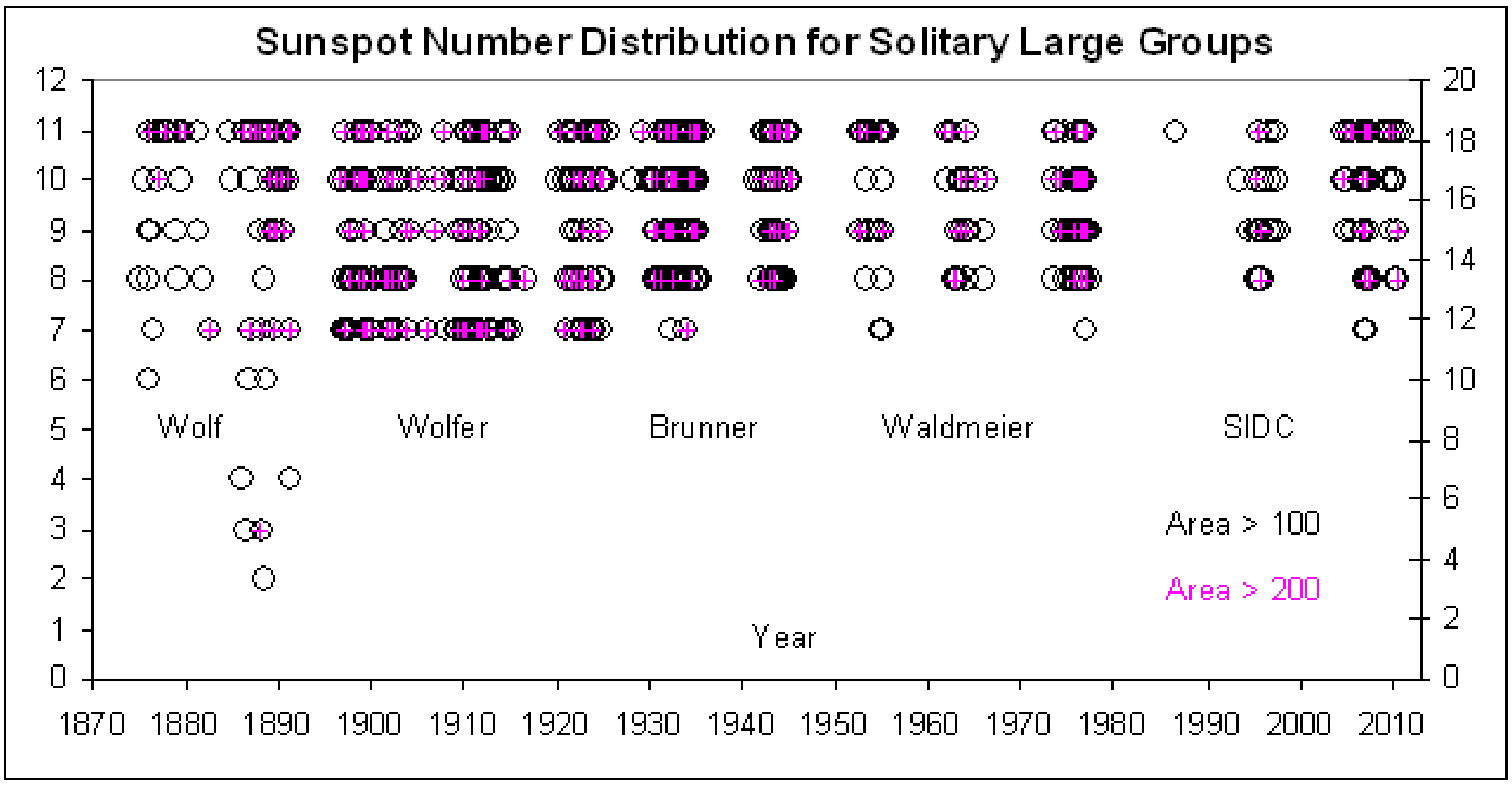

Figure 33: For days where only one group was observed, the sunspot number (if less than 12) for that day (i.e. for that solitary group) is plotted if the projected area of the group is larger than 100 µHemisphere (circles) and larger than 200 µH (pink "+" symbols). The right-hand scale is for sunspot number divided by 0.6, i.e. on the original Wolf scale.

Brunner himself writes in 1936 (Brunner 1936) that "The subjective method of counting may also have an influence. In large centers of activity one is inclined – and this perhaps rightly – to give some single spots according to their sizes a different weight", but then continues "In the spot-statistics, introduced for our observatory by Rudolf Wolf 80 years ago, all these circumstances have been considered as far as possible by introducing a reduction-factor on Wolf's unit. The latter is determined by comparison of corresponding observations. In determining the Wolf relative-number a weight of ten is given for the groups of spots and a weight of one for the number of single spots or nuclei".[4] This seems to indicate that spots were not weighted, although Brunner at times might be inclined to do so. His assistant Max Broger (observed 1897-1936) appears to have weighted some of his counts, so it is conceivable that discussion was going on at Zürich about the preferred counting method.

The long-time observer Herbert Luft (1908, Breslau, Germany; 1988, New York, USA) was a corresponding contributor to the Zürich series. As a teenager, he joined various Amateur Associations and was mentored by the slightly older Wolfgang Gleissberg who suggested Luft concentrate on sunspots. Luft's notebooks are archived at AAVSO (http://www.aavso.org/herb-lufts-notebooks-new-science-aavso-archives) and L. Svalgaard recently digitized the observational material. The nearly 12,000 pages yielded 10,434 usable observations (when image quality was good enough) during 1924-1987. Interesting enough, Luft started using the weighting scheme on the 24th February 1947, but abandoned the scheme April 5th the next year. Figure 34 shows two pages from March 1948. Luft also started to use the Zürich Group Classification System at the same time (and did not later abandon the classification). The spot count for the A-groups were the same with (number to the right of the class letter) and without weighting (number to the left of the class), while the H-I, D, and E class groups had a weighted spot count on average 45% higher.

[4] Presumably meaning umbrae (spots) within each penumbra

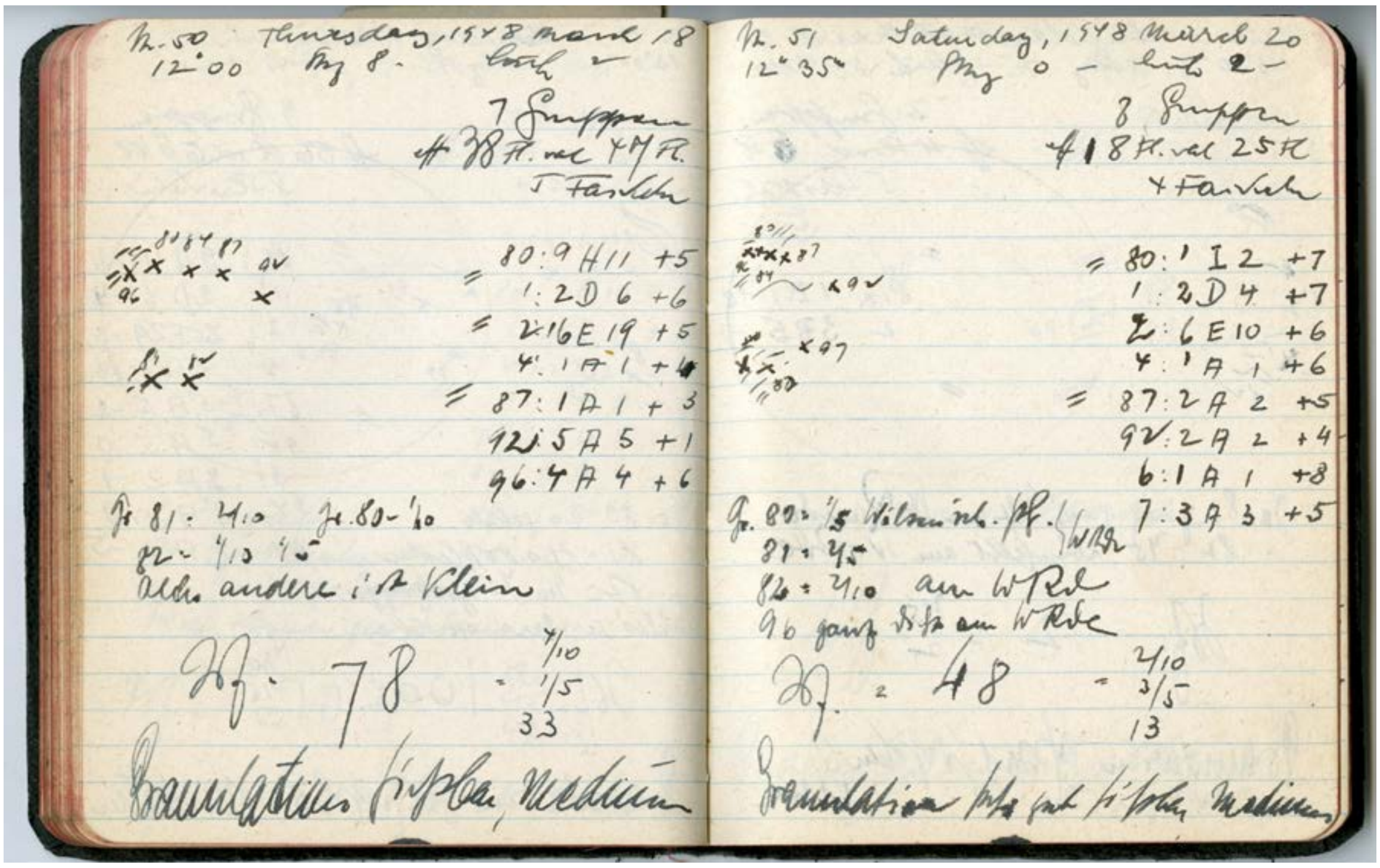

Figure 34: Pages from Herbert Luft’s notebook for March 18th and 20th, 1948. South is up and West is left. The letters in the columns at the right on each page show the Zürich Group Class for each numbered group, flanked left and right by the raw and the weighted count of spots in each group. The telescope was a superb 54mm aperture Merz used at 96X magnification. In spite of the crude-looking drawings, the counts of groups and spots are of high quality, as can be seen from comparisons with MWO.

We interpret this to indicate that Waldmeier was trying to get other observers to adopt the weighting scheme, but with little success. To our knowledge, the weighting was only adopted on a continuing basis by the Zürich observers and not by any others. Wolf and Wolfer published all raw observations from corresponding observers. Brunner reduced that to only the Zürich observers, and Waldmeier stopped the publication of all raw observations completely, noting that all data would be available in the archives of the observatory. As noted above, however, the post-1937 archives are apparently lost. We make a plea here that anybody who has archived correspondence with the Zürich observatory since 1925, to send copies of the material to us so that we can recover at least some of the raw data in order to investigate the effect of the weighting on the SN record.

One such case is that of Harry B. Rumrill (1867-1951) who was a friend of Rev. Quimby (American observer 1897-1921 whose data were utilized in the Group Backbone Construction). Rumrill continued Quimby’s observations of sunspots through to 1951. His data and notebooks were considered lost until L. Svalgaard, with the help of “The Antique Telescope Society” (Bart Fried, Jack Koester), located most of them in early 2012. Rumrill used 2” telescopes early on, but from 1942 employed exclusively a 4” Brashear refractor (Figure 35, left). The ratio between the Zürich Sunspot Number and Rumrill’s indicates an increase of the Zürich values from ~1945, by about 20%.

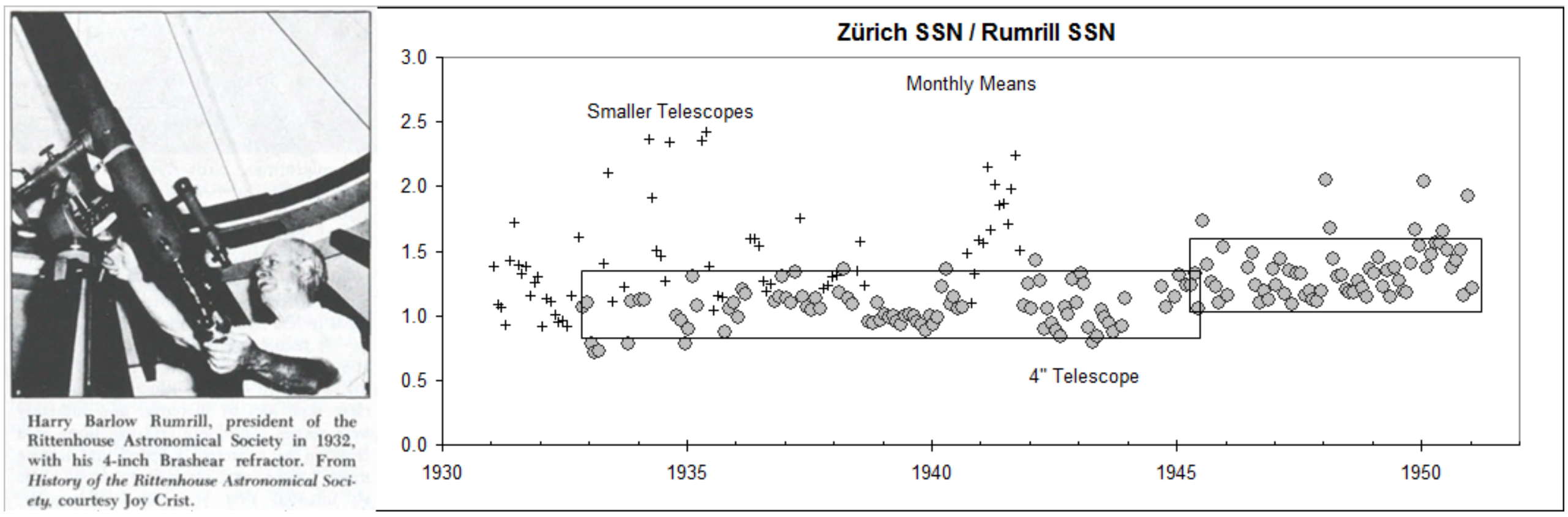

Figure 35: Ratio of monthly means $R_Z$/(Rumrill SSN). Data taken with small telescopes are plotted as small "+" symbols.

The total sunspot area data available over the time of the Waldmeier jump can also provide a useful external reference. There is indeed a strong (slightly non-linear) relationship between the SN, $R_Z$, and the projected (i.e. observed) sunspot area, $S_A$. On average: $R_Z \approx K\ S_A{}^{0.732}$ with no linear offset, so we can meaningfully form the ratio between the quantities (Figure 36); it seems clear that the ratio is lower than average before ~1947 and higher thereafter. We ascribe the difference to introduction of the full weighting scheme, as there is no metadata indicating a change in derivation of the RGO sunspot areas at that time.

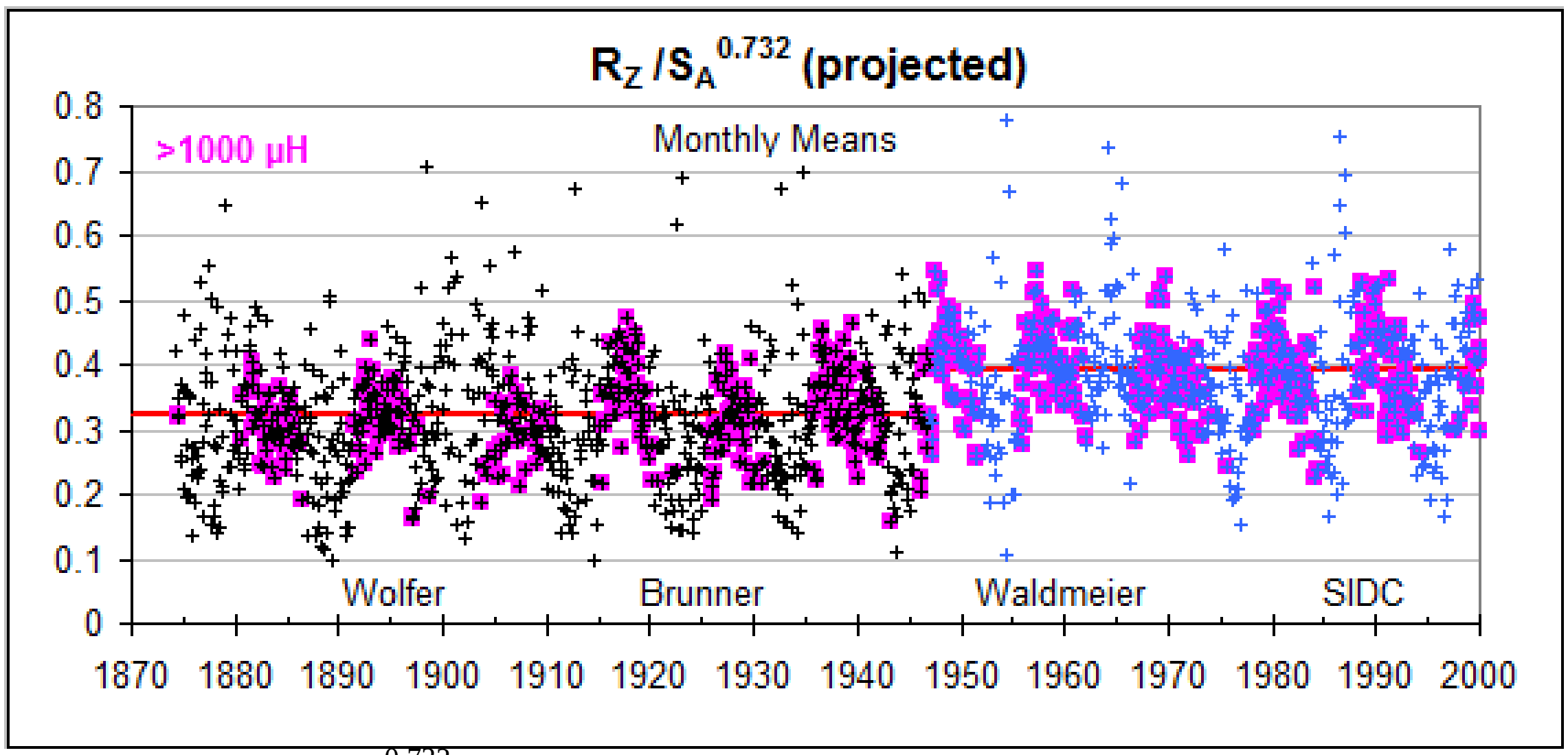

Figure 36: The ratio $R_Z/S_A{}^{0.732}$ (see text) for monthly means. If the mean area, $S_A$, is in excess of 1000 μHemispheres, the data point (+) is marked by a pink square. Before 1947, a horizontal red line shows the average ratio at 0.3244. After 1947, the red line shows the average ratio at 0.3931. In addition, the individual monthly values are shown as light blue plusses.

Using the value for the factor K derived from the pre-1947 data to calculate the monthly mean $R_Z$ from $S_{A,}$ we get excellent agreement before 1947 (Fig. 37) but a definite discrepancy thereafter, with the observed $R_Z$ being larger by a factor of 1.22 than the calculated value, call it $R_C$, we would expect from $S_A$. This suggests that the weighting was fully implemented by 1947.

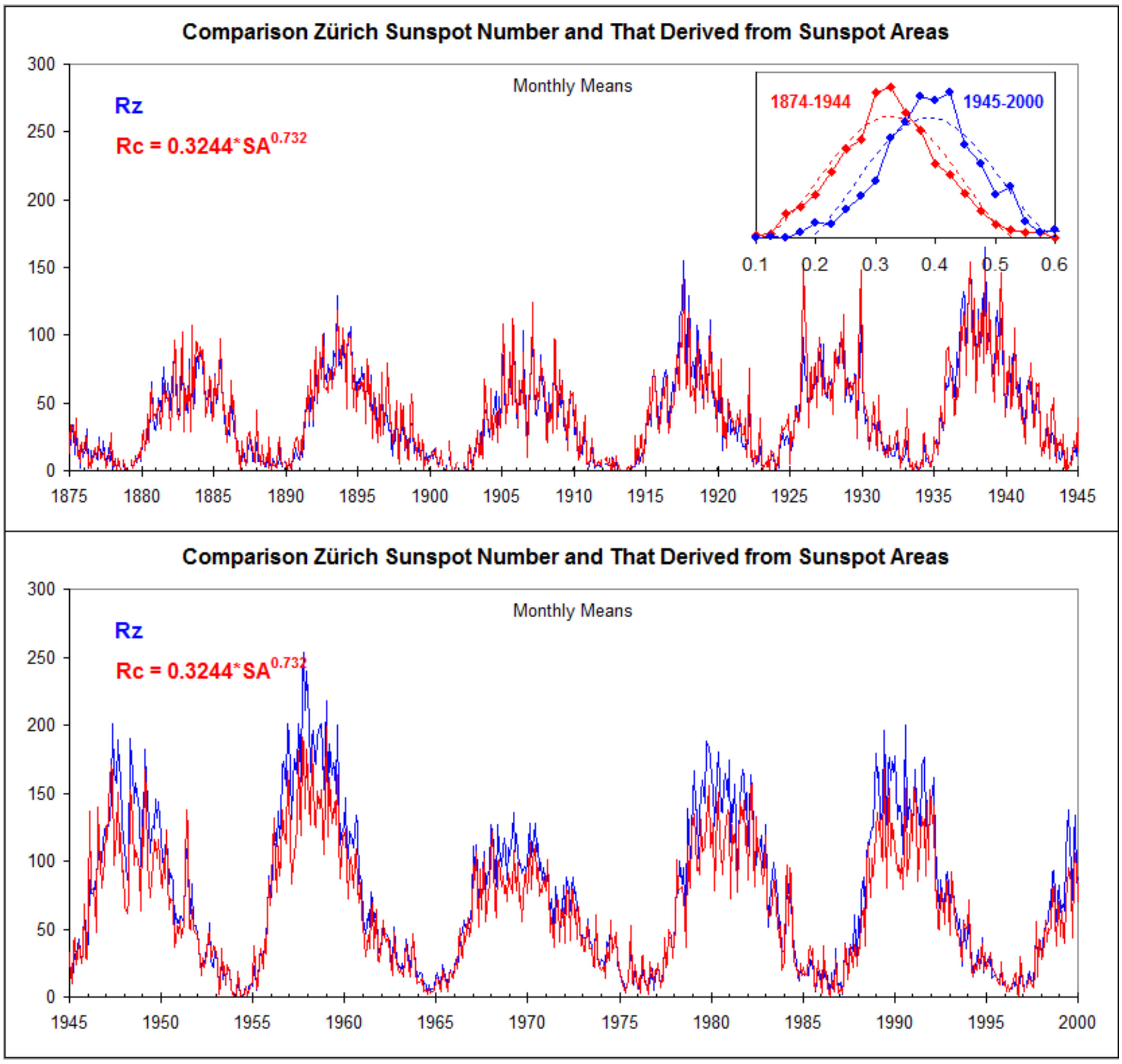

Figure 37: Calculated $R_C$ red, and observed, $R_Z$ blue, Zürich Sunspot Number using the relationship derived before 1945. The constant K = 0.3244 is determined from the best fit before 1945. The insert shows the distribution of data points in Figure 36 partitioned by the year 1945.

From ~40,000 Ca-K spectroheliograms taken at the 60-foot tower at Mount Wilson between 1915 and 1985, a daily index of the fractional area of the visible solar disk occupied by plages and active network has been constructed (Bertello et al. 2010). Monthly averages of this index are strongly correlated ($R^2$=0.8) with the sunspot number. Using the correlation based on the Wolfer-Brunner era, we can calculate the expected sunspot number for the Waldmeier era from the Ca-K index (Fig. 38). On average, the observed sunspot number after 1945 is a factor 1.21 higher than the expected value, again showing the influence of the weighting of sunspots according to size, coinciding with the tenure of Waldmeier:

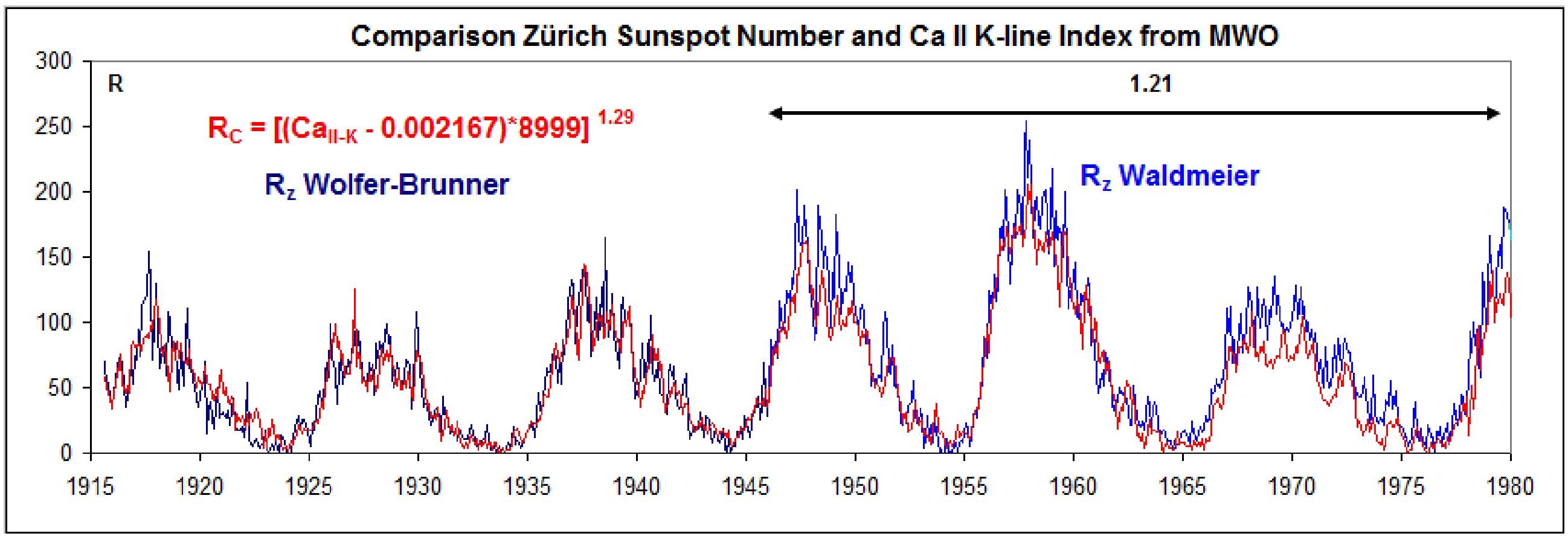

Figure 38: Comparison of the MWO Ca II K-line index with the Zürich sunspot number.

The effect of the weighting of individual spots can also be inferred by considering ionospheric measures of solar activity. The ionospheric F2-layer critical frequency, f$^o$F2, is the maximum radio frequency that can be reflected by the F2-region at vertical incidence (that is, when the signal is transmitted straight up into the ionosphere). The critical frequency has been found to have strong solar cycle dependence (Sethi et al. 2002). Back in 1952, Ostrow and PoKempner (1952) compared the dependence of f$^o$F2 on the Zürich sunspot number and concluded that there are “differences in the relationship between f$^o$F2 and sunspot number for the current (18$^{th}$) and preceding (17$^{th}$) sunspot cycles” (Figure 39). In Figure 39, it is instructive to follow the dashed line (March 1944 to June 1947) from the low sunspot numbers of cycle 17 to the high values of cycle 18. Adapted after Qstrow and PoKempner (1952). The shift (red arrow) in sunspot number to bring the curves for cycles 17 and 18 to overlap is the now familiar ~20%. Today, we can ascribe their further conclusion that “the sunspot number is therefore not entirely satisfactory as an index for ionospheric variations” to the result of the introduction of Waldmeier’s weighting scheme. The “fault” is not with the relationship, but with the sunspot number.

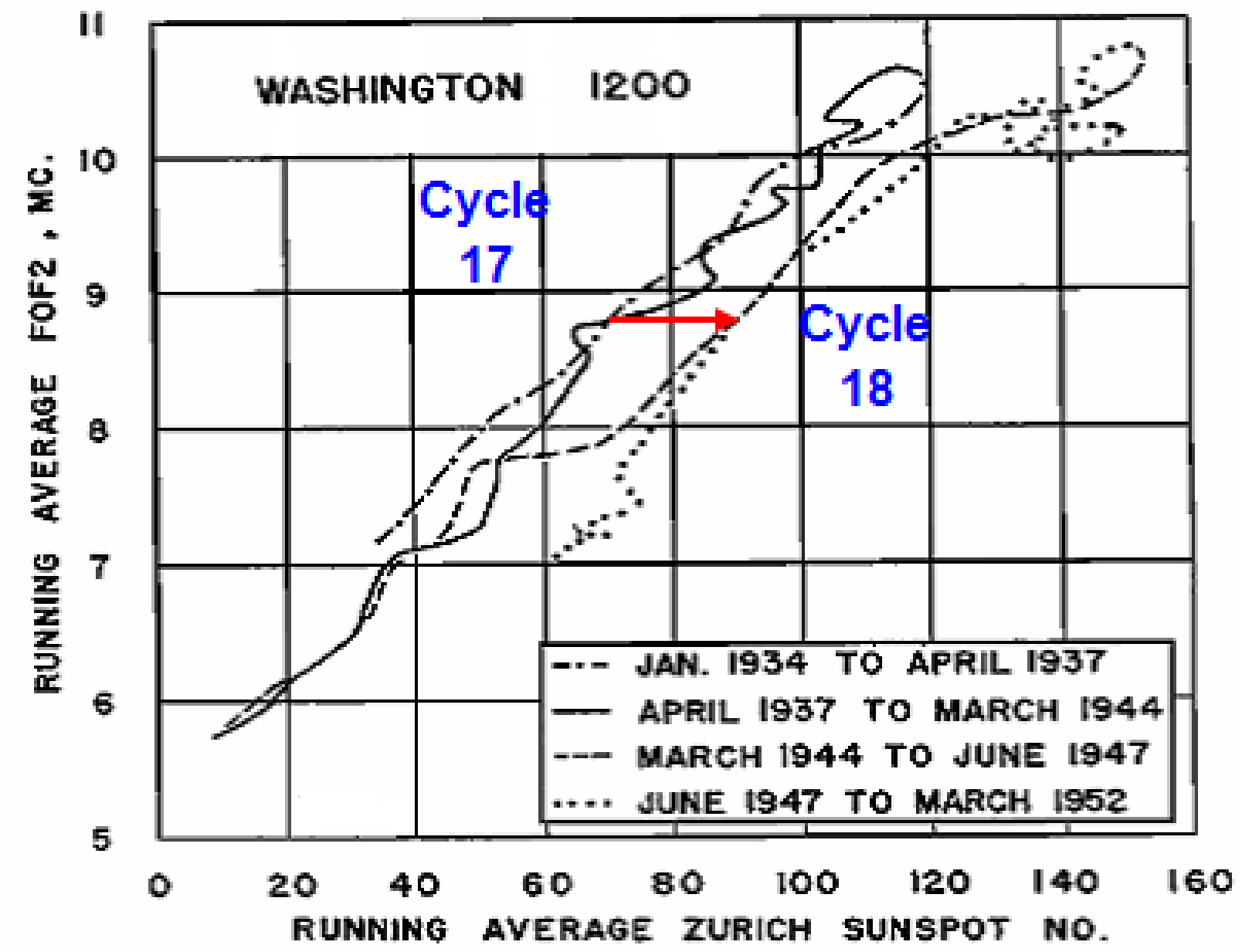

Figure 39: 12-month running average of the monthly median foF2 at local noon against 12-month running average of the monthly Zürich sunspot number.

Another independent comparison can be done with the range rY of the East component of the geomagnetic field, described in section 2. Considering the yearly averages of the rY index for geomagnetic stations in the latitude range 20º-60º, the residual differences from station to station are small and mainly due to local inhomogeneities in underground conductivity. Normalizing the range for a given station to a reference station (POT-SED-NGK) eliminates those small variations and allows us to make a composite of all stations. The result for observations since 1890 from a large number of observatories is shown in Figure 40.

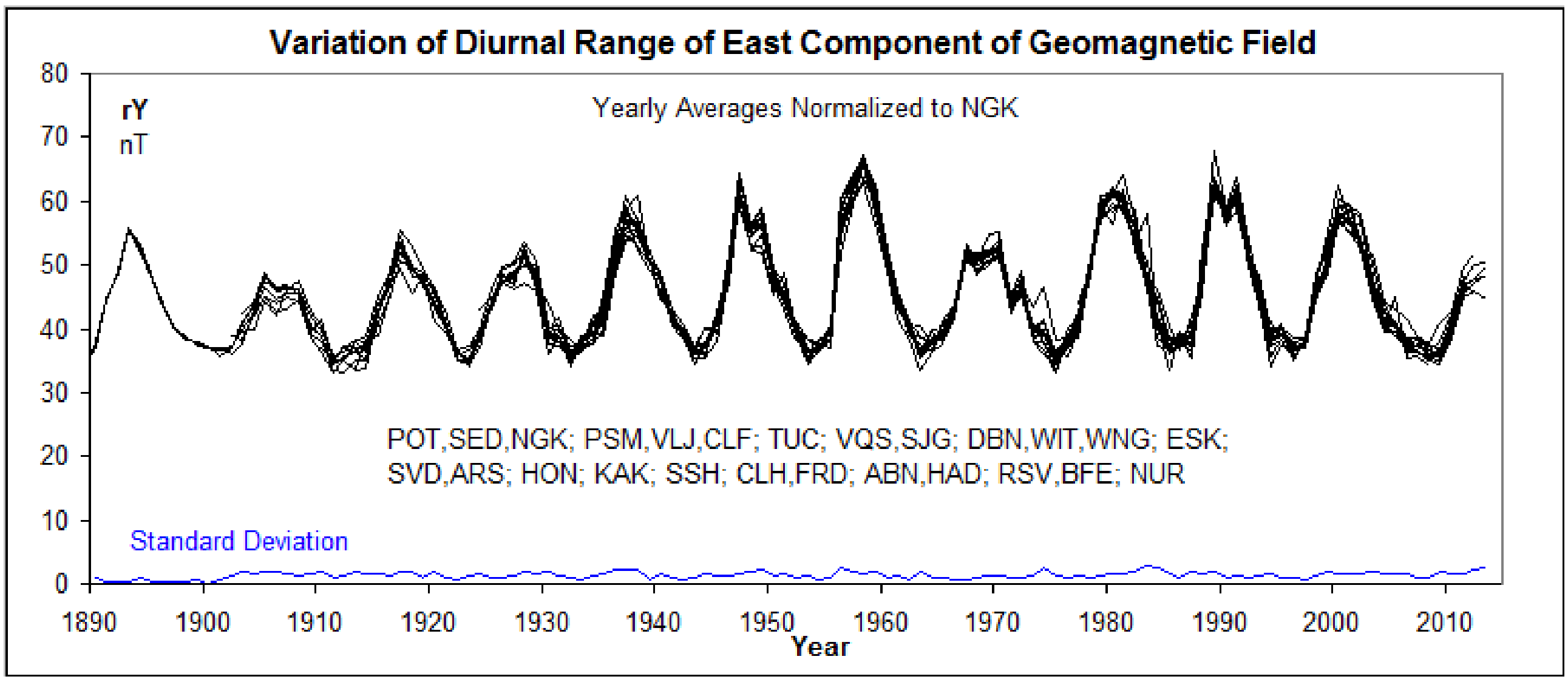

Figure 40: Composite record of the ranges of the diurnal variation of the East Component of the geomagnetic field. The composite is the average of long series of observations from 14 “chains” of stations (identified by their standard station codes), each chain plotted with a thin black line. A chain is the combined record of a station and its replacement stations that were necessary over time to escape urban development, normalized to the POT-SED-NGK chain. The very small standard deviation is shown in blue at the bottom of the figure.

That Figure 41 looks very much like a plot of the sunspot number is, of course, not a surprise as the linear correlation coefficient is in excess of 0.97 (see Figure 29) . With such high correlation, it should be possible to see the influence of weighting. Indeed, Figure 41 shows that the slope of the regression line is different before and after 1947, and that in the ratio 6.217/5.005 = 1.24, again suggesting that weighting increases the sunspot number by about that amount.

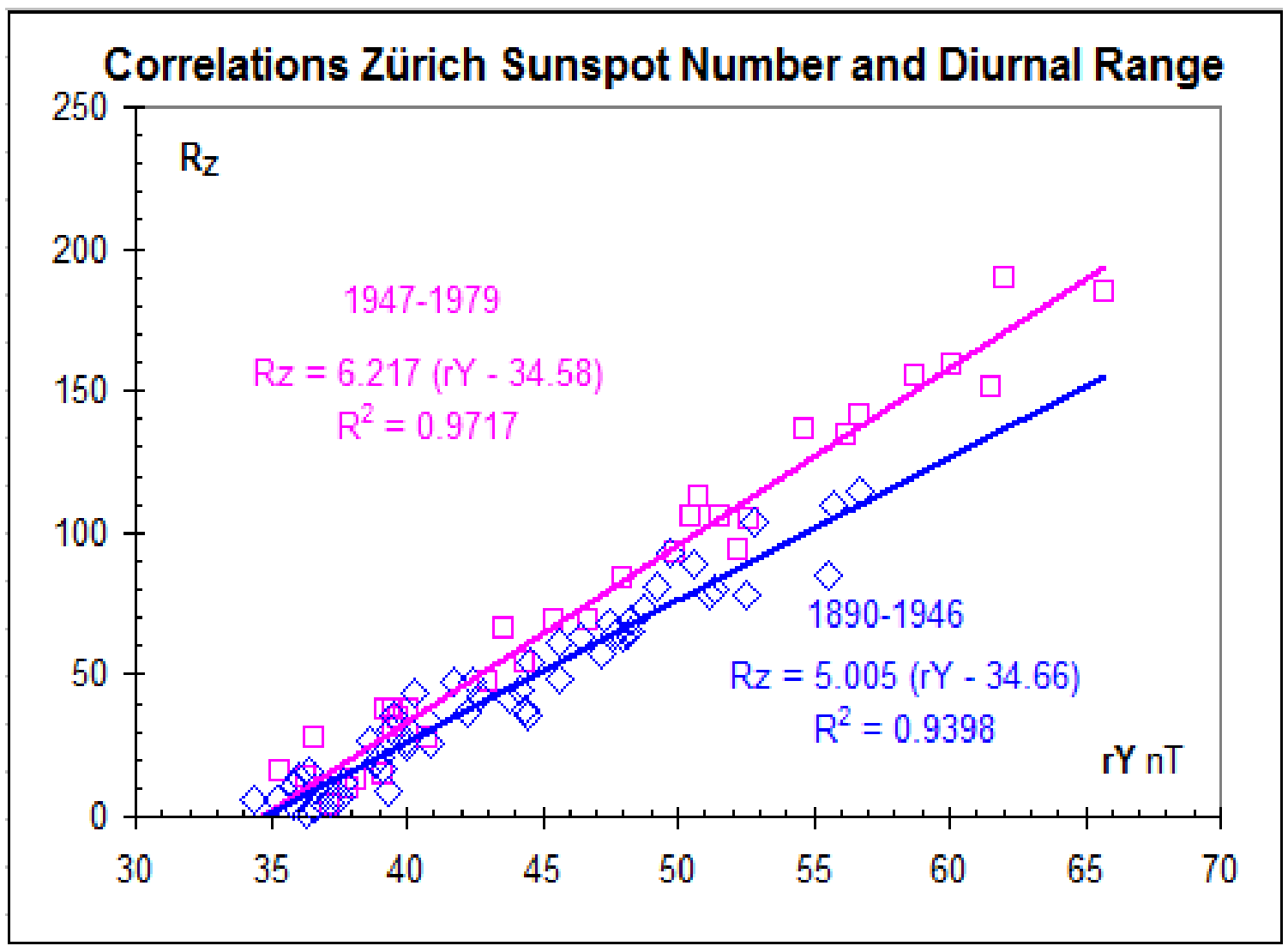

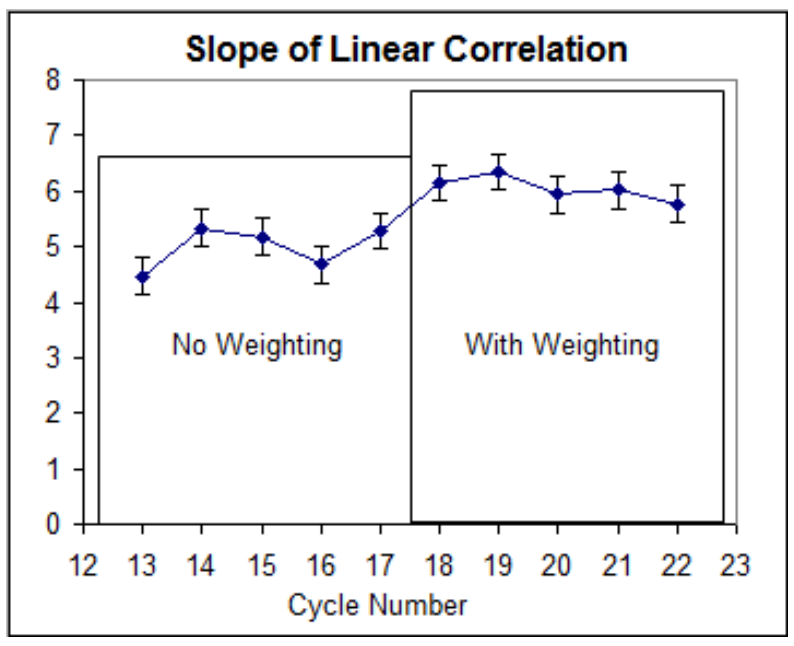

Figure 41: (Left) Linear correlations $R_Z$ as a function of rY for time before and after 1947. (Right) Variation of slope with time.

Calculating the slope of the correlation for each sunspot cycle (Figure 41, right), we find, as before, that the increase in slope takes place between the 17th and 18th sunspot cycle (see also CAWSES Newsletter: http://www.bu.edu/cawses/calbrating_sunspot_number_using_mag_needle.pdf). There is therefore little doubt that Waldmeier introduced the weighting scheme in full force in or about 1947.

At the reference station Locarno, weighting has been used since the beginning in 1957, closely following Waldmeier's prescription (Sergio Cortesi, personal communication). To assess the magnitude of the increase due to weighting, Leif Svalgaard undertook to examine all the drawings and individual counts of groups and spots made at Locarno for the past decade and re-count the spots with and without weighting. There were 3229 observation days with 9532 groups containing 49,318 un-weighted spots at the time of writing. The weighted spot count was 72,548, for an excess of 47%. The counts translate into an average sunspot number of 26.88 [(10*9532 + 49318)/3229*0.6] without weighting and 31.19 with weighting, for an excess of 16% for this rather low solar activity. It is, perhaps, noteworthy that the average number of (unweighted) spots per group for this period (2003-2014) is low, only 5.17.

To verify that the re-count is valid, i.e. that Svalgaard has understood and applied correctly the Waldmeier weighting scheme, the observer Marco Cagnotti in Locarno agreed to maintain a parallel count of unweighted spots at a continuing basis since January 1st, 2012, following a brief trial in August 2011. We remind the reader that the sunspot count that Locarno is reporting is done visually at the telescope and not from the drawings. It is rare, though, that there is a difference.

Figure 42 shows that Svalgaard and Cagnotti very closely match each other in applying the weighting scheme, thus sufficiently validating the approach.

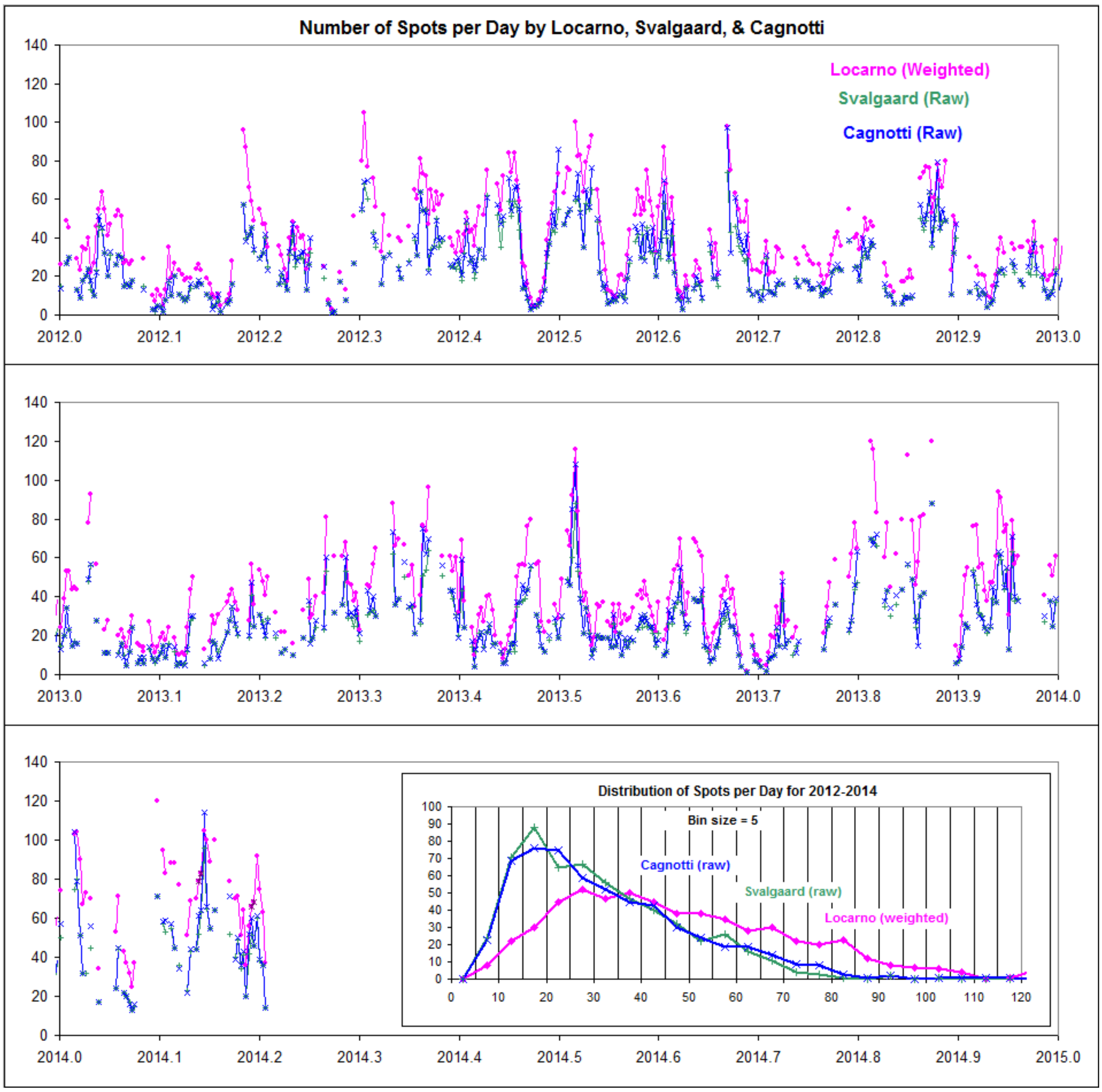

Figure 42: Comparison of the number of sunspots per day determined by Cagnotti (blue) and Svalgaard (green) without weighting, i.e. by counting each spot singly as prescribed by Wolfer and Brunner, with the number reported by Locarno (pink) employing the Waldmeier weighting scheme. The insert shows the distribution of counts in bins of five.

To determine the effect on the relative SN of the weighting, we evaluate the Relative SN = 10 Groups + Spots for Locarno, Cagnotti, and Svalgaard and compare that with the International SN $R_i$ divided by the k-factor of 0.6 (Figure 43). Ideally the ratio between the Locarno SN and $R_i/0.6$ should be unity, which it very closely is (0.99).

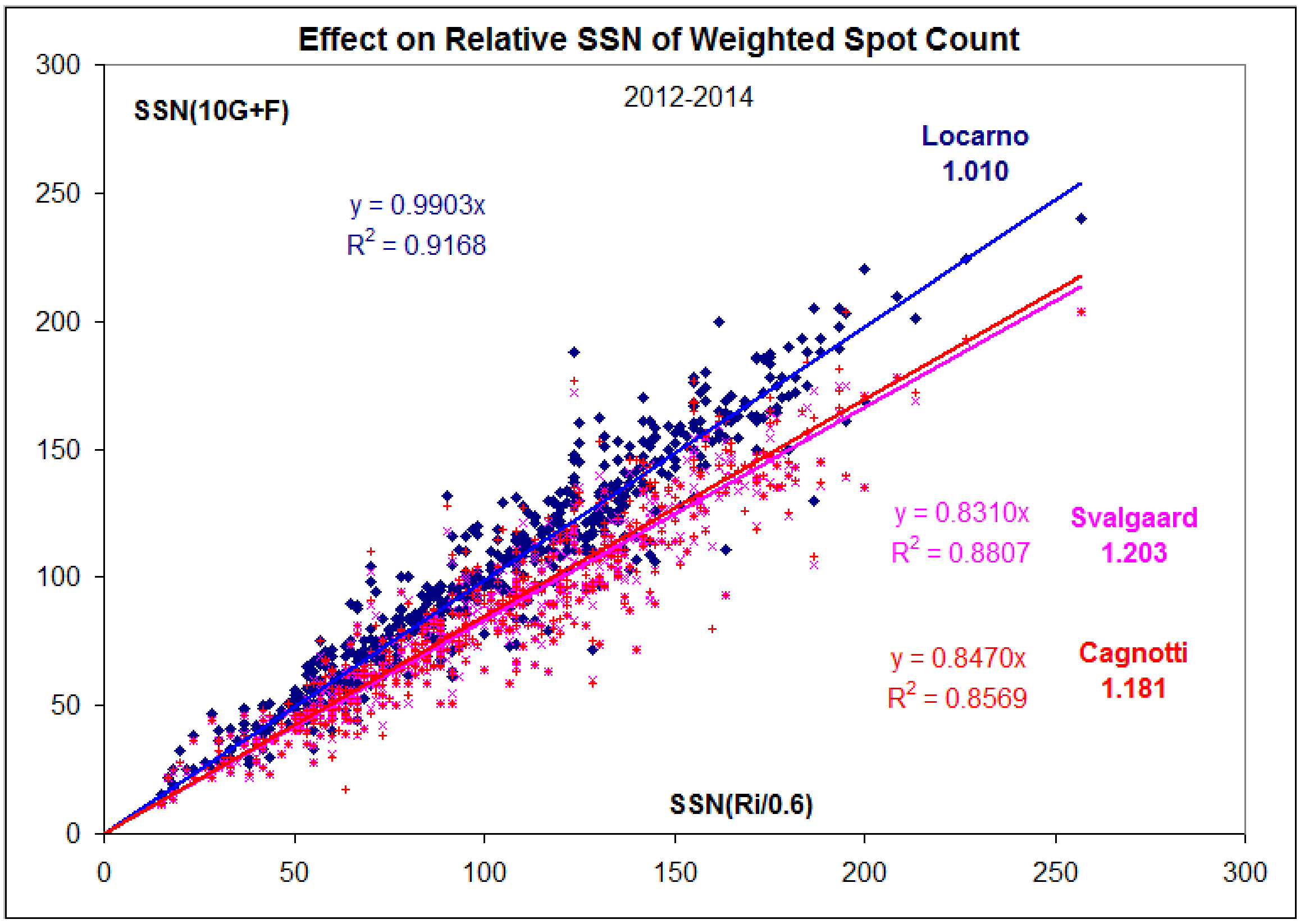

Figure 43: The relative SN calculated from the weighted spot count reported by Locarno (blue) compared to the International SN without the standard 0.6 factor. The unweighted counts by Svalgaard (pink) and Cagnotti (red) agree very well with each other and translate into a SN that is only 0.839 of the International SN (which then is higher by a factor of 1.19 on average).

Determining the weight factor w, for years with different levels of solar activity allows us to quantify the relationship between w and the sunspot number as reported. Treating the values for the deep minimum 2008-2009 as outliers (as also the weight factor for very low sunspot numbers does not matter) yields $w = 1.123 \pm 0.006 + R_i/(1416 \pm 140)$, with a range of 1.126-1.264 for $R_i$ in the range 4-200. The effective average weight factor for 1947-2014 is then readily determined to be 1.20 (Figure 44). The bottom panel of Figure 44 shows what the sunspot number would be [pink squares] if we divide the International sunspot number by the weight factor just determined to correct the sunspot number for the over-counting caused by the weighting.

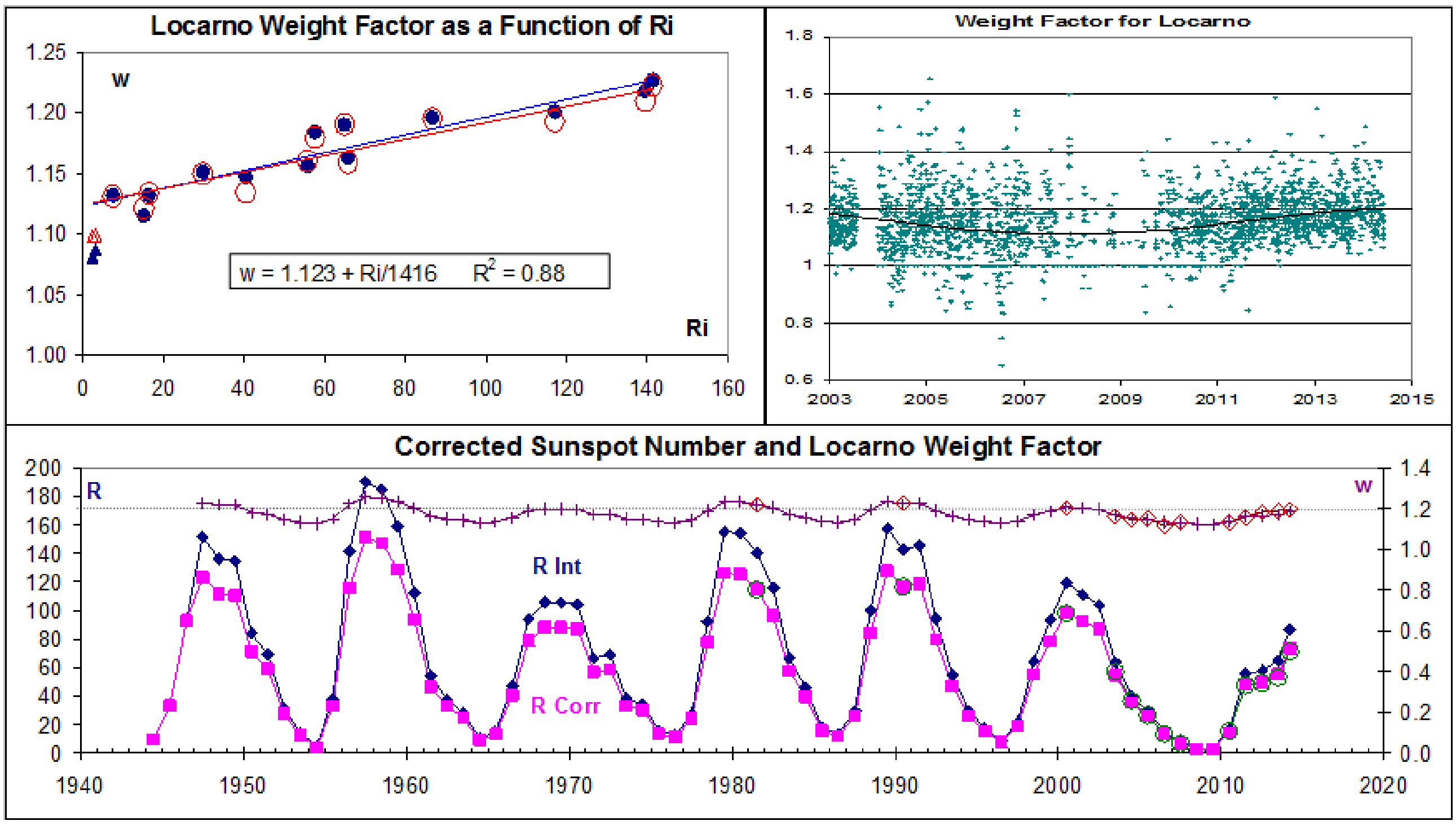

Figure 44: (upper left): the weight factor for the years 1981, 1990, 2000, 2003-2014 (other years still in preparation) as a function of the International Sunspot Number. The blue filled circles show the average weight factor for each year based on Locarno drawings. The red open circles show the ratio between the average weighted sunspot number and the average un-weighted sunspot number. Outliers are marked by small triangles. The best-fit regression lines (omitting the outliers) are shown with the equation giving the average coefficients. (Upper right): The weight factor at Locarno for each day since 2003. (Bottom panel): The weight factor calculated from the International Sunspot Number (upper curve and scale at right). Red diamonds show measured values. The blue diamonds and blue curve show yearly averages of the International Sunspot Number as reported by WDC-SILSO. The pink squares and curve show the resulting values when corrected for weighting. Values for years when the raw un-weighted averages were actually measured, rather than calculated, are marked by open green circles.

Is the weight factor observer dependent? With a novice one might be inclined to think so, but with training, observers tend to converge to agreement. We can compare the weighted counts made by the veteran Cortesi and the new observer Cagnotti from 2008 to the present (Figure 45): there does not seem to be any systematic difference.

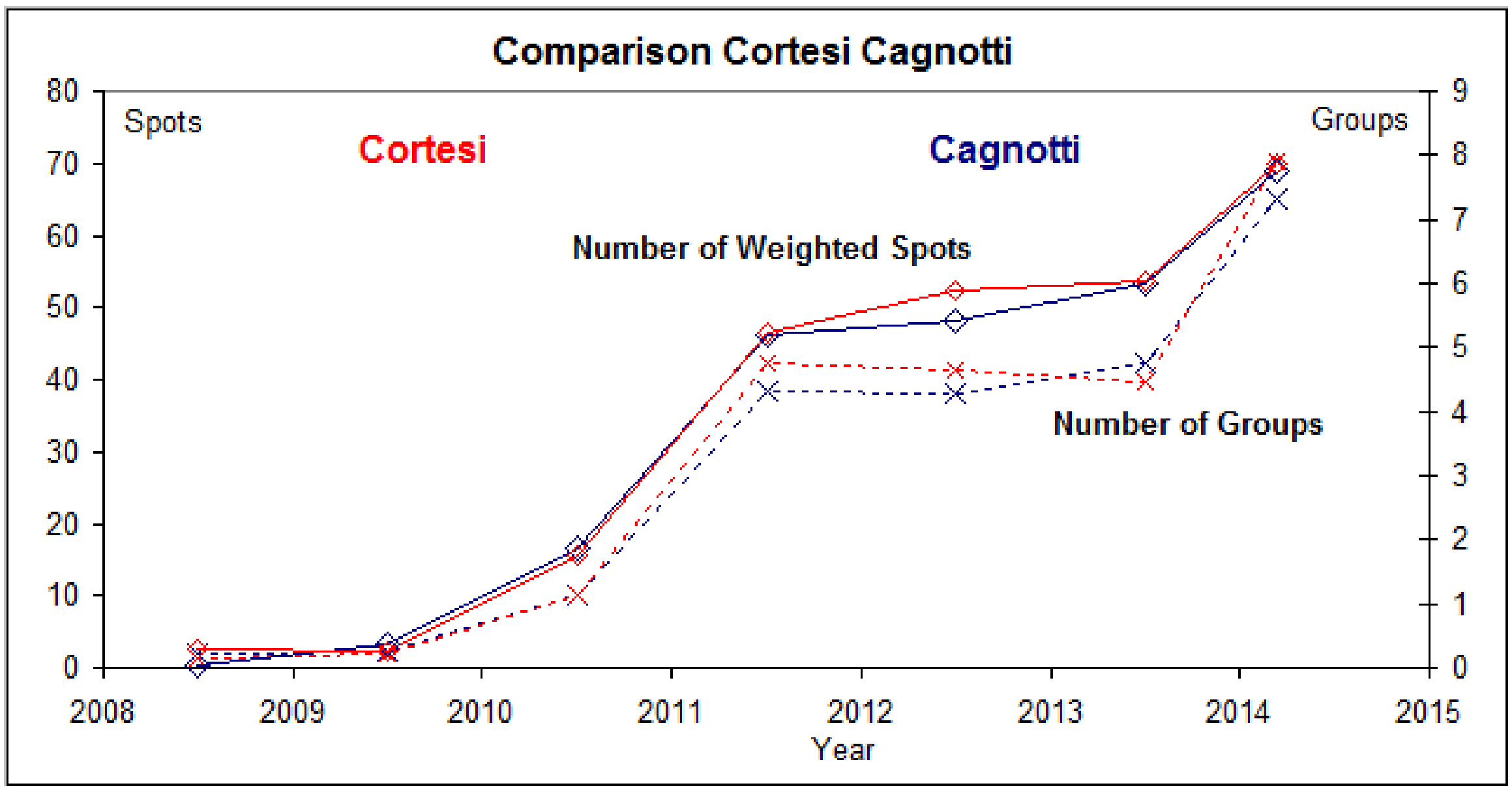

Figure 45: The yearly average group count (dashed lines and crosses) and weighted sunspot count (full lines and diamonds) for observers Cortesi (red) and Cagnotti (blue).

Waldmeier also introduced a new classification of groups (the Zürich classification) based on understanding of the evolution of the group rather than mere proximity of the spots. This tends to increase the number of groups over that that mere proximity would dictate. We find that, on average, on a fifth of all days, an additional group is reported over what is observed at MWO, which means that the relative sunspot number increases by about 3% due to this inflation of the group count brought about by the better understanding of what constitutes a group. Kopecky et al. (1980) quote the observer Zelenka suggesting a possible influence of the new Zürich classification of groups. This problem deserves a full, future investigation.

Figure 46 illustrates the problem. Today, we may use the magnetic field information to discriminate between groups, but such information was not available to observers in earlier times so proximity and longitudinal extent were the primary criteria for groups.

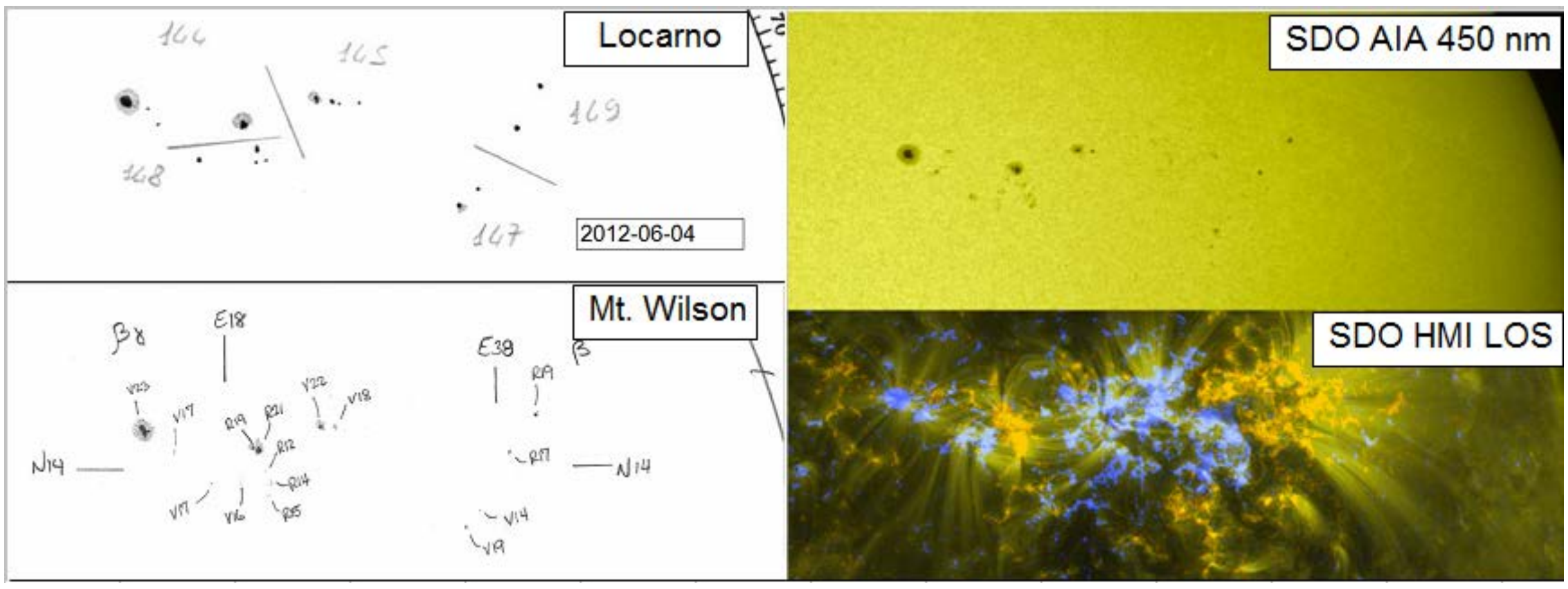

Figure 46: Groups and spots observed 4th June 2012 at Locarno (five groups; upper left), MWO (two groups; lower left with polarities indicated), SDO AIA at 450 nm (upper right), and their magnetic fields (blue positive, orange negative; lower right). The two MWO groups are marked by the latitudes and longitudes of the centroids (N14, E18 and N14, E38).

Recently, Lockwood et al. (2014) have suggested that the magnitude of the Waldmeier jump in $R_Z$

in ~1947 is ~12% rather than the ~20% that we find in several indices and deduce from an analysis of the effect of the weighting. The difference is due to their use of annual averages, rather than monthly means, of the SN and sunspot area in their determination of the non-linear relationship between these two parameters. This understates the size of the discontinuity and the corresponding correction factor.

### 5.3. The RGO-SOON jump (1976)

As the WDC-SILSO data set contains the raw sunspot and group counts, it is possible to compute a GN by the same method as the International SN, using a statistics over many stations with one pilot station (Locarno). This GN offers the following advantages: it is built on a single continuous set of data and the base data are the same as the ones used for producing the SN, allowing a direct unambiguous comparison. Moreover, the group count is unaffected by the sunspot weighting adopted specifically at the Locarno station (cf. previous section).

We can then compare this reconstructed GN with the original GN from Hoyt and Schatten (1998a), which is based on RGO photographic data until 1975 and on sunspot drawings from the 4 stations of the USAF SOON network afterwards. As shown in Figure 47 (straight group counts without the 12.08 scale factor), both series match rather well, with a global average ratio close to unity, although the monthly mean ratio is rather variable (~10% rms). However, this average ratio shows a clear 10% upward trend occurring over the interval 1974 - 1982, rising from 0.97 to 1.08. This change of scale occurs around the time of the transition from the RGO to the SOON data set and is probably due to the different underlying data and methods.

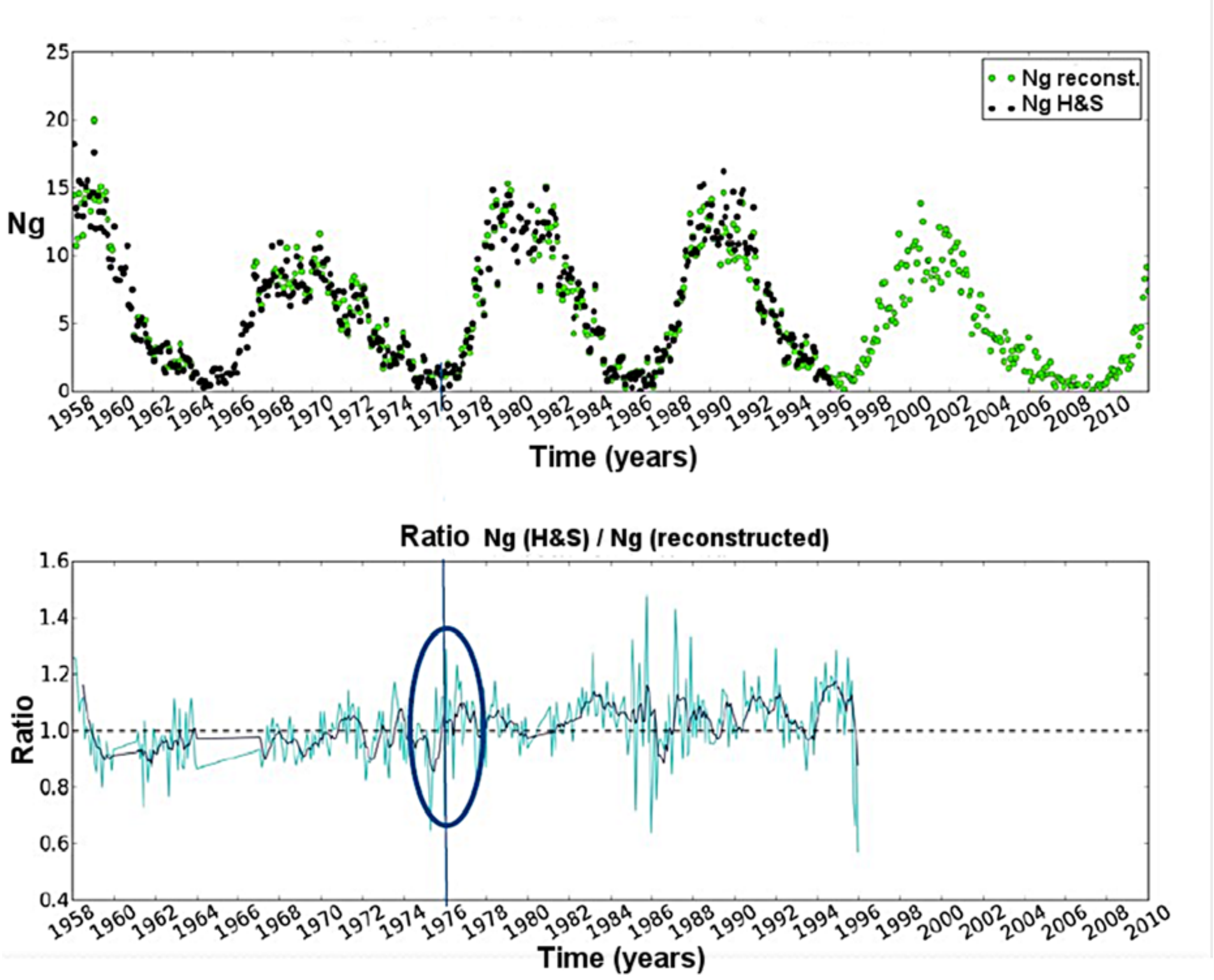

Figure 47: Top panel: monthly mean group counts ($N_g$): in black, the original series from Hoyt and Schatten (GN divided by the 12.08 constant) and in green, new reconstructed average group counts based on the WDC-SILSO observations. Lower panel: ratio between the original and the new group counts: monthly mean ratio (blue) and 12-month smoothed ratio (black). The time when the original group number switches from the Greenwich to the SOON data set is marked by the ellipse.

As the reconstructed GN is not affected by such a transition and its scale is expected to remain constant, the jump must therefore reside in the original GN series and be caused by a scaling mismatch between its two base data sets. Such a scaling shift parallels an equivalent mismatch in the RGO-SOON sunspot areas already identified by other analyses (Hathaway et al. 2002, Foukal 2013). Although the RGO scale is slightly lower than the reconstructed GN, it seems that the main bias is an overestimate of the group counts in the SOON catalog after 1975. This may be due to different group splitting rules applied for this data set (meant primarily for the real time numbering of active regions) versus the earlier RGO catalog, which also uses another group classification scheme than the McIntosh classification in SOON data (Willis et al. 2013).

Therefore, a 10% reduction of the GN must be applied after 1975 to bring it to the same scale as the GN and SN in the first part of the 20$^{th}$ century.

# 6. The $R_i$-Locarno drifts (1980-2014)

Following the unexpected low activity during the cycle 23-24 transition and the anomalous evolution of various solar, heliospheric or ionospheric indices, different comparisons were made recently between the SN and the $F_{10.7}$ radio flux (Svalgaard and Hudson 2010, Lukianova and Mursula 2011, Lefèvre and Clette 2011, Clette and Lefèvre 2012). They show an unprecedented divergence between those two indices starting around 2000, i.e. just after the maximum of cycle 23, with the SN falling 20% below its standard $F_{10.7}$ proxy (Fig. 48). This motivated an investigation about a possible bias in the SN series.

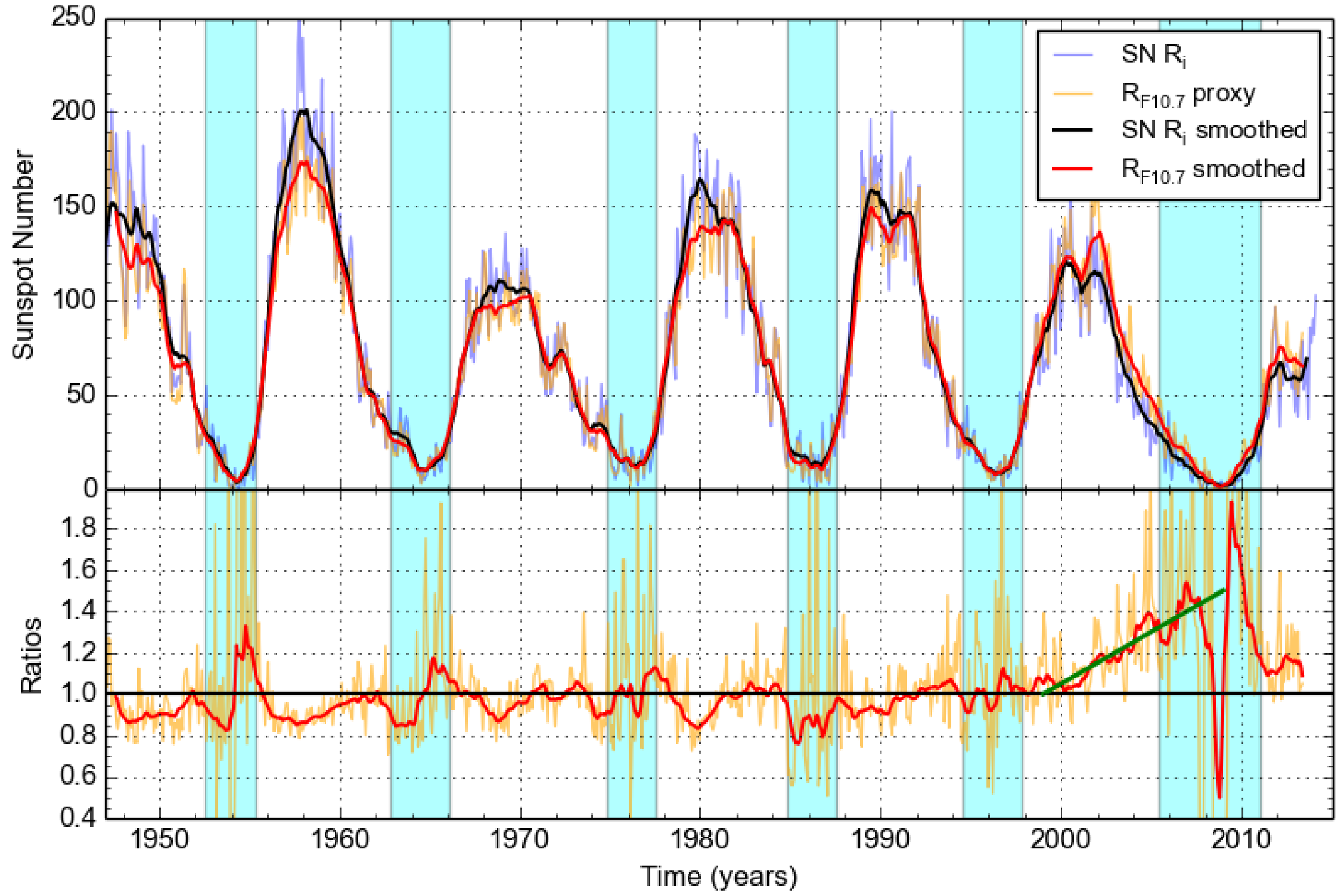

Figure 48: Plot of the SN $R_i$ (blue line) and of the SN proxy $R_{F10.7}$ (red line) based on the $F_{10.7}$ radio flux (Johnson 2011). Smoothed curves use the classical 13-month Zürich smoothing function. In the lower panel, the monthly mean ratio $R_{F10.7}/R_i$ (yellow) and its 13-month smoothed equivalent (red line), showing the unprecedented drift in cycle 23, after the year 2000.

### 6.1. The drift diagnostics: the global network-averaged k coefficient

For this analysis, we used the entire 32 year archive of raw sunspot observations available at the WDC-SILSO. In order to ensure a good overlap between individual series, we restricted the data to the 80 stations that contributed continuously for more than 15 years. Among them, 16 stations provided data for the full 32 year period and sometimes even earlier, including e.g. Locarno, Uccle (Brussels) or individual observers like Kenichi Fujimori (Japan).

Our analysis consists in computing the monthly average k coefficient, i.e. the average of the daily ratios between the International SN and the raw Wolf number from each station for the same date. In order to filter out the monthly variability and extract the long-term trends, we smoothed this ratio with a Gaussian smoothing function with a FWHM of 13 months. As an example, Figure 49 shows the resulting k series for the K. Fujimori station, one of the long-duration stations.

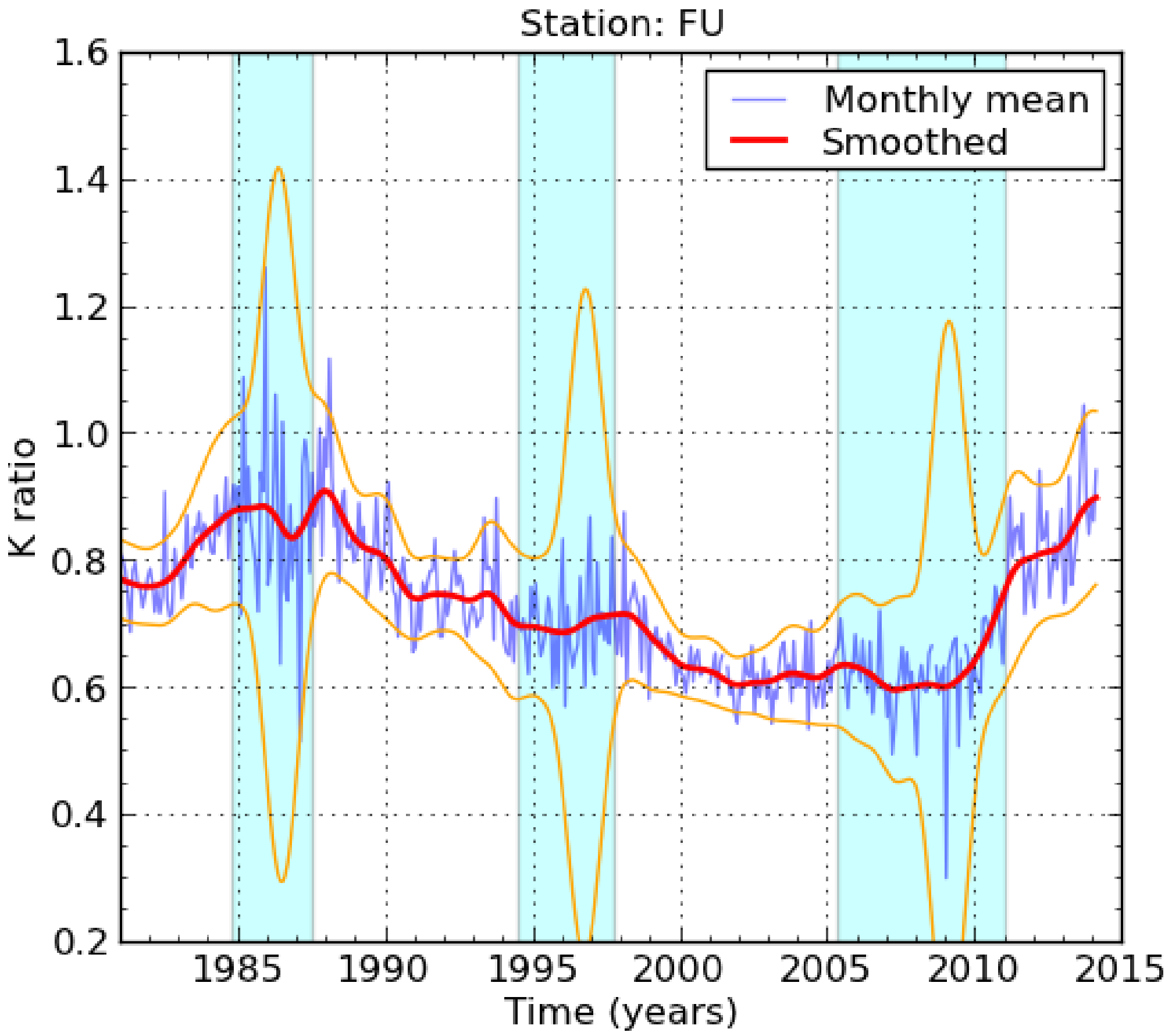

Figure 49: Sample plot for one station (K. Fujimori, code FU) showing the variation of the k ratio relative to the Locarno-based SN: monthly means (green line), and 13-month smoothed values (red line). The standard rms error is indicated by the orange lines and reaches maxima at the times of solar cycle minima, when the SN becomes small (blue-shaded bands).

Knowing that the long-term scale of $R_i$ is defined by the Locarno Wolf numbers (cf. Fig. 7), we wanted to check if a drift could be detected in the reference Locarno series. Therefore, we made global comparisons between network-wide averages of the raw k data and this reference series. We base our analysis on the assumption that scale variations of individual stations are uncorrelated. In this case, random individual errors cancel out in a global average and any remaining trend in the average k ratio can only reside in the common reference, i.e., the Locarno series. The absence of correlation among our sunspot observers is a reasonable assumption as they observe independently and have no way to communicate globally in real time and thus to influence each other, as they are dispersed all over the globe. This negligible mutual correlation is confirmed by simple cross-correlations between the actual individual k series.

In order to build the network average, in a first step, we scale all series so that their average k ratio is equal to unity over the time interval 1987-1995 (cycle 22). As the average k coefficient for different stations can differ by as much as a factor of 3, this first normalization ensures that all stations are included with the same weight in the subsequent global average k series. The time interval for this first normalization was chosen because this is when a maximum number of stations were simultaneously observing and it corresponds to a maximum of solar activity, thus providing more accurate average k ratios.

Then, in a second step, all normalized k profiles are averaged to obtain a first average k profile (Figure 50). As this profile now spans the entire 32-year interval, all k series can be normalized more accurately by a least-square fit to the average profile for the entire observing period of each station. Some additional stations that were not observing in the 1987-1995 interval can also be added at this stage. Then, a final average k series can be computed by averaging all individual normalized r series.

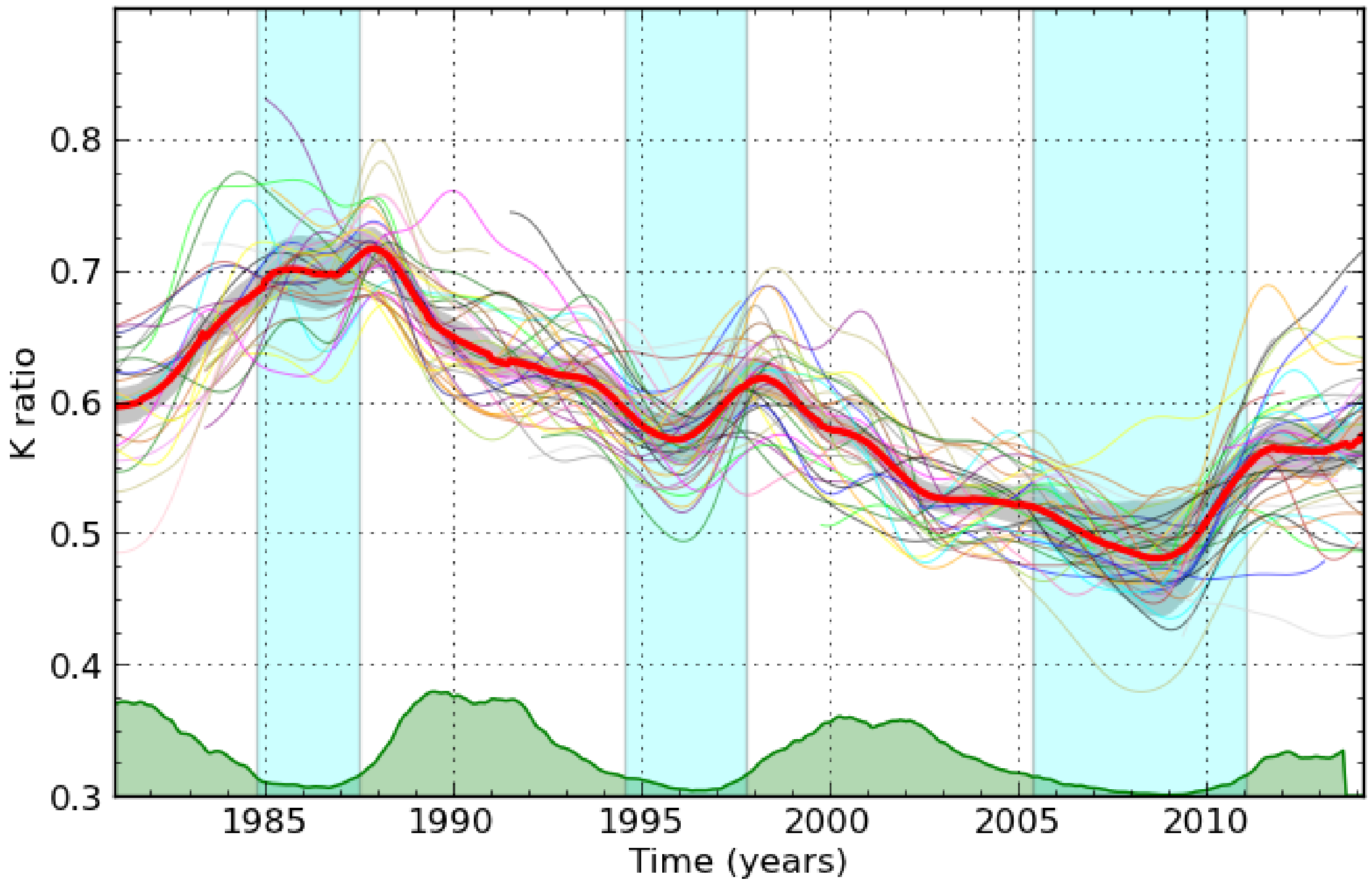

Figure 50: Plot of the normalized 20-month Gaussian smoothed k ratios for all stations used for producing the average k profile (thick red curve). It shows the dispersion between individual series. All series were normalized by a least-square fit over the entire time interval 1981-2013 (single constant scaling factor for each series). The gray shading around the red curve gives the standard error on the average k value.

In the process, we can check the quality of fit of each individual station relative to the average profile (linear correlation coefficient, intercept with origin). Figure 51 gives an example of the linear regression for the Uccle station. This fitting step also allowed us to rank the stations according to two criteria: length of the observations and stability of the k coefficient. Note that we used here a standard regression of the station k versus the average network k, thus neglecting the error in the average network k, which is a fair approximation as the latter has a much smaller error than the k of an individual station.

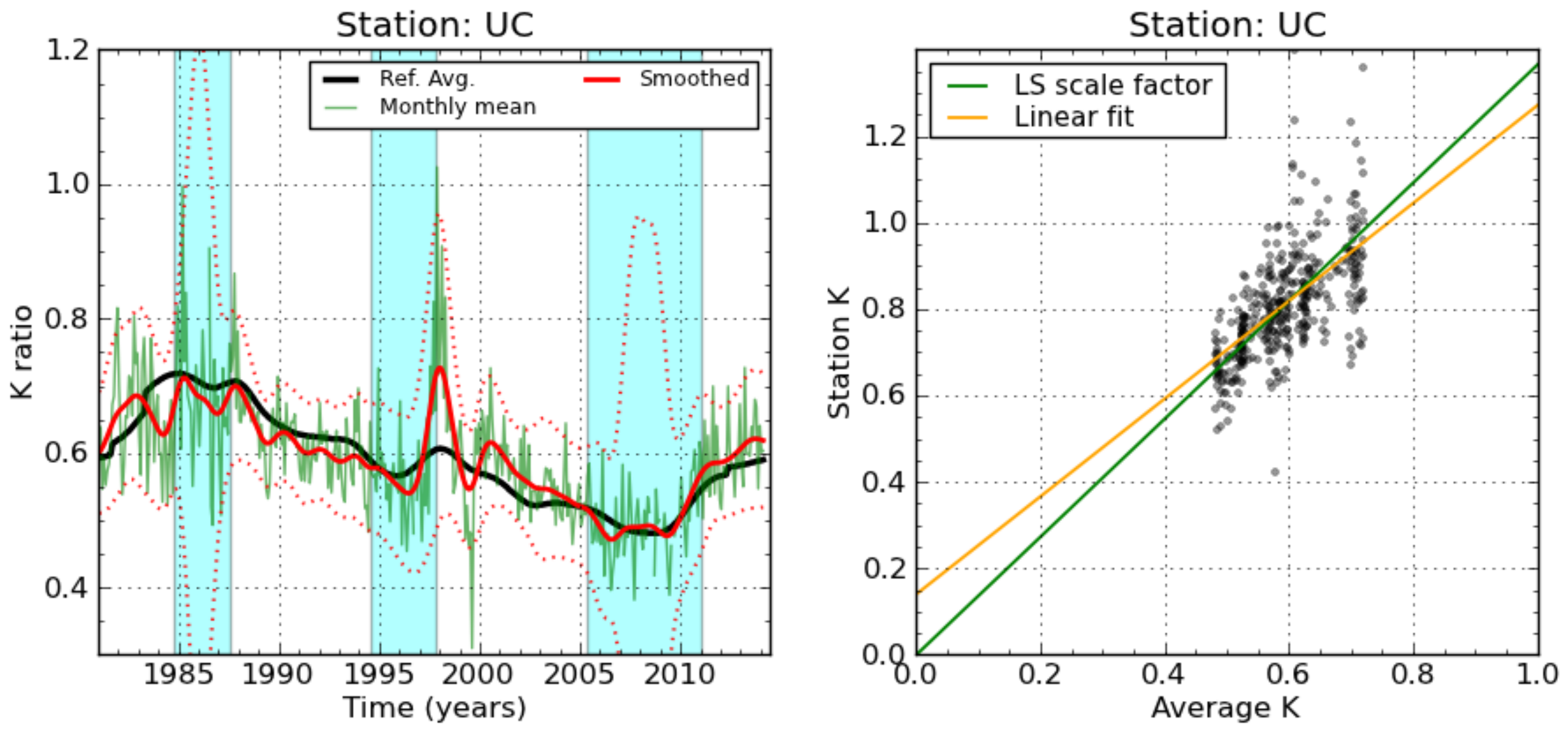

Figure 51: Sample plot of the k coefficient series for the Uccle station (ROB, Brussels). Left panel: monthly means (green line), and 13-month smoothed values (red line) are compared to the average k curve based on multiple stations (black line; see text). The original UC k series has been rescaled only by using a constant factor to fit the average K curve. Right panel: scatterplot of the Uccle k coefficient versus the average multi-station k coefficient. Two linear fits are shown: linear fit with intercept at the origin (strict proportionality, black line) and normal linear fit (orange line). The close match between the two fits indicates that the Uccle k coefficient is closely proportional to the global multi-stations average k variations.

In order to check the influence of the station quality, we repeated the above normalization and averaging process for various overlapping and non-overlapping subsets of stations. The resulting average k series are shown in Figure 52, for successively larger sets of stations (but integrating progressively stations with shorter duration or lower stability). Except for local discrepancies of maximum 10%, they all show the same reclining “Z” profile: a first rise from 1981 to 1987 is followed by a quasi monotonous decline up to 2008 and by a recent return to a value close to the initial 1981 value. The most reliable profiles (green and orange) give an amplitude of those drifts of +/- 15%. As this variation of the ratio is common to all stations, the source of the trends can only be attributed to the Locarno station. Given the close consistency of all average k profiles and assuming that the initial Locarno k coefficient in 1981 was matching the Zürich scale, we rescaled our average k profiles to start at 0.6 in 1981.

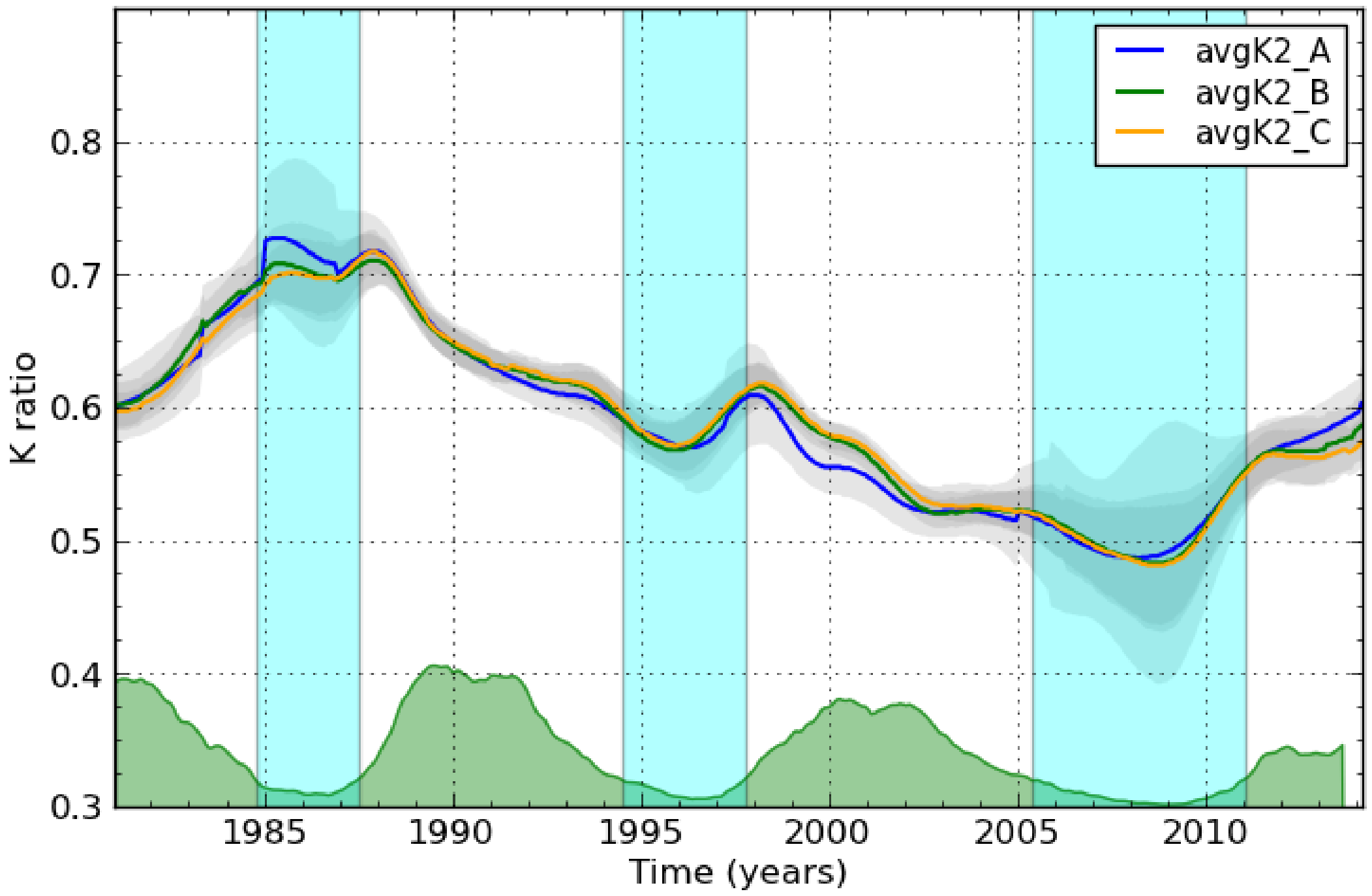

Figure 52: Plot of three different multi-station average k ratios. Three subsets of stations have been chosen, starting with the most reliable and long-duration stations (set A; 9 stations, in blue) and then adding more stations of shorter duration and lower stability (set B, 37 stations, in green; set C: 79 stations in orange). The gray shading gives the standard error of the average k ratio, which increases at cycle minima (blue shading). Although all three sets are very different, all curves display the same drifts and reversals, including local features.

Finally, another feature of the average k profile are the dips coinciding with the minima of solar activity, which are superposed on the overall reclining “Z” shape. As we found a significant cycle dependence of the weight factor (Fig. 44) in our parallel-counts study over the 2003-2013 period and as we know that none of the SILSO stations except Locarno itself are using weighted counts, this ~15% modulation can most probably be attributed to the Locarno weighting effect. Such a cycle modulation can be expected as during cycle minima, most of sunspot groups are small and contain small spots. There are almost no groups of Zürich type E, F, G and H, which contain spots with large penumbrae. The weighted counts should then hardly differ from unweighted counts, while over the rest of the solar cycle, they should lead to inflated k values, causing the observed dips. Further analysis will be required to fully confirm this connection.

## 6.2. Extended study (1955-2013)

In order to verify if any similar scaling drift was present before 1981 and if the scaling experienced a jump during the 1981 Zürich-Brussels transition, we extended the analysis back to 1955, by pre-pending raw data collected during the late Zürich period. For this purpose, we encoded original paper reports that were provided to the WDC-SILSO through the Specola Solare Observatory. Of course, the number of stations is much lower than after 1980 (22 long-duration stations). There are also gaps in the data series, in particular for years before 1965. The recovered data are only from auxiliary stations and from Locarno, as we were unable to locate original detailed counts from the Zürich station itself. Notably, a few stations provided very long series spanning almost the entire 60 year period, including Locarno, Catania, Uccle, Kanzelhöhe, etc. and a few other stations straddle the critical 1980 transition, providing a key reference for the continuity at that time.

By applying the same normalization and averaging method, we obtained an extended k series for four combinations of stations (Figure 53). The series show larger random variations as can be expected from the lower sample of stations. However, the k ratio does not show any long-term trend before 1980, while the same reclining "S" profile is again found after 1981. The drift thus starts once the Zürich reference is replaced by the Locarno station. As the average k ratio oscillates around 0.6 until 1981, without any significant discontinuity on the transition year, we have also a confirmation that the initial scale of the new Locarno-based International SN matched the scale of the preceding Zürich SN and that no significant scaling jump was associated with the Zürich-Brussels transition.

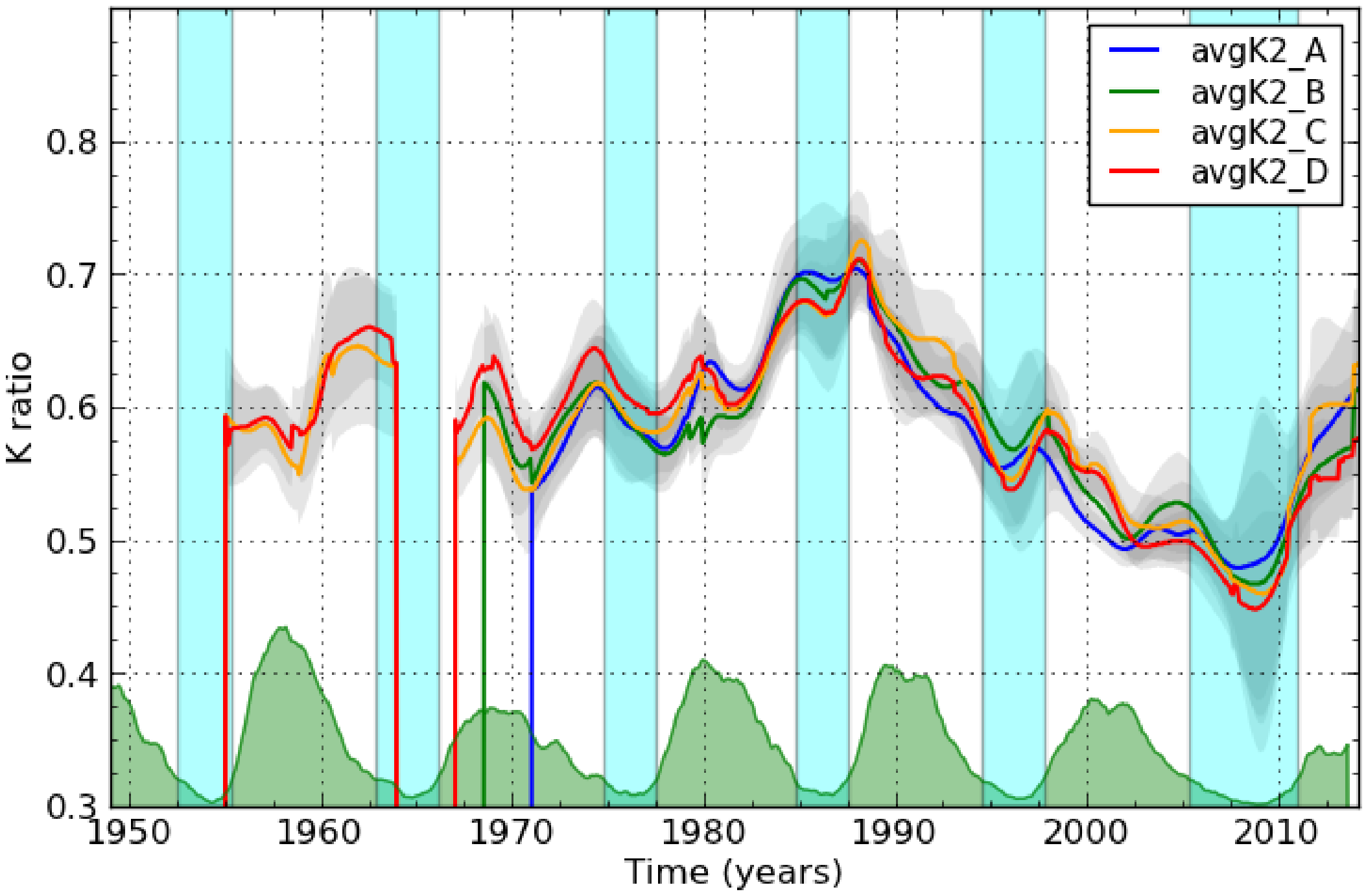

Figure 53: Plot of different multi-station average k ratios spanning the 1955-2013 interval, equivalent to Figure 52 but here for 4 different subsets of stations (A: 3 best stations; B: 6 stations; C: 16 stations; D: 22 stations). The gray shading indicates the standard error on the average k. All curves overlap well the interval 1981-2013, confirming the drifts over that period, while before 1981, no significant trend is present. The data gaps correspond to missing years in the recovered Zürich archives.

## *6.3. Possible causes: current interpretation*

As the above detected trends must be due to changes in the Locarno counts, we investigated directly the possible causes of such biases using the information in the Locarno observing logs. A first possible cause of long-term trends in the Locarno Wolf number is a change in the instrument. We had the confirmation that no modifications were made to the instrument since its installation and that there were no changes in the observing method, i.e., visual count on the aerial image at the eyepiece with the aperture stopped down to 80mm (Sergio Cortesi, private communication).

Another factor may be a systematic change in the image quality (local seeing conditions). As a quality index is attributed to each observation (Kiepenheuer scale, decreasing from 5 to 1 for improving sky conditions; Kiepenheuer 1962), the yearly average quality index can be plotted as a function of time (Fig. 54). This curve shows mainly a 0.25 jump around 1969. This sharp degradation was clearly associated with the construction of a large villa just south of the observatory. Thereafter, there is only a slight degradation by 0.15 mainly over the interval 1989-

2000.

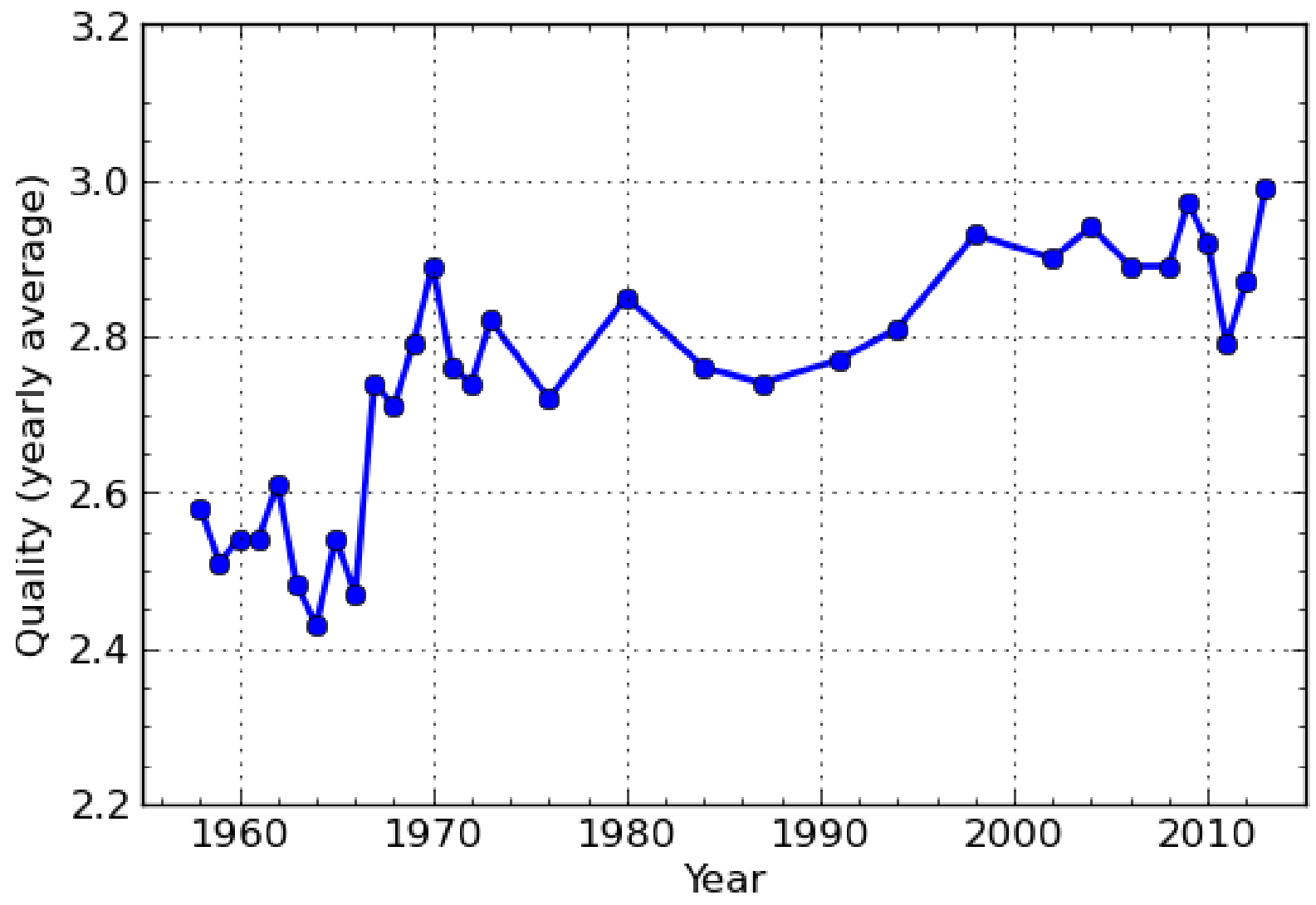

Figure 54: The variation of the annual average quality index of the visual observations at the Specola Observatory in Locarno (Kiepenheuer scale: higher values correspond to lower image quality). A sharp jump occurs in 1968, followed by a smaller progressive degradation.

Overall, the changes in seeing conditions do not match the pattern of changes in the k ratio (times of inflexion and reversal of the trends). Moreover, the largest change takes place before 1980 and did not produce a significant jump in the Locarno k coefficient, making it unlikely that the smaller variations of the average seeing after 1980 could explain larger deviations of the Locarno scale.

A last factor could be a change in the eyesight of the lead observer, Sergio Cortesi, who observed during the remarkable duration of more than 55 years. The only material available to check this is a comparison between the Wolf numbers obtained by S. Cortesi and those obtained by auxiliary observers who occasionally replace him. Those auxiliary observations were often made for single days interleaved with those of the main observer. Figure 55 shows the k personal coefficient for the observer who accumulated the longest series (Michele Bianda). Just like similar k series for 3 other auxiliary observers, those k values only show very weak mutual trends. This comparison thus seems to contradict a possible change in the counts made by S. Cortesi.

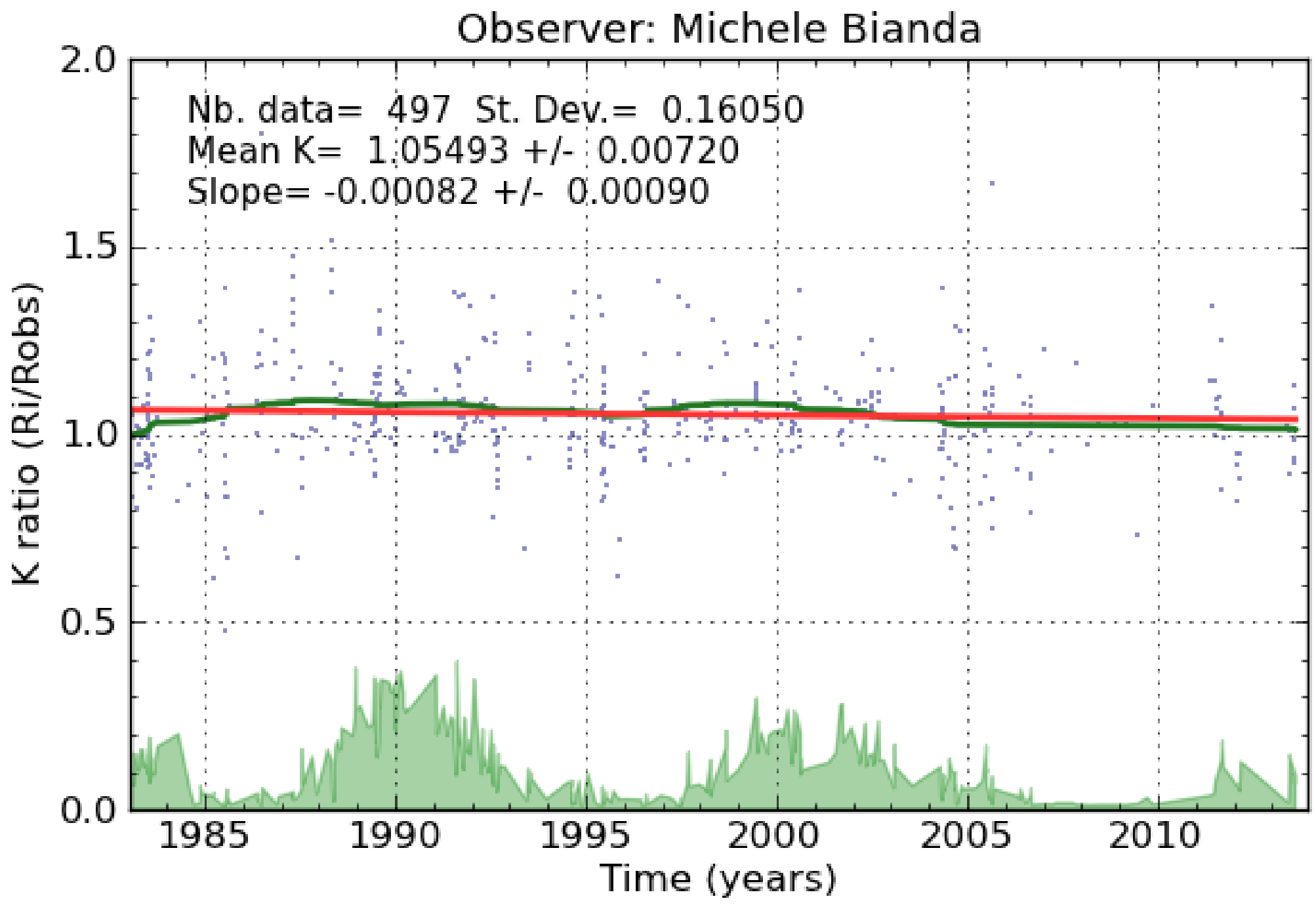

Figure 55: Sample plot of the relative k coefficient of an auxiliary observer of the Specola station in Locarno (Michele Bianda) versus the SN (tied to the prime Locarno observer, Sergio Cortesi). The red line shows the linear fit to the data, while the green curve results from a 5-year running mean. No significant linear trend is found over the whole 1983 - 2009 interval. The parameters of the linear regressions are listed in the plot and the green-shaded curve gives the evolution of the solar cycle, as time reference.

However, when considering the high consistency between counts from all Locarno observers, we must take into account the fact that all Locarno observers are in close daily interaction, by contrast with the widely dispersed observers in the international sunspot network. Therefore, the observations made by auxiliary observers may be influenced by preceding observations made by other observers, primarily S. Cortesi, when trying to remain consistent with the counts of past days (M. Cagnotti, private communication).

Moreover, a well-documented change in the observing routine took place progressively since 2005. As a careful preparation of the future replacement of S. Cortesi by Marco Cagnotti, the new Director of the Specola Observatory, the latter has been trained and is progressively taking over a larger fraction of the observations since 2005. Currently, M. Cagnotti has effectively become the prime observer for the Locarno station. This change matches well the observed recent reversal of the network-averaged k ratio, which increased systematically since 2005, indicating a progressive rise in the Locarno counts relative to the network average. This favors the following interpretation: over the last decades, the Wolf numbers reported by S. Cortesi decreased slowly due to subtle aging effect and eyesight degradation. The Wolf numbers started to increase again with the progressive contribution by a younger observer, who now essentially records the same sunspot counts as S. Cortesi when he was younger, more than 30 years ago.

In regard to the initial increase of the network k ratio between 1981 and 1984, we speculate, in the absence of direct quantitative evidence, that when Locarno took over as reference station, it started to overcount. Indeed, while in the previous 25 years, Locarno could take the Zürich Wolf numbers as a direct reference to maintain a common average scale on a daily basis, after 1980, it was suddenly in a stand-alone situation. Then maybe, a primary concern of missing small spots led to an opposite effect: more spots were included in the counts. The effect may have been amplified because it happened in the declining phase of cycle 21, when sunspots were actually decreasing in numbers.

## 6.4. The need for a full $R_i$ re-calculation

In order to assess the consequences of the extracted drift, we first applied a correction to the current standard $R_i$ number by multiplying the series by the average network k ratio. Looking now at the network statistics after applying this drift correction, we can take advantage of the fact that the new k coefficients relative to the network average instead of Locarno will be much more constant over the full 32 year period. It then becomes meaningful to compute an average k over the full duration of the observations from each station. We can then obtain the overall distribution of k coefficients for all stations in the SILSO network (Fig. 56). Surprisingly, the distribution is highly non-uniform with three prominent peaks centered on 0.77, 0.88 and 1.00.

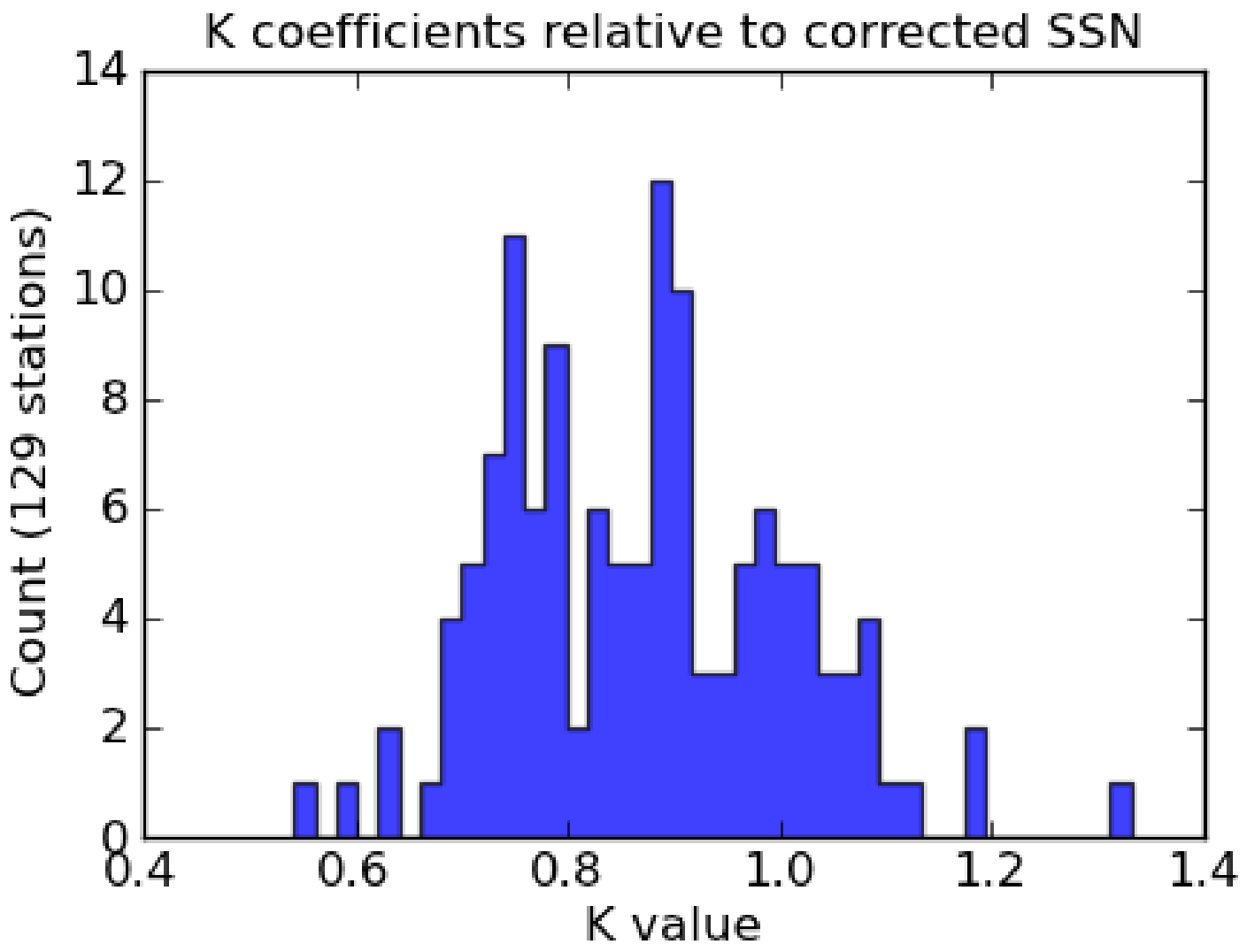

Figure 56: Histogram of the k personal coefficients of all stations after correction by the average correction for the Locarno drift. It shows a multimodal distribution with several well-marked peaks.

The first peak corresponds to a sub-population of stations that provide the highest sunspot counts. Their Wolf numbers are about 20% lower than the Locarno values, probably reflecting the effect of the sunspot weighting used only by the Locarno station.  The other peaks correspond to higher k values and thus to stations that give lower sunspot counts, with the upper end of the distribution near 1, i.e., stations that report SN numbers similar to the Wolf's original values in the $19^{th}$ century. Further investigations at the level of individual stations will be required to fully understand what defines the well-marked sub-populations of observers appearing in the k distribution. At this stage, it just illustrates the prospects that are opened by the corrected $R_i$ for refined analyses of the visual sunspot counting process based on the huge data set at our disposal at WDC-SILSO.

However, the corrected sunspot number used here is only an approximation of the true sunspot number that would be calculated using the standardized method described in section 2.3 but based on a new unbiased reference. Fortunately, such an end-to-end re-calculation is possible, as the WDC-SILSO maintains a full archive of all raw observations that were used to derive the $R_i$ number. As in our present analysis, our reference was the average over all available long-term stations, we probably get a close approximation of what would be this new standard $R_i$ series. However, there are two main restrictions:

- Firstly, the use of a massive average prevents an in-depth understanding of the reference scaling, as the individual history of each station is diluted over a large sample. For instance, it would prevent the detection of a slow drift associated with the evolution of the network properties (e.g. evolution of telescopes) due to the steady renewal of its composition, with stations entering and leaving the network. In addition, the current analysis provides the relative variation of the k ratio but does not directly give the absolute scaling relative to the preceding Zürich series (NB: here, we attached the k scale to the Locarno numbers in 1981 but as the reference interval is very short, the statistical uncertainty is high).
- Secondly, by contrast with this retrospective bulk analysis, this approach is not really applicable for the progressive monthly computation of the future sunspot numbers. Diagnostics of systematic station flaws that can be easily established over a 30 year interval, cannot be done instantaneously and are lagging behind by months or even years, as information about the future evolution is lacking.

Our experience thus indicates that our new reference when fully re-calculating the sunspot number from all raw observations should be based on one or a few well-identified high-quality stations, for which we can obtain a good record of their daily practices.

We thus face the question: which reference can we use? Three main options can be envisioned:

- **Correcting the Locarno series**: the prospects for correcting the past observations are limited, because there is no material record for visual counts. We can consider instead using the Locarno sunspot drawings, but those drawings are made independently at full aperture (15cm) and include all spots, even short lived ones that are neglected in the official Wolf number. Therefore, it will lead to a different Wolf number with a different k personal coefficient relative to the preceding Zürich-Locarno cross-calibration. We thus lose the direct scaling link that was intrinsic to the equivalence between the Zürich and Locarno SSNs. Still, this alternate Locarno series may be more homogeneous and the counts could be done without the size-dependent weighting, while still retaining a strong affiliation with the preceding Zürich standards,
- **Choosing an alternate pilot station**: our analysis already hinted at rather stable long-duration stations that tracked well the network average, and thus do not suffer from large deviations. A more focused analysis of individual stations should be done to settle this critical selection. This alternate pilot station should however offer the same guarantees of long-term observations as Locarno for future years. This probably restricts the choice to professional observatories. The Uccle station seems then to be a prime candidate for three reasons: absence of drifts, long overlap with the Zürich era (since 1940), and sunspot counts based on the drawings from which a detailed catalog is currently in construction and would allow fully documenting the counts.
- **Building a composite reference** based on an average over a core group of high-quality stations: this option would make it more difficult to understand the factors influencing the scaling, but it would avoid the risk of drifts going unnoticed as in the case of a single station. A continuous mutual comparison of the pilot stations would allow pinpointing and discarding any accidental flaw present in only one series. In order to be manageable, this core group should not include more than 4 to 10 stations. Here, as losing one of the stations does not cause a complete disruption, it relaxes the selection criteria for what concerns the longevity of the stations, opening the door to good and dedicated individual amateur observers.

The last two options will be tested in coming months, by comparing the output of different simulations (i.e., full recalculation of the SN over the last 32 years) based on different combinations of pilot stations. The first option first requires a recounting based on the Locarno sunspot drawings, before this new Locarno series can serve as reference and be tested for its stability.

## 6.5. *Implication for the solar cycle: a variable sunspot population?*

Given the amplitude of the diagnosed Locarno drifts, they can lead to significant biases relative to other solar indices, in particular $F_{10.7}$. In order to assess if this drift can explain the cycle 23 divergence between the SN and $F_{10.7}$ mentioned at the start of this section, we tentatively used the corrected series $R_i$ obtained as above by multiplying the series by the average k ratio.

Figure 57 shows a comparison between the original $R_i$ number, the corrected number and the $F_{10.7}$ proxy $R_{F10.7}$ The corresponding ratios are shown on the lower panel. We find that the corrected series matches better the $F_{10.7}$ proxy up to 2002. This thus gives a confirmation that the correction derived self-consistently only from original sunspot data effectively eliminates a true bias in the SN series. However, while the corrected SN and $F_{10.7}$ come closer to each other after 2002, the difference is only reduced by half and is still significant over the rest of cycle 23. The persistence of a significant deviation suggests that the recent divergence between those indices cannot be simply explained by a flaw in the SN but must be associated with a true change on the Sun.

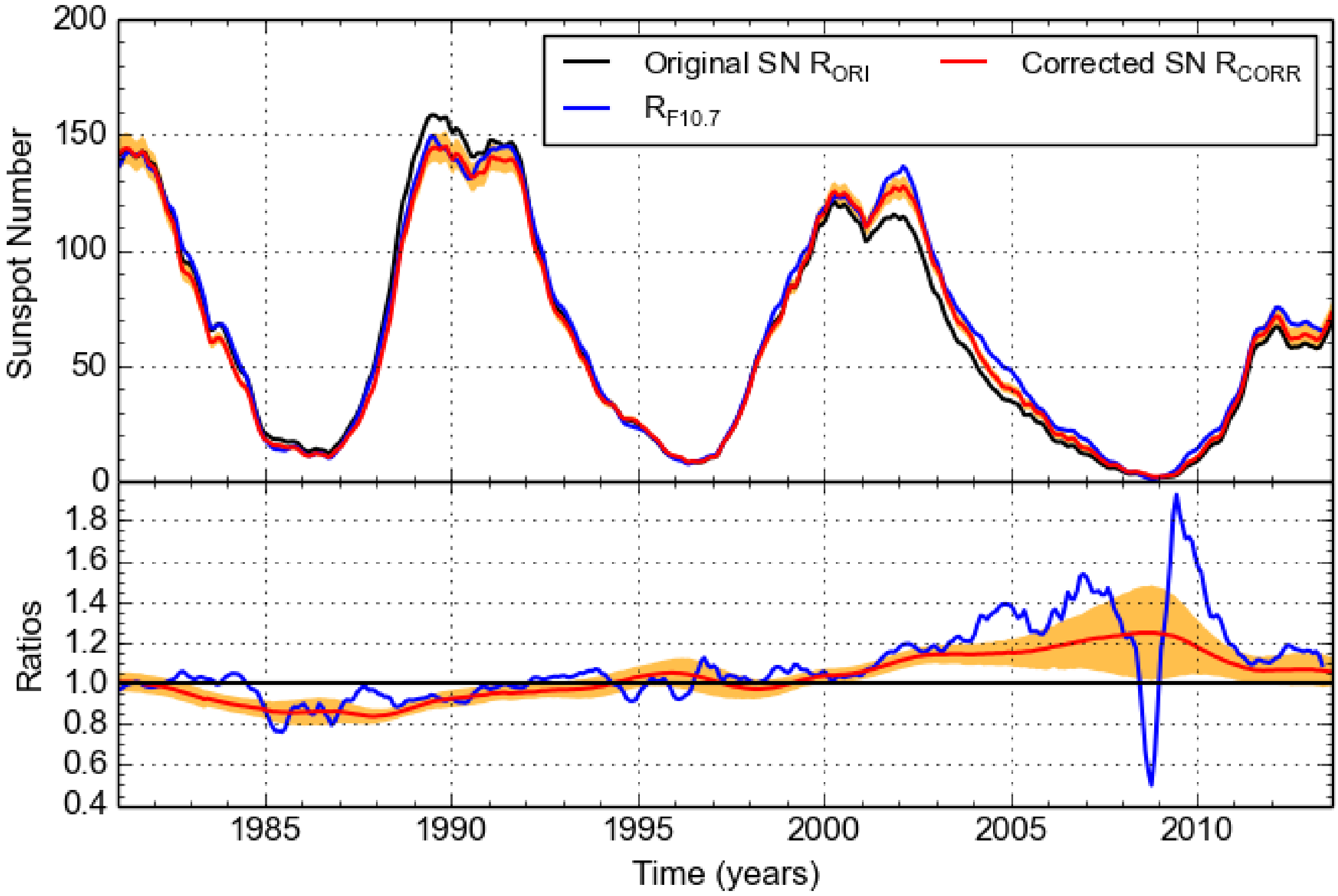

Figure 57: Upper panel: comparison of the original SN series $R_{ori}$ (black) with the new $R_{corr}$ series after correction by the average network-wide k ratio (red) and the $F_{10.7}$-based proxy $R_{F10.7}$ (blue) over the last 3 solar cycles. Lower panel: the ratios $R_{corr}/R_{ori}$ (red) and $R_{F10.7}/R_{ori}$ (blue), with the standard error in the corrected SN (orange shading). They show a better agreement between $F_{10.7}$ and the SN after the correction, but still with a significant though reduced divergence after 2002.

In order to diagnose the possible cause of this change, Lefèvre and Clette (Lefèvre and Clette 2011, Clette and Lefèvre 2012) exploited the most extensive sunspot catalogs available for the last 28 years in order to derive the occurrence rate of groups or individual spots of different sizes. This study, together with parallel results by Kilcik et al. (2011), finds a significant decrease in the number of small sunspots, by about a factor 2, after about 2000, i.e. precisely when the above index deviation sets in (cf. de Toma et al. 2013a). The affected sunspot population consists of pores without penumbra and with a short lifetime of less than 2 days, both isolated (A and B-type groups) or inside groups containing larger sunspots.

Such a scale-dependent change in the sunspot population should be put in relation with two parallel findings derived by completely different measurements. Using spectroscopy, Penn and Livingston (2011) found a systematic decline of the average core magnetic field in sunspot umbrae over the 2002-2013 period. Combined with the determination of a sharp lower threshold in the magnetic field strength allowing the formation of a sunspot (~1500 Gauss), they conclude that, should this trend continue, the formation fraction of sunspot will be decreasing due to sunspots vanishing at the lower threshold, truncating the lower part of the distribution of magnetic fields, i.e., the small sunspots. Although their speculation of a steady long-duration trend is debated (de Toma et al. 2013b), other studies still indicate that the recent trend is part of a particularly deep solar cycle modulation (e.g. Nagovitsyn et al. 2012, Pevtsov et al. 2014).

Using long-duration helioseismic data from the BISON network, Basu et al. (2012, 2013) detect the onset of a similar deviation between the frequency of high-frequency p modes and its SN-based proxy, after 2000. Such modes are confined near the surface and thus record the average magnetization at the surface due to the presence of active regions. They also find that modes having a deeper turning point start to deviate even earlier, already during cycle 22. They interpret this behavior as a thinning of the near-surface magnetized layer. The coincidence of those symptoms may indicate that this size-dependent variation of the sunspot population is real and reflects a physical change in the underlying dynamo processes or in the magnetic diffusion processes leading to the sunspot group decay. Finally, using image data from the RGO and SOON catalog, Javaraiah (2011) finds a significant change in the growth and decay rates of sunspot groups over past solar cycles. Although this analysis did not consider the change in sunspot size population, such global variations of the average rates can be influenced by the change in the relative fraction of small versus large sunspots, i.e. spots with very different lifetimes.

Other solar indicators also tend in the same direction. E.g., for each magnetogram taken at the 150-ft solar tower of Mount Wilson Observatory (MWO), a Magnetic Plage Strength Index (MPSI) value is calculated by summing the absolute values of the magnetic field strengths for all pixels where the absolute value of the magnetic field strength is between 10 and 100 gauss. This number is then divided by the total of number of pixels (regardless of magnetic field strength) in the magnetogram. The magnetic calibration after the instrument upgrade in 1982 is believed to be good, or at least stable (Parker et al. 1997). On average, there is a very nearly linear relationship between MPSI and the sunspot number: $SSN^* = 54.7\ MPSI^{1.0089}$. We can thus calculate a synthetic SSN* for each (monthly) value of MPSI, and form the ratio between the observed sunspot number and the synthetic one derived from MPSI (Figure 58). A 5-month, centered running average is shown by blue diamonds. The ratio is high in the approach to solar minimum and in the very early part of the ascending phase of the cycle (large boxes) before settling down at solar maximum where it is well-defined.

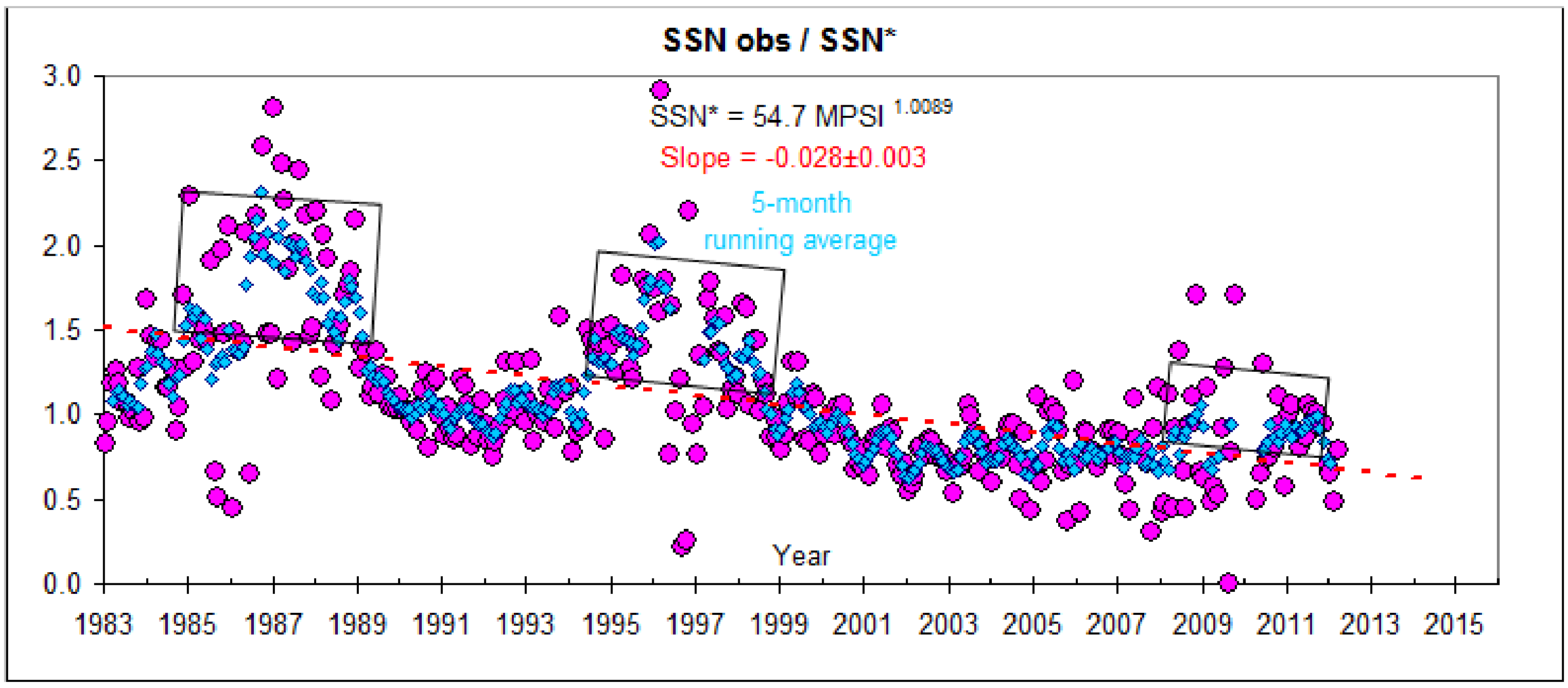

Figure 58: The observed International Sunspot Number (SSN) divided by a synthetic sunspot number derived from the MWO Magnetic Plage Strength Index (pink circles for monthly values). A 5-month, centered running average is shown by blue diamonds. The ratio is high in the approach to solar minimum and in the very early part of the ascending phase of the cycle (large boxes) before settling down at solar maximum. For solar minimum years, the ratio is between two very small numbers and is thus very noisy and at times undefined. There is also a second-order annual variation of unknown origin, phased with solar distance.

Again, we find that the ratio has been declining over the past two cycles, consistent with the similar findings mentioned above. Observations at or after the current maximum should settle the matter of the reality of a secular change. Unfortunately, MWO is currently not taking magnetograms, teaching us the detrimental effect of stopping a valuable synoptic observing program.

As noted by Lefèvre and Clette (2011), such a scale-dependent variation of the sunspot population could match the existence of two dynamo components, e.g. a deep and shallow dynamo. Various recent developments in dynamo theory provide schemes that could be tested against this peculiar evolution of sunspots, Babcock-Leighton near-surface flux diffusion mechanism (Muñoz-Jaramillo et al. 2010, 2011), role of a near-surface shear layer (Brandenburg 2005) or a near-surface magnetic flux aggregation mechanism (Schatten 2009). The selective disappearance of small spots can also have implications on the total and spectral irradiance reconstructions. Indeed, while vanishing sunspots will lead to lower sunspot blocking and thus should increase the TSI, the corresponding weakening magnetic fields should still be present but below the 1500G lower limit found by Livingston and Penn (2009). They may thus contribute to a corresponding excess in the near-UV and microwave emissions by the plage component. The result would be a reduction of the effective irradiance drop for weak solar cycles, which may be related to the existence of a base level in solar flux, as proposed by Schrijver et al. (2011). In any event, these scale and lifetime dependencies of sunspot magnetic fields should be a warning against any simple linear rescaling of the present irradiance proxies by a relation of proportionality.

## *6.6. Variations in the GN/SN scaling ratio*

A scale-dependent deficit of sunspots such as the one found in the above studies, should logically lead to a corresponding change in the average number of sunspots per group. However, inherent in the definition of the Group Number is the assumption that the ratio between the number of spots and the number groups, i.e. the average number of spots per group is constant. Therefore, a new question arises: is the fixed 12.08 constant used in the GN definition a valid assumption?

We can investigate this assumption using data from the German SONNE network of sunspot observers (Sonne 2014; Andreas Bulling Personal Communication) and from the long-running Swiss station Locarno (Locarno 2014) supplemented by observations at Zürich by Waldmeier (1968) and Zelenka and Keller (Keller and Friedli 1995). From each data source, the number *G*, of groups, and the number *S*, of "spots" reported by the observers is extracted and tabulated. "Spots" is in quotation marks because Waldmeier, as we have shown, and to this day Locarno as well, weighted larger spots stronger than small spots. The SONNE observers do not employ weighting: each spot is counted only once. It is important that for both groups of observers, the counting methods (albeit different) have been unchanged over the period of interest.

If the Relative Number, *R*, and the Group count, *G*, are known, the spot count can be calculated as $S = R/k - 10G$, where *k* is the k-factor introduced by Rudolf Wolf to bring observers onto the same scale as Wolf himself, who by definition had $k = 1$. As explained in section 2.2, for the later Swiss observers *k* was set by *adoption* to 0.60. The SONNE series is adjusted to match the Swiss k-factor, which, however, is also applied to the group numbers reported by SONNE so that a composite group count can be computed over many observers, effectively resulting in a spot/group ratio that is independent of the k-value. The published data for Waldmeier and SONNE gives us *R* and *G*, so *S* has to be calculated as detailed above. For Locarno, Zelenka, and Keller, both *S* and *G* are available directly. Given *G* and *S*, either determined directly or calculated from *R* and *G*, the average number of spots per group, *S/G*, can now be computed for each year.

Figure 59 shows that the average number of spots per group has been decreasing steadily for both SONNE and Locarno and is therefore not likely to be due to drifts of calibration or decreasing visual acuity of the primary Locarno observer (Sergio Cortesi since 1957). This is consistent with the conclusion in section 4.2 (Figure 45) where we compare the weighted counts made by the veteran Cortesi and the new observer Cagnotti from 2008 to the present, and find no systematic difference or variation with time.

Because the Locarno observers weight the spot count according to structure and size of spots, they report more spots that the SONNE observers. This is clearly seen in the bottom panel. Also, in Figure 59, we plot (top panel) the variation of the ratio between the number of spots and the number of groups for the more than 431,000 SONNE individual daily observations without any correction for k-factors (green line with plus marks). The "raw" data show the same general variation and decline as the adjusted observations. Such a trend could be a solar phenomenon or due to an increasing group count brought about by sharper determination of what constitutes a "group". There seems to be a solar cycle variation as well.

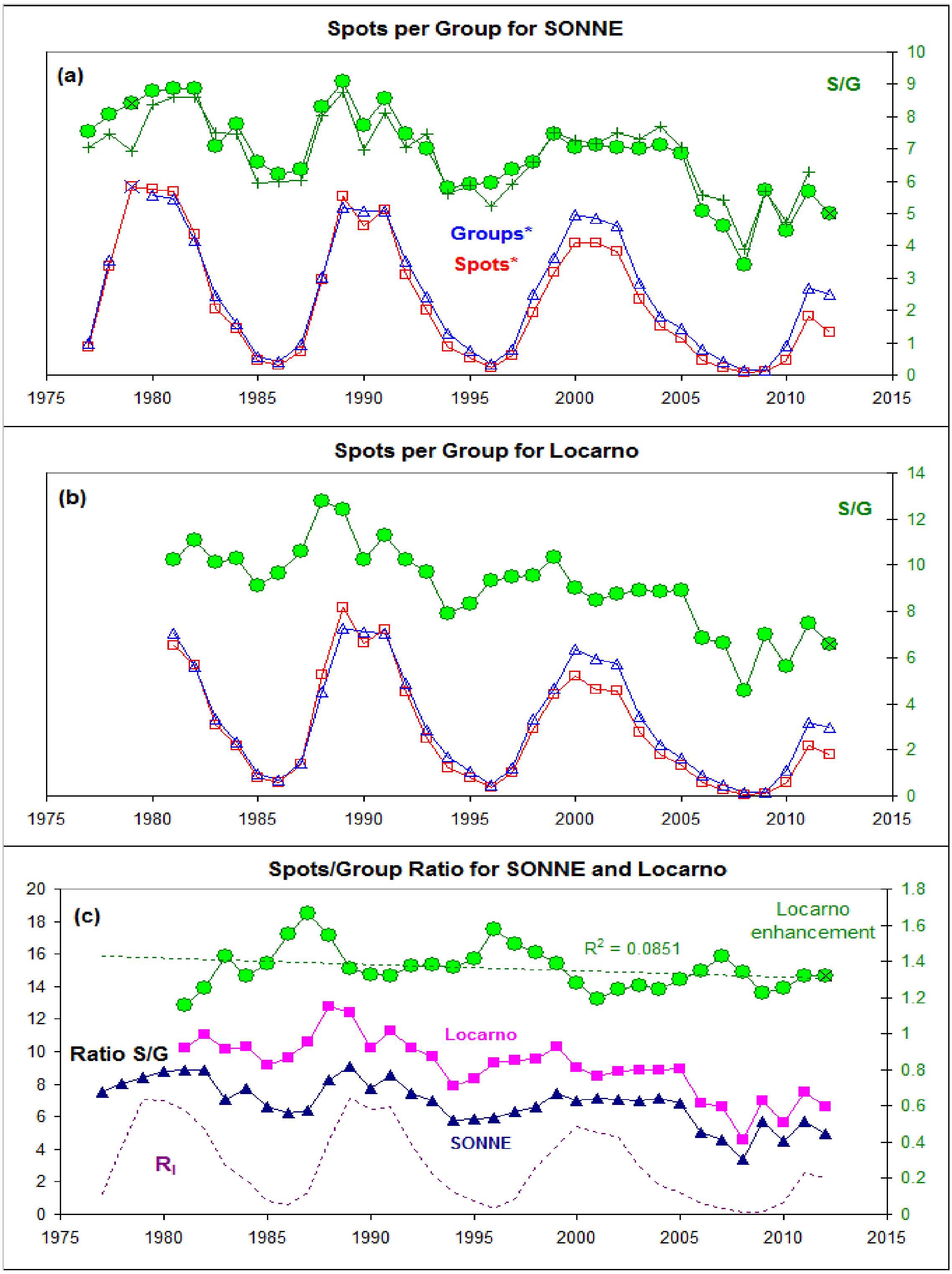

Figure 59: (Top) The number of spots per group as a function of time (green circles) for SONNE. The green curve with pluses shows the ratio derived from the raw counts, not corrected with k-factors. The lower part of the panel shows the variation of number of groups (blue triangles) and the number spots (red squares) both scaled to match each other before 1992. Note the decreasing spot count, relative to the group count. (Middle) Same, but for Locarno. (Bottom) The decrease of the ratio Spots/Groups for Locarno (pink squares) and for SONNE (blue triangles) using the left-hand scale. The enhancement of the Locarno ratio over SONNE (see text) is shown by the green circles (right-hand scale). The trend indicated is not significant.

When Wolf chose 10 as the weight for Groups in his definition of the Relative Sunspot Number, he remarked that he could as well have chosen 9 or 11, but that 10 was certainly “more convenient”. For Wolfer the ratio spots/groups was on the average 9.0 (Figure 60). In Figure 60, the recent decrease of the ratio also seems to be seen at Kislovodsk, supporting a solar cause, rather than a drift due to an individual station.

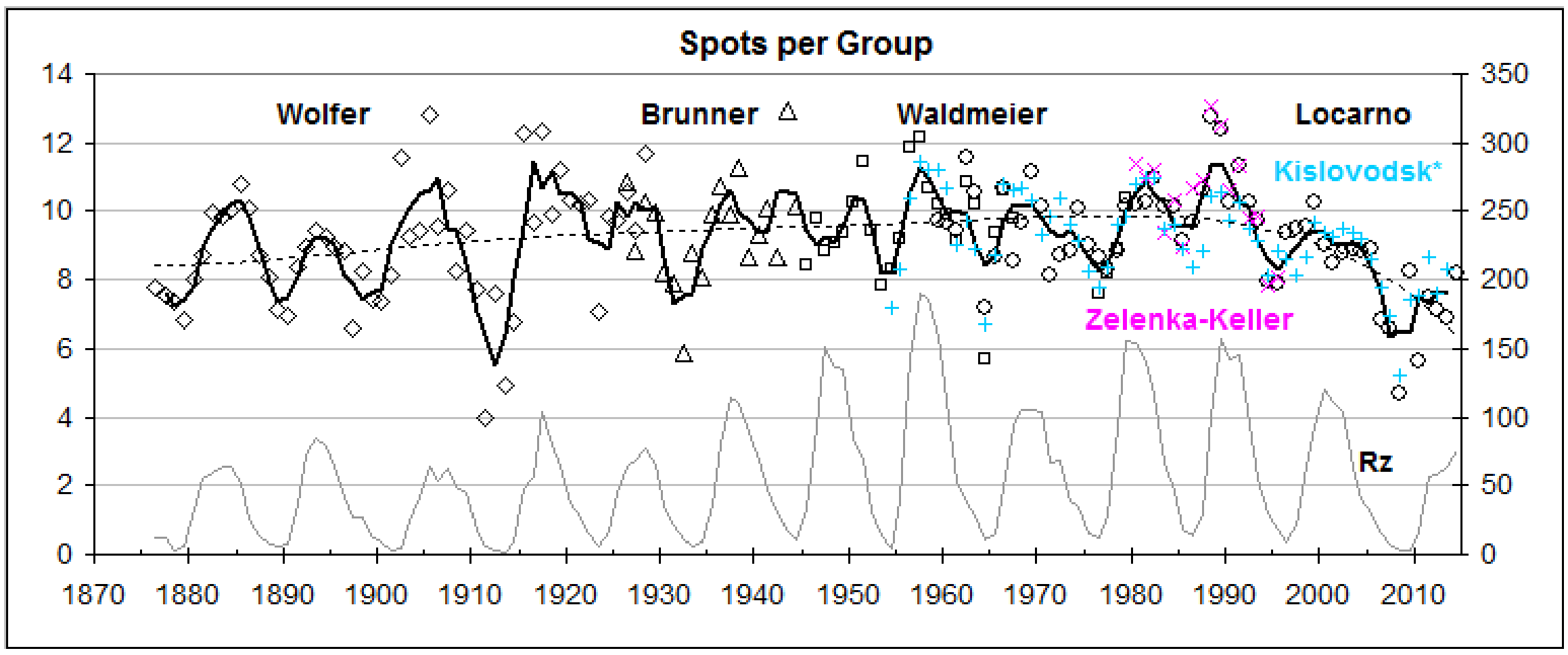

Figure 60: The ratio between the number of single spots and the number of groups as recorded by the Zurich observers for each year of observation. There is a clear solar cycle dependence (sunspot number shown at bottom) with more spots per group at higher solar activity. The ratio observed at Kislovodsk (with no weighting) scaled to Locarno follows the same general trend.

On account of the weighting, one would expect a dramatic increase (~40%) in the ratio when the weighting scheme was introduced. That this is not observed presents a puzzle which at the present time has not been resolved.

In order to confirm the decline over the recent cycles, we derived a reconstructed group number based on the WDC-SILSO archive over the last 6 cycles. The resulting values, given in Table 1 and plotted in Figure 61, the ratio GN/SN stays near 12.5 over cycles 19 to 22, corresponding to an average number of spots per group of 11, i.e. slightly higher than the standard value in the original group number definition. Again, those higher values can be attributed to the higher weighted counts at Locarno, the WDC pilot station. Then the ratio drops sharply between cycles 22 and 23.Thereafter, even during the present cycle, the GN/SN ratio remains at a lower value of 11, i.e. an average number of spots per group as low as 9, which is much below the standard Group Number scale.

A recent long-term reconstruction of the yearly average number of spots per group by Tlatov (2013) suggests that there were rather large variations of this quantity over the past century. In Figure 61 (right plot), we included the last points from this study. It shows that the SONNE values nicely extend Tlatov's values and that all series share the same downward trend, except for the high value for cycle 22 and the globally higher values for the Locarno-based data set.

| Cycle | Ns/Ng | North | South | SN/GN | North | South |
|---|---|---|---|---|---|---|
| 19 | 11.3 | | | 12.8 | | |
| 20 | 10.3 | | | 12.2 | | |
| 21 | 10.4 | | | 12.2 | | |
| 22 | 11.9 | 12.5 | 10.7 | 13.2 | 13.5 | 12.5 |
| 23 | 9.3 | 9.5 | 8.7 | 11.6 | 11.5 | 11.9 |
| 24 | 8.5 | 8.7 | 6.4 | 11.0 | 11.2 | 10.3 |

Table 1: Average ratio between sunspot counts and group counts and average ratio between the SN and GN over the last 6 solar cycles. Cycles 23 and 24 mark a sharp drop relative to earlier cycles.

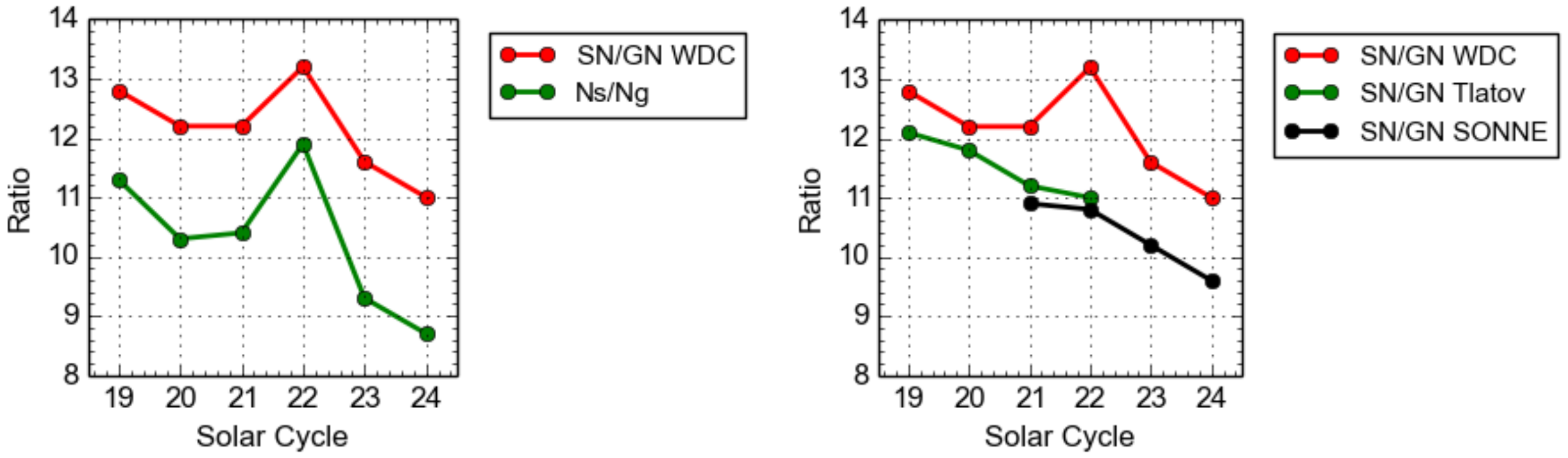

Figure 61: (Left) variations of the average number of sunspots per group (green) and SN/GN ratio (red) over cycles 19 to 24, showing a sharp decrease in cycles 23 and 24. (Right) Comparison of average number of sunspot per group for different data set: this study (red), data from the German SONNE network (black) and values from Tlatov (2013) in green, all showing a consistent decrease over recent solar cycles.

The above converging results give a strong indication that the diagnosed deficit of small spots is indeed directly reflected by a drop in the ratio between the SN and GN. Consequently, any interpretation of past disagreements and drifts of the SN relative to the GN, in which the information about the actual group size is absent, should take this significant variability into account. The above analyses indicate that the amplitude of the variations can reach up to 30%. It may thus prove pointless to try matching the GN and SN series to better than a few percent, in particular for durations shorter than a solar cycle. Instead, the remaining differences may be considered as a useful indicator of true changes in the activity regime of the Sun, like the ones that seem to occur right now during the cycle 23-24 transition, rather than irreducible flaws in either series.

Definitely, more work is needed to better track those changes in the past. It will require the use and probably the construction of improved sunspot catalogs, providing detailed properties of individual sunspot groups.

# 7. The solar cycle in a new 250-year perspective

## *7.1. Combining all corrections: a tentative synthesis*

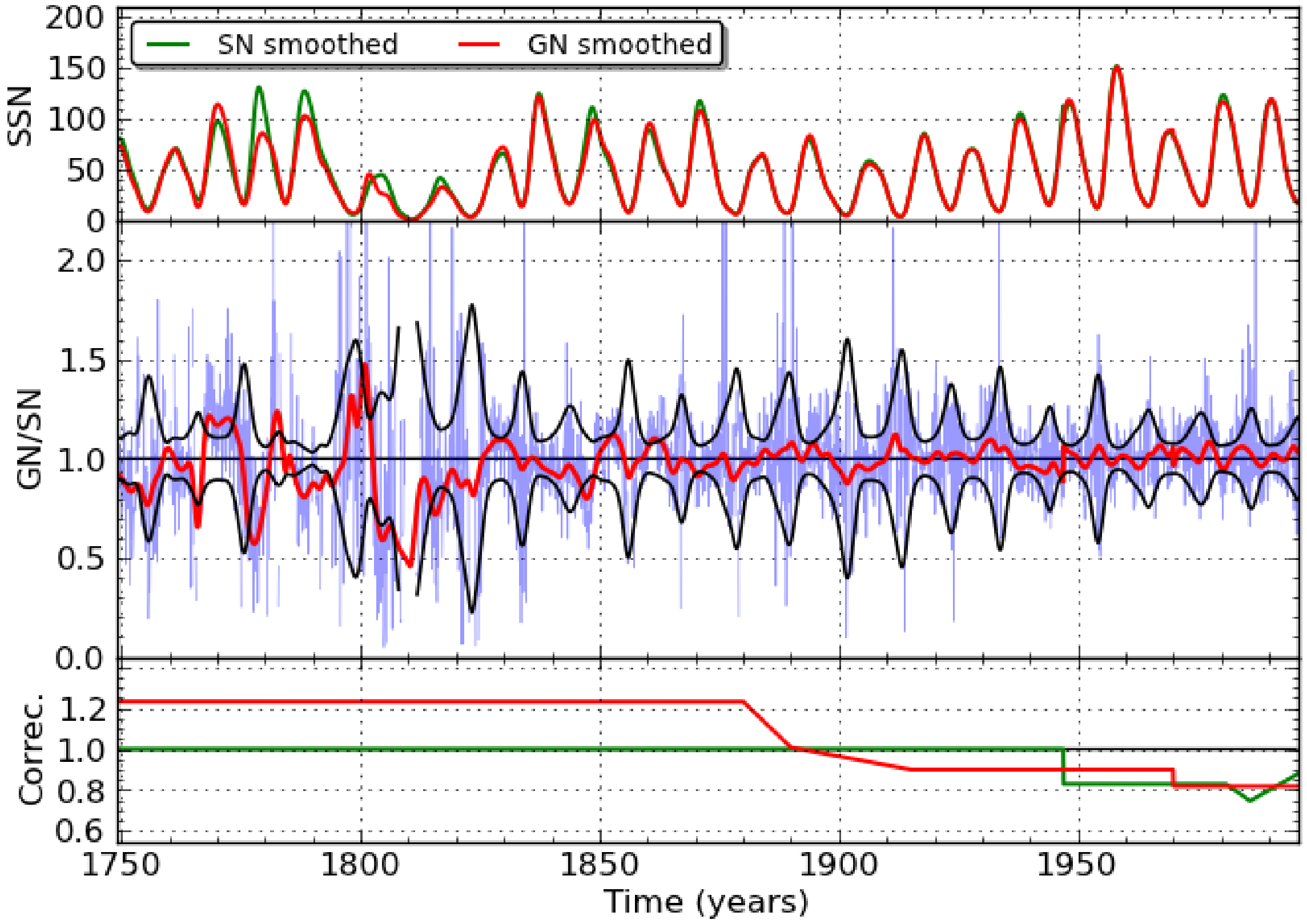

Figure 62: Comparison of the SN and GN series after correction of the main biases identified by recent analyses (see main text). The plot corresponds to Figure 1, with the lower panel showing an initial version of the corrections (red for correction to the GN, and green for the SN). The ratio in the middle panel is largely uniform from the early 19th century onwards, now remaining within the uncertainties. Local significant deviations still remain in the 18th century, when observations are sparse and uncertainties become large and difficult to estimate precisely.

In the previous sections, four primary corrections have been identified:

1. The RGO early drift affecting the group number
2. The Waldmeier sunspot weighting bias affecting the recent sunspot number
3. The RGO-SOON scaling bias occurring after 1975
4. The Locarno scaling drift affecting the recent sunspot number

Here, as a first step, we will apply the first three corrections to synthesize a corrected series from 1750 to the present. At this point, we will ignore the earlier part of the series, which is still affected by larger uncertainties. Based on the measured amplitudes and drift intervals, we applied the following modifications to the series:

1. A 40% progressive ramp between 1880 and 1915, raising all GN values before 1915. The chosen ramp matches the progressive rise in the k ratios between reference visual observers and RGO-based group counts.
2. A sharp 20% jump in 1947, lowering all SN values after that year. Note that this jump falls near the time of the solar minimum between cycles 17 and 18, i.e. a period when the effect of weighting should be limited.

3. The variable 15% Locarno trend in the SN starting in 1981.
4. A sharp 10% jump in 1976, lowering all SOON-based GN values after that year.

As can be seen in Figure 62, with only those four corrections, the resulting series largely match over the entire 1820-2013 time interval, within the limits of statistical uncertainties. Only before 1820, when observations become sparse and the uncertainties grow and also become more difficult to assess, there are still significant disagreements, in particular for cycles 2, 3 and 4, which will require further analyses.

## 7.2. Implications for long-term solar activity: no trend in solar cycle amplitude ?

After corrections, the GN and SN series mostly agree over the last 250 years. Moreover, when compared with the base SN series and original GN series, both the backbone re-calibrated GN, shown in Figure 28 and the equivalent corrected SN series plotted in Figure 63, indicate an important consequence: the secular trend in solar cycle amplitude, shown by the dashed in Figure 63, is strongly reduced after applying the corrections.

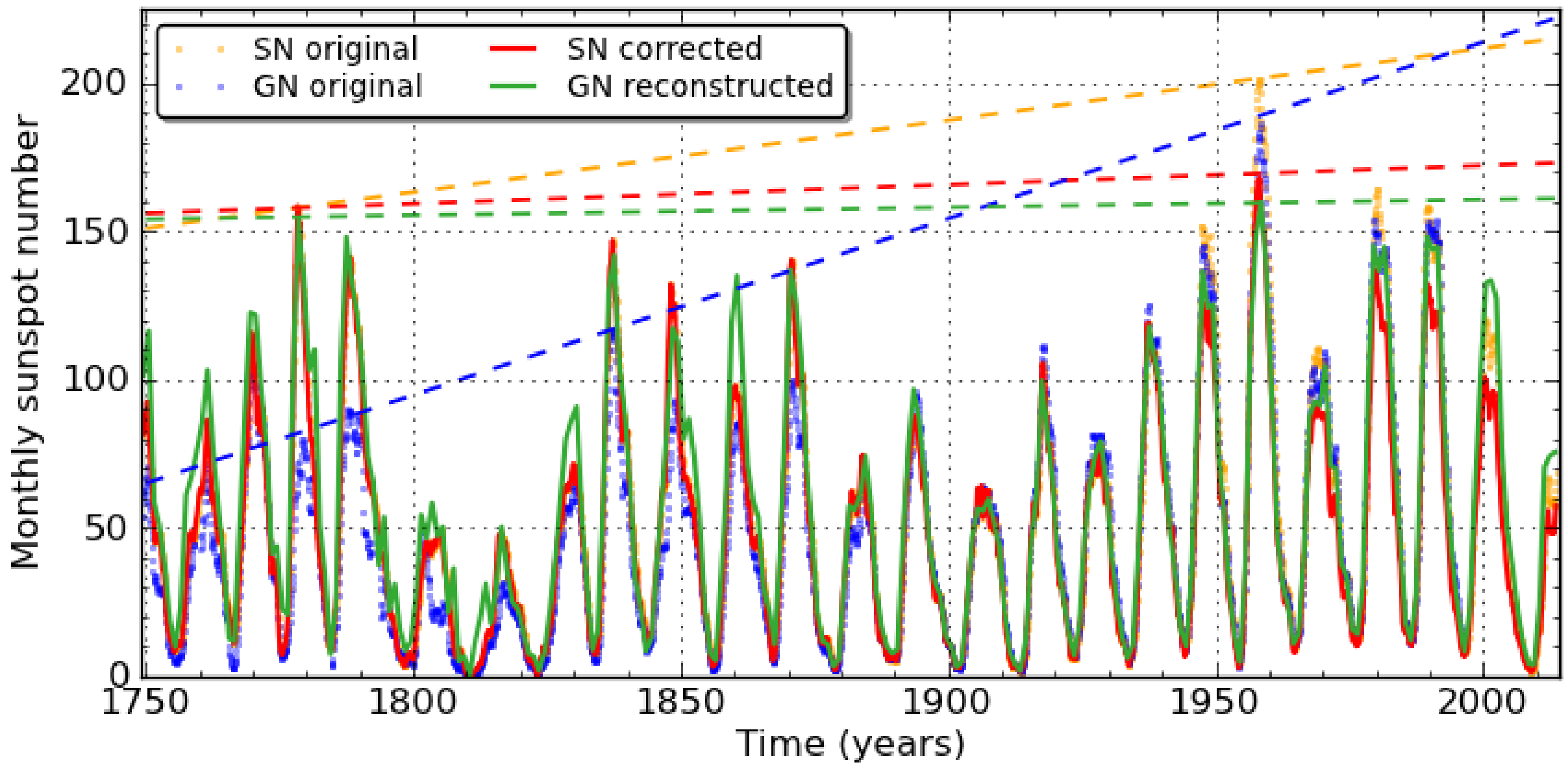

Figure 63: Comparison of the original and corrected SN and GN series over the entire interval 1749 - 2013, showing the limited difference in maximum cycle amplitudes between the 20th century and previous centuries after the new corrections. In order to better visualize the trends, dashed lines connect the highest maxima of the 18th and 20th century, for each series of the corresponding color.

Now, except for the highest recorded cycle (19), the maxima of highest cycles of the past centuries are essentially the same as the recent maxima of the late 20th century. We note that recent independent reconstructions of the Sun's open magnetic field, based on the geomagnetic record, also show a very limited difference of the highest peak 11-year amplitudes occurring in the 19th and 20th centuries over the available 1840-2010 interval (see Figure 30 in Lockwood 2013). Therefore, the upward trend in solar activity levels between the 18th and 20th that was adopted in many past interpretations and models is now questioned, as well as the associated concept of an abnormally high “Grand Maximum” occurring in the second half of the 20th century.

However, although recent cycles do not reach unprecedented amplitudes anymore, the repetition of five strong cycles over the last 60 years (cycles 17 to 22, with the exception of cycle

20) still marks a unique episode in the whole 400-year record. This unique character is also illustrated when considering another sunspot byproduct, i.e. the number of spotless days over each sunspot cycle minimum. As can be seen in Figure 64, this number is strongly anti-correlated with the amplitude of the adjoining cycles (given by the reversed green curve).

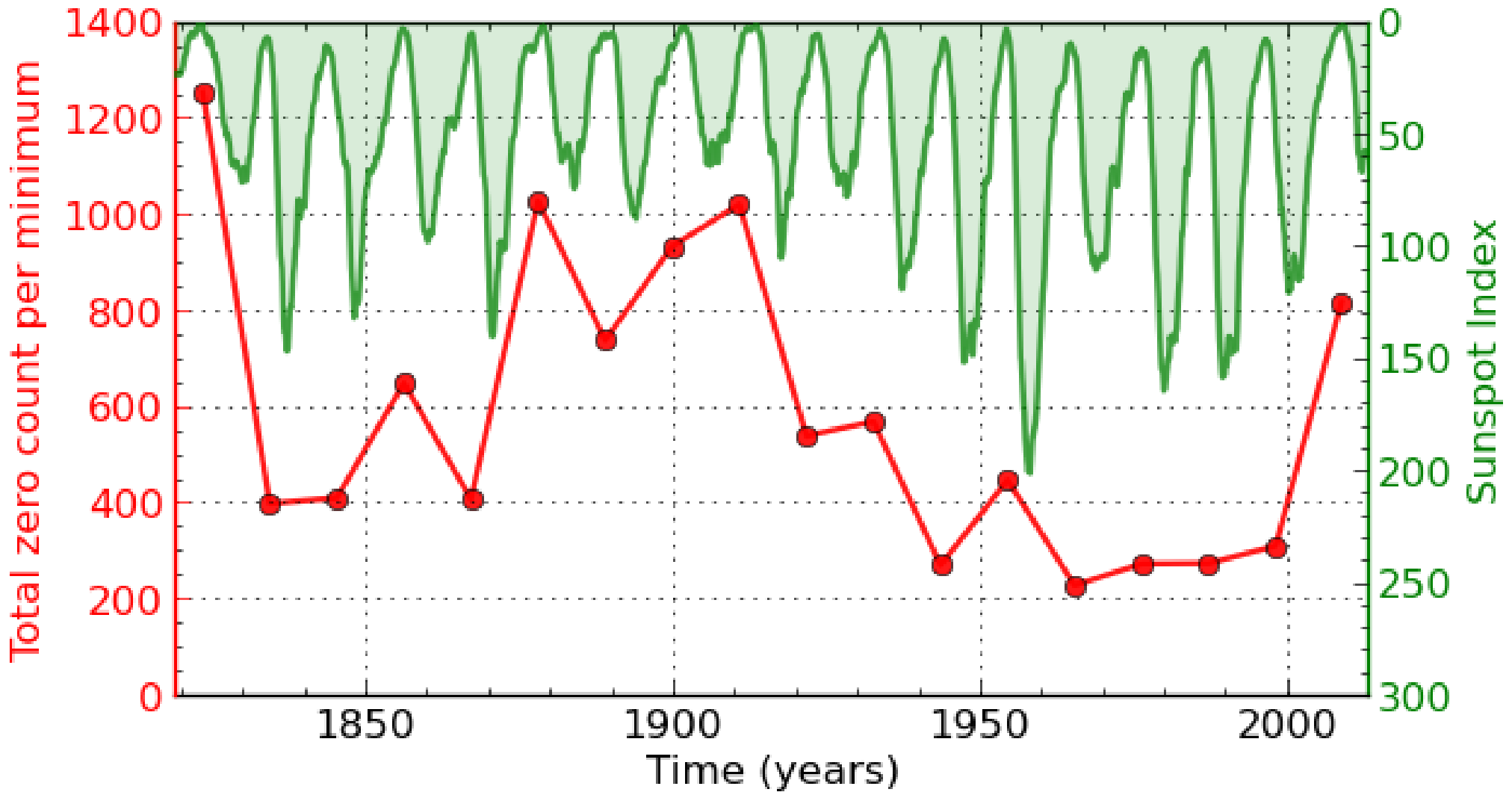

Figure 64: Cycle-to-cycle variation of the total number of spotless days from cycle 6 to 24, for which daily sunspot numbers are available (red curve). The SN series is over-plotted with a reversed scale to highlight the strong anti-correlation between this indicator and the amplitude of the solar cycle. The count for the cycle 23-24 minimum is similar to the late 19th century. The first value for the cycle 5-6 minimum (Dalton minimum) is much larger.

The recent protracted 2008-2010 minimum is marked by a particularly high count of spotless days (~800). Although such a high value was not reached since the early 20th century, the plot shows that it was largely exceeded in the cycle 5-6 minimum belonging to the small Dalton minimum. Therefore, although it marked a contrast with recent minima, this last minimum was not particularly exceptional in a long-term perspective. On the other hand, the uninterrupted series of low spotless day counts during the last 6 cycles stands out as a unique episode over the last 250 years and most probably the last 400 years.

## 7.3. Cycle 24 in a 250 year perspective

As the cycle 23-24 minimum seems to mark a transition in the long-term solar activity, it is interesting to put the current evolution of cycle 24 in context, in order to identify the range of possible scenarios that may develop for the rest of this cycle. Unfortunately, current statistical or physics-based models of the solar cycle can only reproduce global properties of the cycle, with typically a smooth temporal evolution with a single maximum preceded and followed by a monotonous rise and declining phase. This is where the SN series provides unique information about the detailed evolution of the 24 past solar cycles.

In Figure 65, using the original SN series, we aligned all cycles (from 1 to 24) at a common tie point (SN=13) on the declining phase of the previous cycle. As expected, cycle 24 is among late cycles preceded by a protracted minimum. Although the cycles are spread over a continuous range of amplitudes, the plot also shows two concentrations of cycle profiles that share similar rise phases: fast rising cycles reaching high maxima (SN > 130) and slowly rising cycles reaching moderate amplitude (SN ~60). Cycle 24 clearly belongs to this second group.

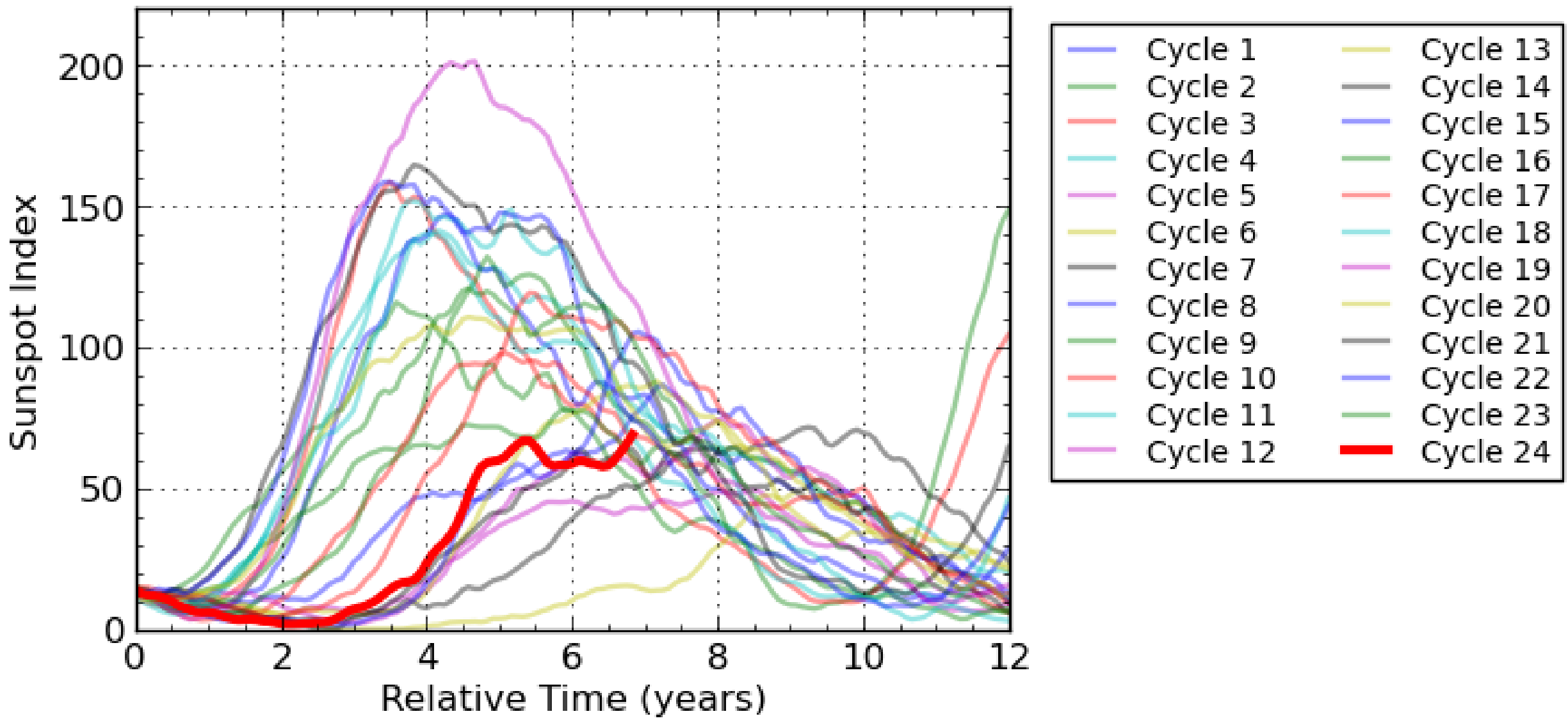

Figure 65: Plot of the 12-month Gaussian smoothed SN for all cycles (1 to 24) after aligning each cycle on the SN=13 crossing point in the final decline of the preceding cycles. Cycle 24 (thick red line) clearly matches cycles that have a late start and moderate amplitude.

In Figure 66, we specifically compare the current cycle with weak cycles 5 and 6 belonging to the Dalton minimum. The rise of cycle 24 is much steeper and even though it is still unsure whether the maximum has actually been reached, the recent SN values clearly exceed the maxima of cycles 5 and 6. Considering now cycle 4 that preceded the onset of the Dalton minimum, while cycle 23 had a similar amplitude, cycle 24 is again strikingly different. Therefore, the peculiar evolution of the current cycle does match the characteristics of the Dalton minimum and cannot be interpreted either as heralding a subsequent extended minimum.

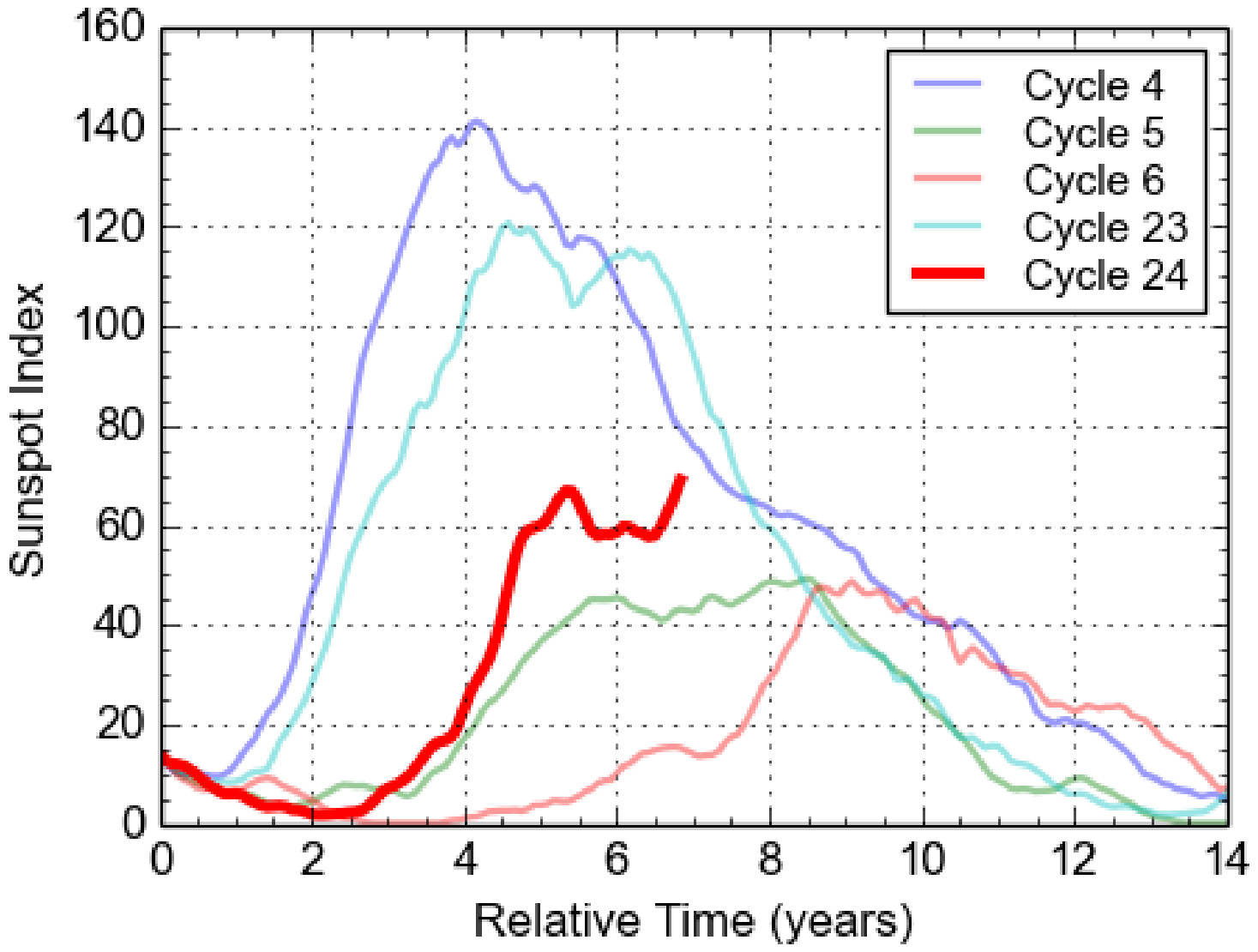

Figure 66: Plot of the 12-month Gaussian smoothed SN for cycles 4, 5 and 6, compared to cycle 24 (thick red line). Like in Figure 65, cycles have been aligned on the SN=13 crossing point in the final decline of the preceding cycle. Cycles 5 and 6 belong to the Dalton minimum, while cycle 4 was the last cycle preceding the Dalton minimum.

Actually, the only cycles that closely match cycle 24 are shown in Figure 67, where we aligned cycles in a tie point (SN=40) in the rise phase of the cycles. Those cycles (12, 14, 15 and 16) all belong to the late 19th and early 20th century. Cycle 24 thus seems to be a return to an average level of activity that prevailed during a 60 year period. Those moderate cycles are characterized by extended and rather flat maxima, lasting up to 4 years. The maximum phase is typically marked by a plateau, on which several peaks are superimposed. Consequently, the absolute maximum of the cycle can then occur quite late, like in cycles 12 and 16, when it occurred more than two years after reaching the plateau phase and more than 5 years after the start of the cycle.

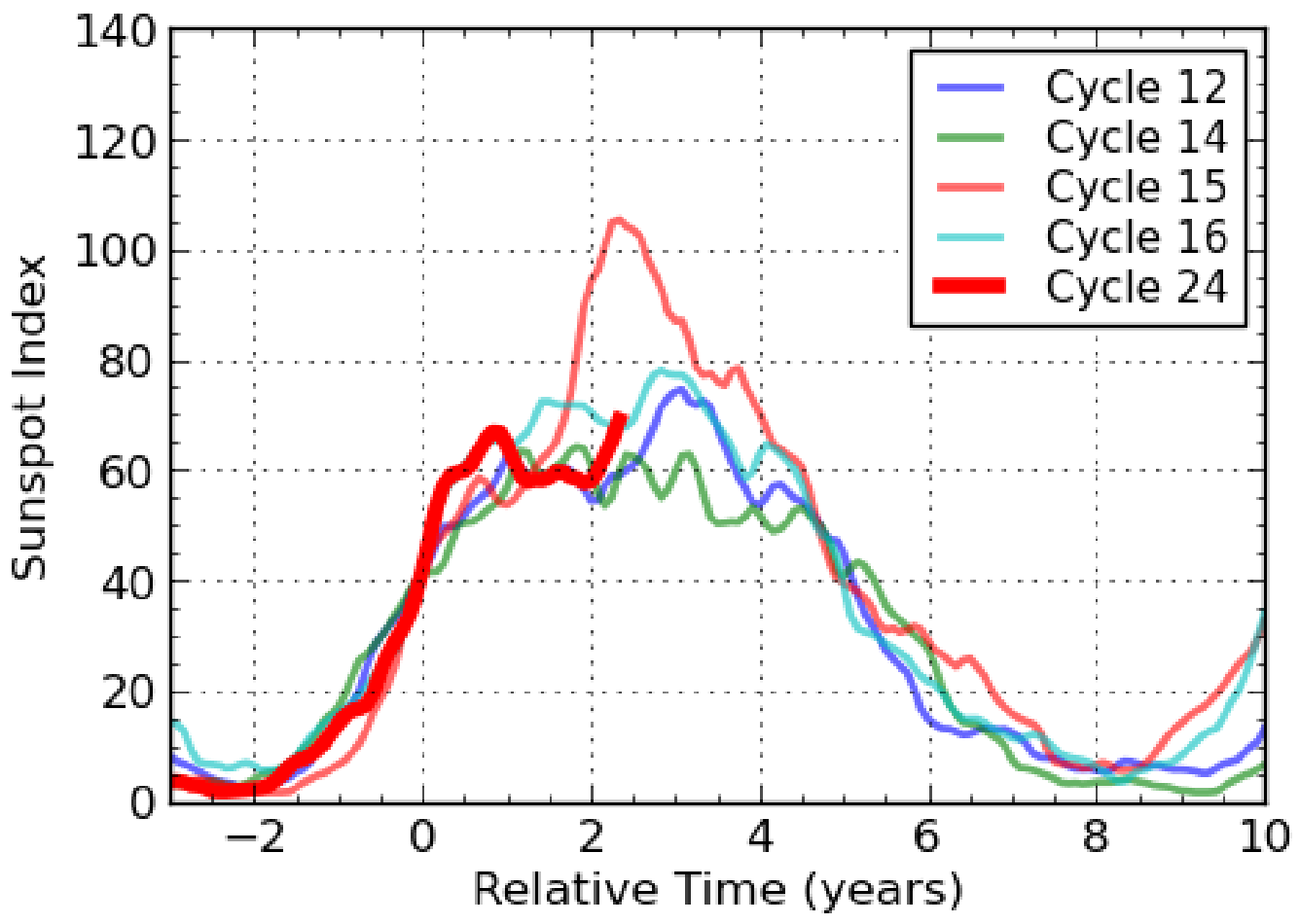

Figure 67: Plot of the 12-month Gaussian smoothed SN for cycles 12, 14, 15, 16, which provide the best match with cycle 24 (thick red line). Here, cycles have been aligned on the middle of the rise phase (at SN=40). Except for the sharp peak in cycle 15, the other cycles all show the same evolution with a rather long and flat maximum.

This is better shown if we avoid any smoothing. Although the overall trends then become more difficult to discern, the detailed progression of the activity is fully preserved. As illustrated for cycle 14 (Fig. 68), the culmination of moderate cycles is actually formed of a succession of multiple peaks or activity surges, up to 6, making the determination of an actual maximum quite elusive.

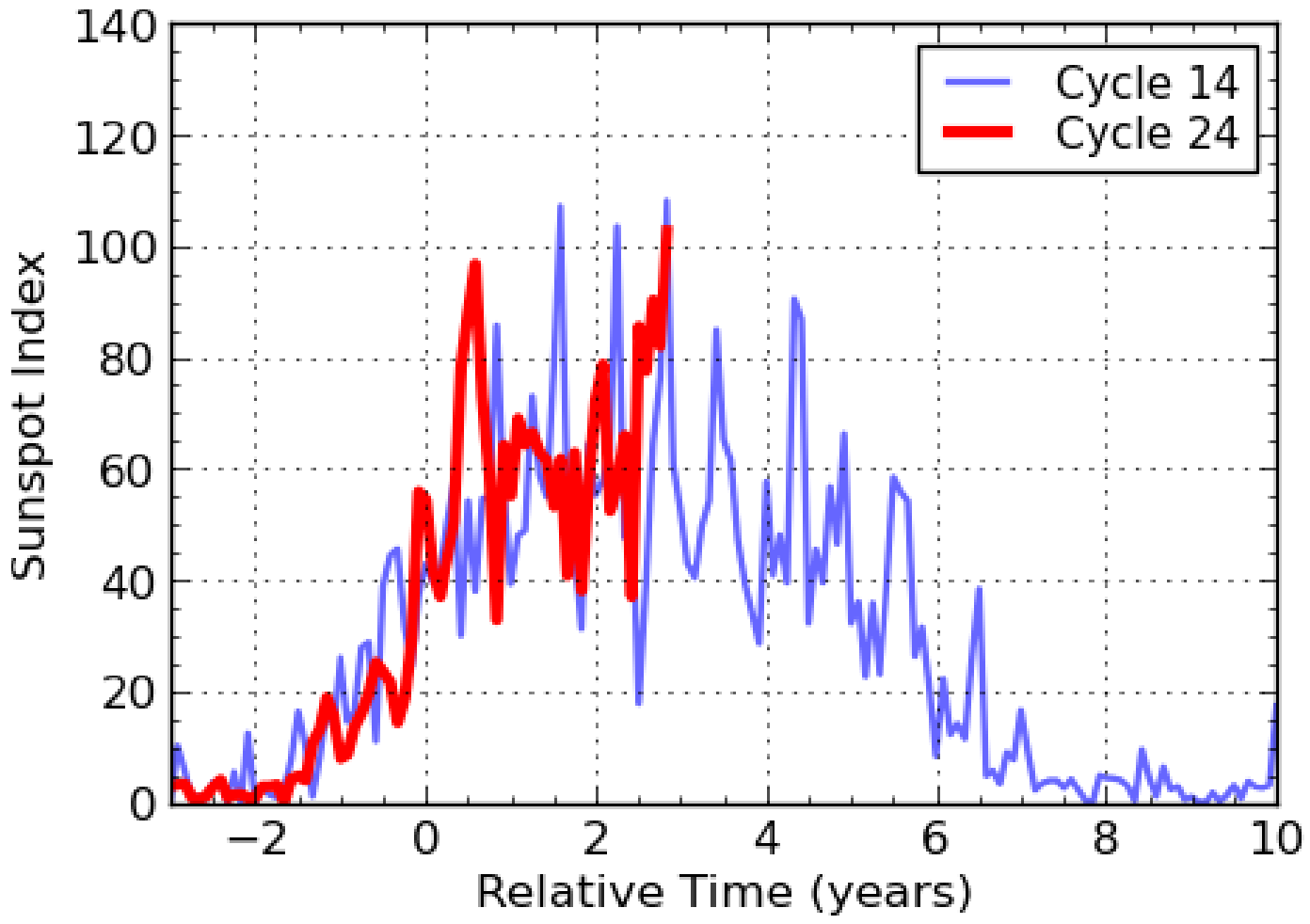

Figure 68: Plot of the monthly mean SN for cycles 14 and 24, after aligning them according to the SN=40 crossing point of the 12-month smoothed SN. Both cycles show a rather flat maximum formed by successive short peaks of activity, lasting each only about 2 to 4 months.

Cycle 15 is of particular interest, as it shares many characteristics with cycle 24 (Fig. 69): a steep rise to a first peak followed by an extended plateau near $R_i$=60, lasting for almost 2 years. Cycle 15 then had a second rise towards a much higher peak, at twice the level of the initial plateau.

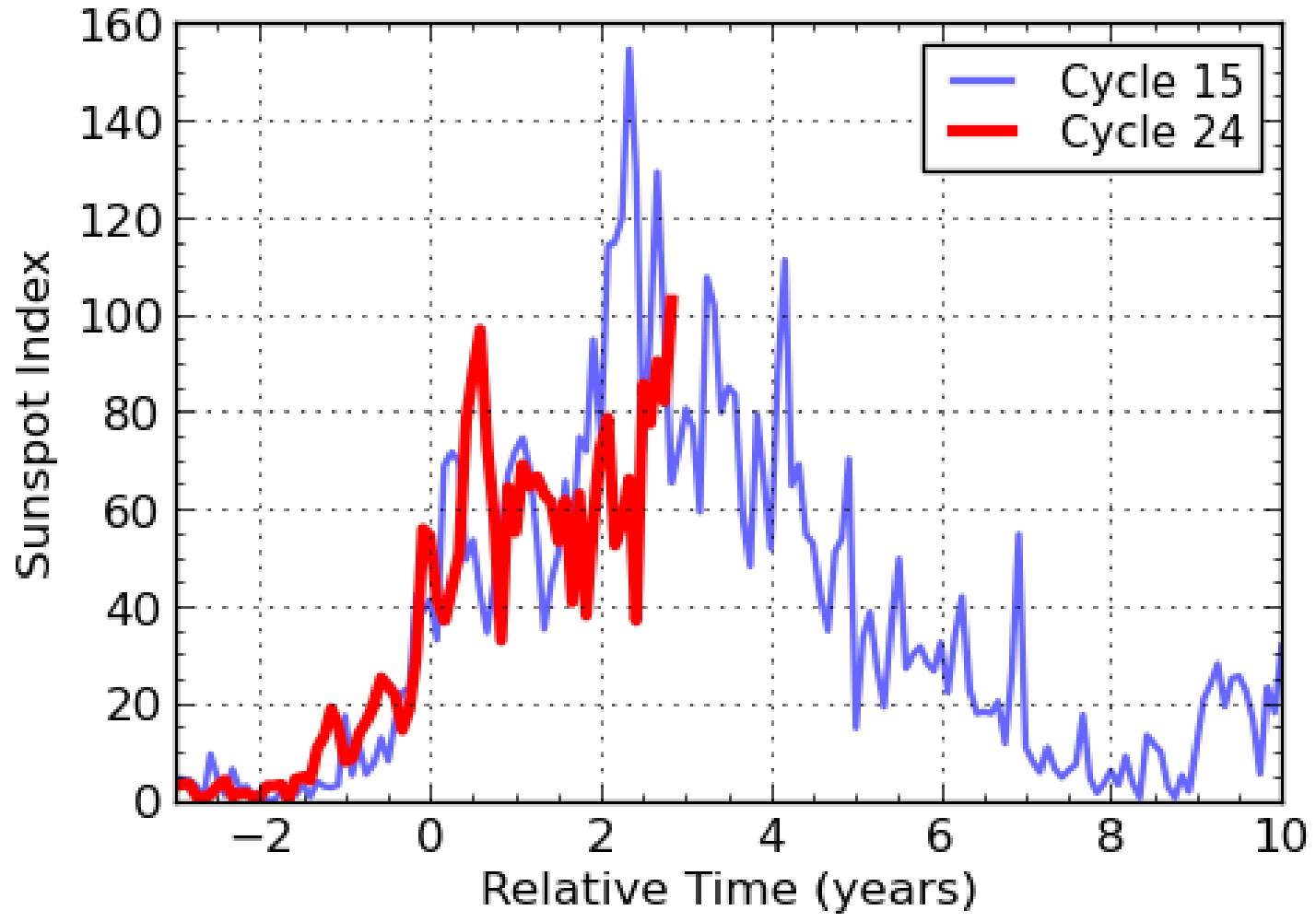

Figure 69: Plot of the monthly mean SN for cycles 15 and 24, after aligning them according to the SN=40 crossing point of the 12-month smoothed SN. Both cycles were characterized by an initial steep rise that is abruptly interrupted and is followed by a long constant plateau lasting for about 18 months. Cycle 15 then entered a second late rise phase to the actual maximum.

The broad sample of solar cycles recorded in the SN series thus brings unique information that should remind us of evolutionary scenarios that should not be overlooked when attempting solar cycle predictions.

# 8. Conclusions

## *8.1. New series and new insights*

In the course of the preceding sections, we tried to show that the sunspot record provided by the SN and GN series is currently the focus of revived interest, far from the frozen image evoked in the introduction. This revival is not only due to a renewed interest in the sunspot record itself, to the availability of newly recovered historical data or to new data processing capabilities. The prime drivers are the recent realization of our very limited capacity to predict the long-term evolution of the solar cycles and the need to quantify the role of the main natural forcing on Earth's climate. Although the solar physics community now clearly identified those important motivations (development of "Space Climate" as a new discipline), the awareness at the level of science funding organizations is still lagging behind. This is why the first initiative aiming at a revision of the SN took the form of Sunspot Number Workshops, an informal and unfunded coordinated effort gathering a community of experts in the field. Let this be our plea for increased support to these fundamental long-term observations and studies of solar and Sun-Earth processes.

Here, we reported about the main advances that were brought by those SN Workshops in just three years, since September 2011. The main biases and drifts w been identified, affecting either the sunspot number or the group number. This first end-to-end recalibration of the SN series since its creation by R. Wolf thus marks an epochal step. It should be pointed out here that only intrinsic diagnostics were used, i.e. analyses based exclusively and directly on the sunspot observations themselves and on original information about the procedures used to create those indices. Comparisons with other parallel and more indirect indices were used when possible but only to provide support and confirmation to the primary diagnostics and not as calibration references. This allowed us to establish the robustness of the diagnosed corrections and to provide an essential validation, given the scarcity of original sunspot data. While the main corrections are now well established, various results discussed in this chapter are still "work in progress". Therefore, some of the discrepancies and puzzling findings identified here point to areas where more effort will still be needed in the future.

Using the new corrections, we show that the sunspot number and group number series can be largely reconciled after applying mutually-independent corrections to each series. Still, discrepancies remain for the early series, before the 19$^{th}$ century. More work is thus still needed to recover new information and improve the interpretation of this early part. Still, as the scaling of the sunspot series is largely established by a backward reconstruction of k personal coefficients, the new improvements to the last two centuries provide a new sounder base for calibrating the early sunspot number, back to the Maunder Minimum and first telescopic observations. New approaches like the "backbone" method presented here can now lean on robust SN values over the last 200 years.

Reconciling the SN and GN series is also essential because almost only group counts can be recovered from the early part of the telescopic sunspot record. The GN series assembled by Hoyt and Schatten will thus remain a reference, though now as the basis for an extension of the SN series for all years before 1750. Here however, we showed that various revisions and corrections are still needed, due to incorrect interpretations of original manuscripts, especially in the early part of the series (e.g., from 1610 to 1750 approximately). In particular, significant revisions based on the work of Vaquero and others are needed for observations within the Maunder Minimum (1645-1715 approximately). We also demonstrated that part of the early sunspot measurements were only meant for the astronomy of position (e.g. solar meridian altitudes), making accurate sunspot counts from such measurements (even spotless dates) extremely difficult. In fact, a "modern" example (using data recorded in the 19$^{th}$ century) has shown that the solar meridian altitudes should be used with extreme caution for the reconstruction of solar activity.

Regarding the impact of the new SN recalibration, the most prominent implication is the significant reduction of the upward trend in the average amplitude of solar cycles that was present in the original GN series between the 18$^{th}$ and the late 20$^{th}$ century. The recalibrated series indicates that 11-yr peak SNs during the 18$^{th}$ and 19$^{th}$ century were comparable to those observed during the recent interval of strong activity during the second half of the 20$^{th}$ century. The scenario of the initial post-Minimum recovery is still uncertain, as the exact amplitude of the first cycles of the 18$^{th}$ century remains difficult to establish given the scarcity of observations over that period. Still, the vanishing upward trend over the last 250 years questions the existence of a modern "Grand Maximum" in the 20$^{th}$ century (Solanki et al. 2004, Abreu et al. 2008, Usoskin et al. 2012, 2014), which resulted primarily from the erroneous transition between Wolf and Wolfer in the Hoyt and Schatten GN time series. As this "Grand Maximum" concept rests on the occurrence of out-of-range amplitudes of the solar cycle, it is definitely contradicted by the re-calibrated and reconciled SN and GN series.

Still, although the levels of activity were not exceptional except maybe for cycle 19, the particularly long sequence of strong cycles in the late 20$^{th}$ remains a noteworthy episode. Indeed, the 400-year sunspot record and one of its by products, the number of spotless days, show that such a tight sequence of 5 strong cycles over 6 successive cycles (from 17 to 22, except 20), which we can call the "Modern Maximum", is still unique over at least the last four centuries. Given the inertia of natural systems exposed to the solar influences, like the Earth atmosphere-ocean system, this cycle clustering could still induce a peak in the external responses to solar activity, like the Earth climate. However, we conclude that the imprint of this Modern Maximum (e.g. Earth climate forcing) would essentially result from time-integration effects (system inertia), since exceptionally high amplitudes of the solar magnetic cycle cannot be invoked anymore. In this suggested revision, the estimated or modeled amplitude of the effects, including the response of the Earth environment, can be quite different, necessarily smaller, and should thus be re-assessed.

The recalibrated series may thus indicate that a Grand Maximum needs to be redefined as a tight repetition/clustering of strong cycles over several decades, without requiring exceptionally high amplitudes for those cycles compared to other periods.

The residual differences between the SN and GN may be considered as random noise. However, based on the abundant sunspot data from recent years, recent statistics indicate that a substantial change in the sunspot population occurred during cycle 23 and continues into the current cycle 24. This change is reflected by a 30% decrease of the average number of sunspots per group. This recent and well documented evolution is concomitant with parallel changes in other solar proxies and suggests that in different regimes of activities, the mutual relation between solar reference indices and proxies may vary. Therefore, the ongoing changes in sunspots invite us to consider the possibility that past variations in the ratio between the SN and GN may reflect true changes in the solar dynamo. This also warns us that many standard proxies used nowadays and mostly based on the last 3 to 5 cycles may not be valid for earlier times and should not be simply extrapolated from recent solar conditions.

Finally, the sunspot number series also provides us a direct picture of the actual evolution of a large sample of past cycles. A direct comparison of the current cycle 24 to equivalent mid-amplitude cycles, indicate that this evolution can strongly differ from the smooth profiles typically delivered by statistical or physical models of the solar cycle. Prominent features like multiple surges in a broad maximum or a two-step rise should not be overlooked when attempting forecasts and should ultimately be reproduced by theoretical models, before we can claim any reliability in future cycle forecasts. In that sense, the sunspot number series does not only provide a numerical reference for the global analysis of past cycles, it also feeds the current and future quest of new concepts to explain the complexity of the solar dynamo. Moreover, by providing a direct measure of sunspot activity spanning multiple centuries, the sunspot number provides a link to millennia of cosmogenic nuclide data. If we are able to understand the relationship between these time series for the last 400 years, then we have a potential key to unlock a much longer record of solar activity.

## *8.2. Future work and prospects*

Now, in order to move forward and assemble a fully revised and extended SN series, three main steps are required:

- Combining all corrections obtained independently and verifying the global consistency of the resulting series.
- Determining more accurately the magnitude and time domain of each correction, and deriving estimates of the uncertainties.
- Extending the original SN series, so far limited to 1750, back to the first telescopic observations. As pointed out earlier, the scarcity of data will most probably impose the use of the group number for this early part. It is thus essential that the GN and SN series are brought first in full agreement for the whole interval following 1750.

As pointed out on several occasions in this chapter, several issues remain open and require deeper analyses that may still span many years. The revised series will thus be open to future improvements as new results are published and new historical documents are progressively recovered. Therefore, in order to properly document future occasional modifications, the WDC-SILSO will implement a versioning system, with an incremental description of changes added to each version.

As the recovery of past observations remains a critical element for future progress, we issue here a general call: readers and their hosting institutions who have raw sunspot observations in their archives are invited to send copies (or even originals if their local preservation is not guaranteed) either to the WDC-SILSO (F. Clette; http://www.sidc.be/silso) or to the newly created Historical Archive of Sunspot Observations (HASO) at the Universidad de Extremadura (J.M. Vaquero; http://haso.unex.es). In particular, past reports or correspondence with original sunspot data that were sent to Wolfer or Waldmeier at the Zürich Observatory are of crucial importance.

Looking further into the future, the next step will be to complement the SN series, which only provides 1-dimensional temporal information, with spatial information though probably over a more restricted period starting only in the mid-18$^{th}$ or early 19$^{th}$ century. This could include the global distributions in latitude (hemispheric separation) and sizes, the topology of the magnetic dipoles (width, tilt), and properties of individual groups (growth, decay, morphology). Recent reconstructions of the Butterfly diagrams based on original Staudach and Schwabe drawings by Arlt (Arlt 2009a, Arlt and Abdolvand 2011, Arlt et al. 2013) are tracing the way for future such analyses. New recovered information on sunspot positions in 18$^{th}$ and 19$^{th}$ centuries by Arlt (2009b), Cristo et al. (2011) and Casas and Vaquero (2014) should soon allow further extensions.

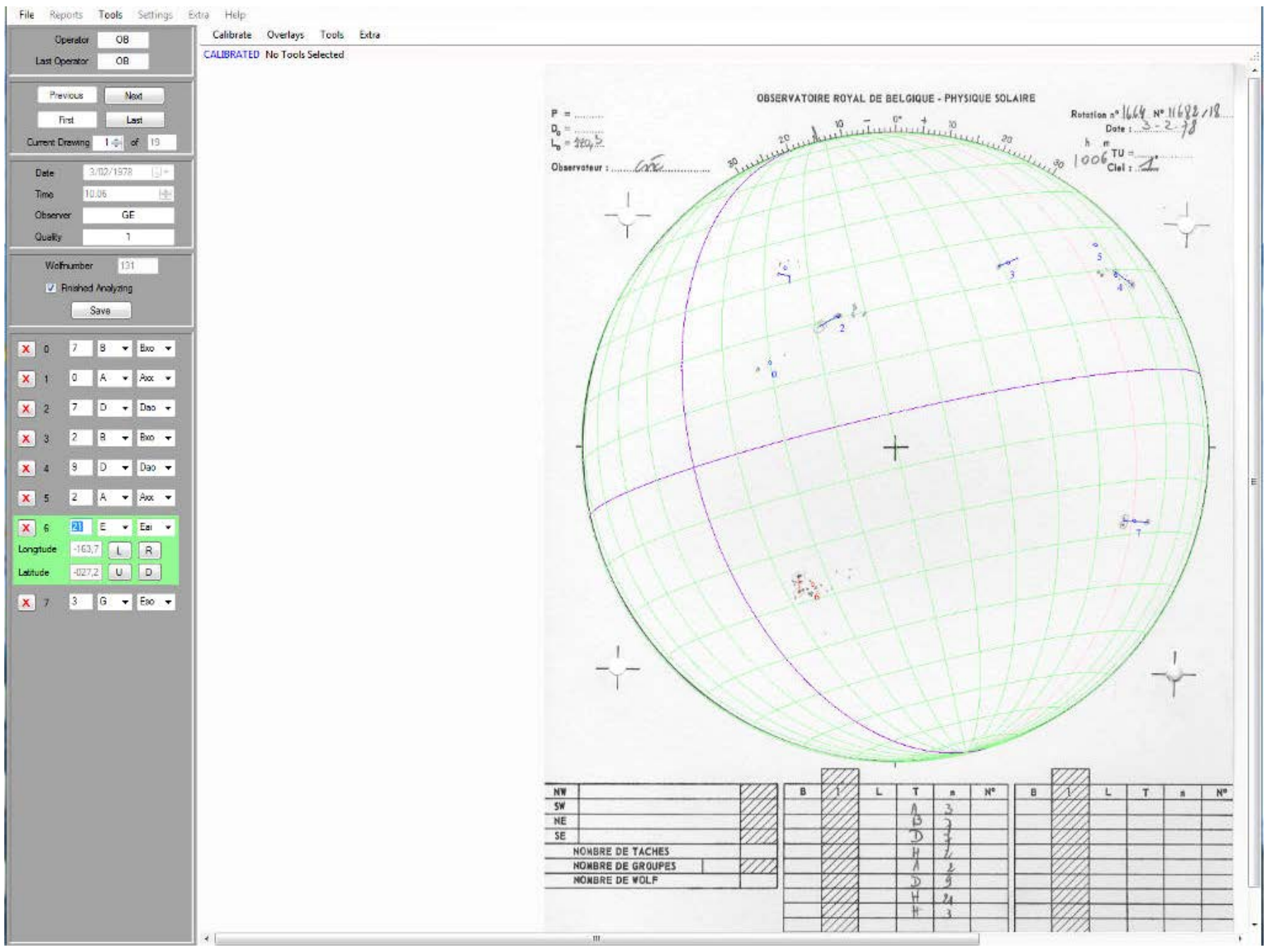

Figure 70: Screenshot showing the main user interface of the DigiSun application developed at the WDC-SILSO (Brussels) for the measurements of individual sunspot groups in sunspot drawings. This tool will lead to a new detailed sunspot catalog spanning the last 70 years, which is currently under construction.

The base data exist (drawings, photographic images) although they are dispersed and often not accessible in digital form. Recently, prototypes of sunspot drawing analysis tools have been developed (e.g. Cristo et al. 2011, Tlatov et al. 2014; DigiSun, Figure 70). Therefore, a major endeavor for the coming years will be the bulk digitization of thus-far unexploited observations and the development of automated feature extraction software. The ultimate goal will be the construction of standardized long-term catalogs of individual sunspots and/or sunspot groups. Such catalogs would allow the creation of new advanced sunspot-based indices and proxies, extending back well before the 20$^{th}$ century.

Thanks to the diversity of the recovered information, various application-oriented indices could be envisioned, complementing the base SN standard reference, for e.g., improved solar irradiance reconstructions or the identification of precursors of extreme solar activity events. Therefore, historical sunspot observations will thus continue to play a central role for our understanding and our forecasting capability of the long-term solar activity and Sun-Earth relations.

**Acknowledgements**

Part of the work contributed here by F.Clette was developed in the framework of the TOSCA project (ESSEM COST action ES1005 of the European Union; http://lpc2e.cnrs-orleans.fr/~ddwit/TOSCA/Home.html) , of the SOLID project (EU 7th Framework Program, SPACE collaborative projects; http://projects.pmodwrc.ch/solid/) and of the Solar-Terrestrial Center of Excellence (http://www.stce.be). F.Clette would like to acknowledge the personal contributions from Laure Lefèvre and Laurence Wauters.

L. Svalgaard acknowledges support from Stanford University.

J.M. Vaquero acknowledges support from the Junta de Extremadura (Research Group Grant No. GR10131) and the Ministerio de Economía y Competitividad of the Spanish Government (AYA2011-25945).

The authors also wish to thank Sergio Cortesi, Marco Cagnotti and Michele Bianda for their hospitality and help for our various investigations on the Locarno and Zürich observations.

The Sunspot Number Workshop series has benefitted from the support of the Royal Observatory of Belgium, the National Solar Observatory, Stanford University, the Air Force Research Laboratory and the Specola Solare Ticinese.

Finally, the authors are also grateful to Rainer Arlt for his detailed reviewing work and constructive comments.